\documentclass[final,5p,times,twocolumn]{elsarticle}

\usepackage{graphicx}
\usepackage{amssymb,amsmath}
\usepackage[stable]{footmisc}
\usepackage{float}

\newcommand\textlcsc[1]{\textsc{\MakeLowercase{#1}}}

\journal{Astroparticle Physics}

\begin{document}

\begin{frontmatter}

\title{Monte Carlo simulations of alternative sky observation modes with the Cherenkov Telescope Array}

\author{M.~Szanecki\corref{cor1}}
\ead{mitsza@uni.lodz.pl}

\author{D.~Sobczy\'nska}
\ead{dsobczynska@uni.lodz.pl}

\author{A.~Nied\'zwiecki}
\ead{niedzwiecki@uni.lodz.pl}

\author{J.~Sitarek}
\ead{jsitarek@uni.lodz.pl}

\author{W.~Bednarek}
\ead{bednar@uni.lodz.pl}

\cortext[cor1]{Corresponding author}

\address{Department of Astrophysics, University of {\L}\'od\'z, Pomorska 149/153, PL--90--236 {\L}\'od\'z, Poland}

\begin{abstract}
We investigate possible sky survey modes with the Middle Sized Telescopes (MST, aimed at covering the energy range from $\sim$100 GeV to 10 TeV) subsystem of the Cherenkov Telescope Array (CTA). We use the standard CTA tools, {\fontfamily{pcr}\selectfont CORSIKA} and {\fontfamily{pcr}\selectfont sim\_telarray}, to simulate the development of gamma-ray showers, proton background and the telescope response. We perform simulations for the H.E.S.S.\--site in Namibia, which is one of the candidate sites for the CTA experiment. We study two previously considered modes, parallel and divergent, and we propose a new, convergent mode with telescopes tilted toward the array center. For each mode we provide performance parameters crucial for choosing the most efficient survey strategy. For the non-parallel modes we study the dependence on the telescope offset angle. We show that use of both the divergent and convergent modes results in potential advantages in comparison with use of the parallel mode. The fastest source detection 
can be achieved in the divergent mode with larger offset angles ($\sim 6^{\circ}$ from the Field of View centre for the outermost telescopes), for which the time needed to perform a scan at a given sensitivity level is
shorter by a factor of $\sim$2.3 than for the parallel mode.
We note, however, the direction and energy reconstruction accuracy for the divergent mode is even by a factor of $\sim 2$ worse than for other modes.  Furthermore, we find that at high energies and for observation directions close to the center of the array field of view, 
the best performance parameters are achieved with the convergent mode, which favors this mode for deep observations of sources with hard energy spectra.
\end{abstract}


\end{frontmatter}

\section{Introduction}
\label{sec:intro}
Imaging Air Cherenkov Telescopes (IACT) detect gamma rays using the Cherenkov images of their electromagnetic showers developing in the atmosphere. The IACT technique has rapidly progressed over the last 20 years (see, e.g., a review in  \citep{buckley08}) and, with the current generation of IACT instruments \citep{Aleksic11,Hofmann00,Veritas08}, it is now the most accurate and sensitive detection technique in the very high energy gamma-ray astronomy.
The  Cherenkov Telescope Array (CTA), the next generation of IACT detectors currently in final stages of design, is expected to improve the sensitivity of present observatories by an order of magnitude, covering the energy range from a few tens of GeV to hundreds of TeV \citep{cta}. 
The experiment will consist of two arrays, one in the northern and one in the southern hemisphere, each including a large number of telescopes ($\sim 30$ for northern and $\sim 100$ southern hemisphere),
with a large field of view (FOV) of $5^{\circ}$--$10^{\circ}$ (depending on the telescope type).
This will allow for several schemes of observation:
\begin{itemize}
\item deep observations -- all telescopes pointed onto one object (intensive data taking);
\item normal observations and monitoring -- a few telescopes oriented towards each of several potentially interesting sources;
\item sky scans -- all telescopes scan a large area of sky in the long-term observations to detect new or transient sources.
\end{itemize}
In this work we thoroughly investigate the last one, i.e.\ sky scans. 

On the experimental ground, scans were first performed by 
the HEGRA system of IACTs in search for a TeV gamma-ray signal from one quarter of the Galactic plane \citep{Aharonian2001, Aharonian2002}. At the same time, the Whipple telescope collected data from nightly calibration scans which covered a sky region  $12.5^{\circ}$-wide in declination  \citep{Kertzman2008}.
Neither of these observations detected TeV emission. Nevertheless upper limits were derived for a large number of sources and  the sky-scan techniques and data analysis methods were developed. 
Later, during the 1400 h-long  survey of the Galactic plane conducted by the H.E.S.S.\ telescope system \citep{Aharonian2005, Aharonian2006, Renaud2009}, over a dozen new sources were detected \citep{Chaves2009}. 

For CTA, an improved  Galactic plane survey should be a major objective and it will also be capable of performing an all-sky survey in unprecedentedly short time at high sensitivity; the scientific rationale and feasibility of both survey types are thoroughly discussed in \citep{Dubus2013}. As also discussed in \citep{Dubus2013}, such surveys can be performed in various modes of observation, in particular,
large number of high-performance IACTs allows for using non-parallel modes with an enlarged FOV. The proper adaptation of such a  mode for a specific telescope array can be a non-trivial task. The optimization of the pointing strategy, taking into account numerous characteristics of an array, e.g.~distance between telescopes, FOV, energy threshold etc,   can significantly reduce the observation time  needed to achieve a given sensitivity.

In this work we consider the array of middle sized telescopes (MST) working in various, parallel and non-parallel,  modes. By performing high-statistics Monte Carlo (MC) simulations of the sky-survey observations, we derive  for each mode the basic performance parameters at both trigger and analysis levels, which then allow us to compare efficiencies of the modes.
Our study is a part of an intensive work within the CTA Monte Carlo Work Package aimed at optimizing the CTA observation scheme. Whereas we consider in detail different modes with the MST array, independent investigations are currently performed for the divergent mode of Large Sized Telescopes (LST) sub-array and the full CTA array working in divergent modes.

\section{Sky survey modes}
\label{sec:modes}

\begin{figure}[t!]
  \begin{center}
    \includegraphics[trim = 3mm 0mm 0mm 70mm, clip,scale=0.25]{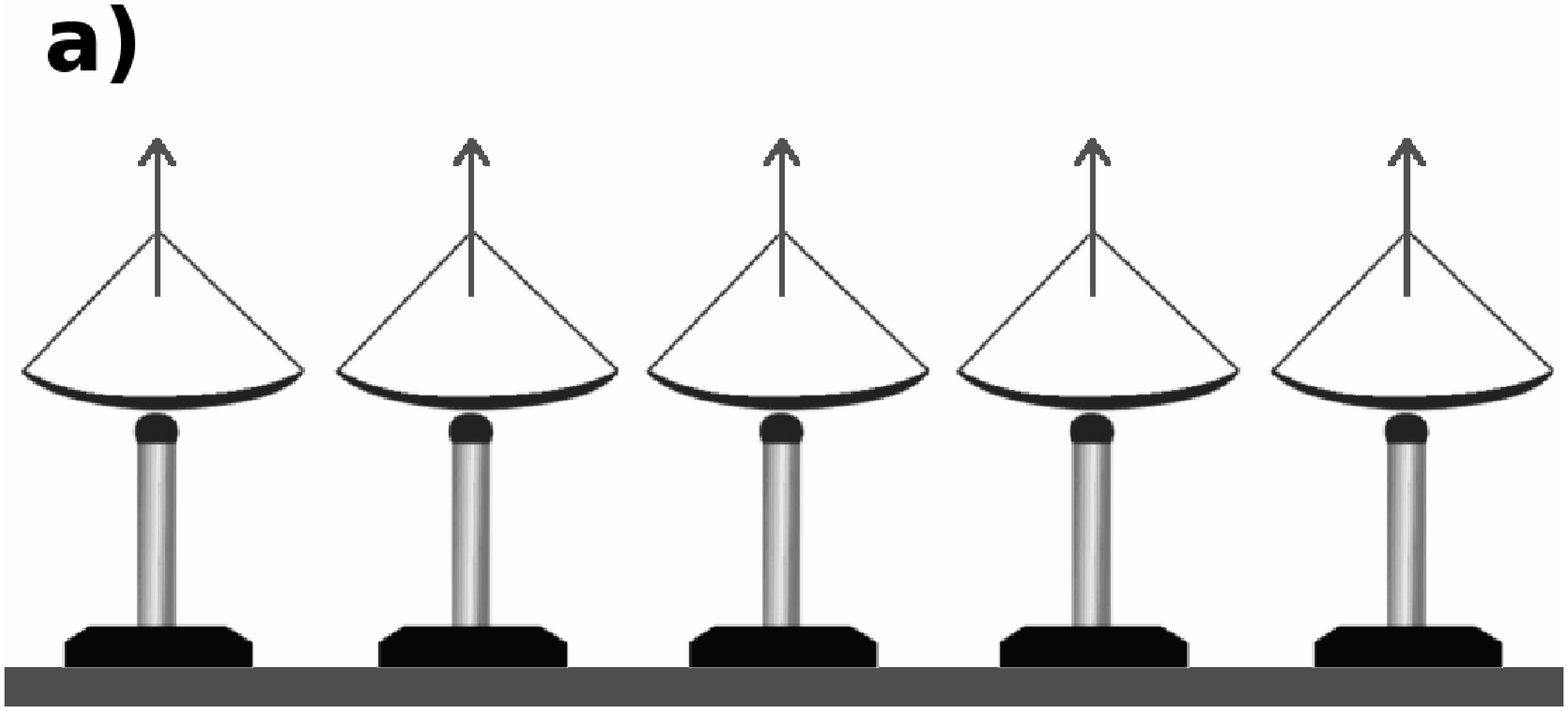}
    \includegraphics[trim = 3mm 0mm 0mm 70mm, clip,scale=0.25]{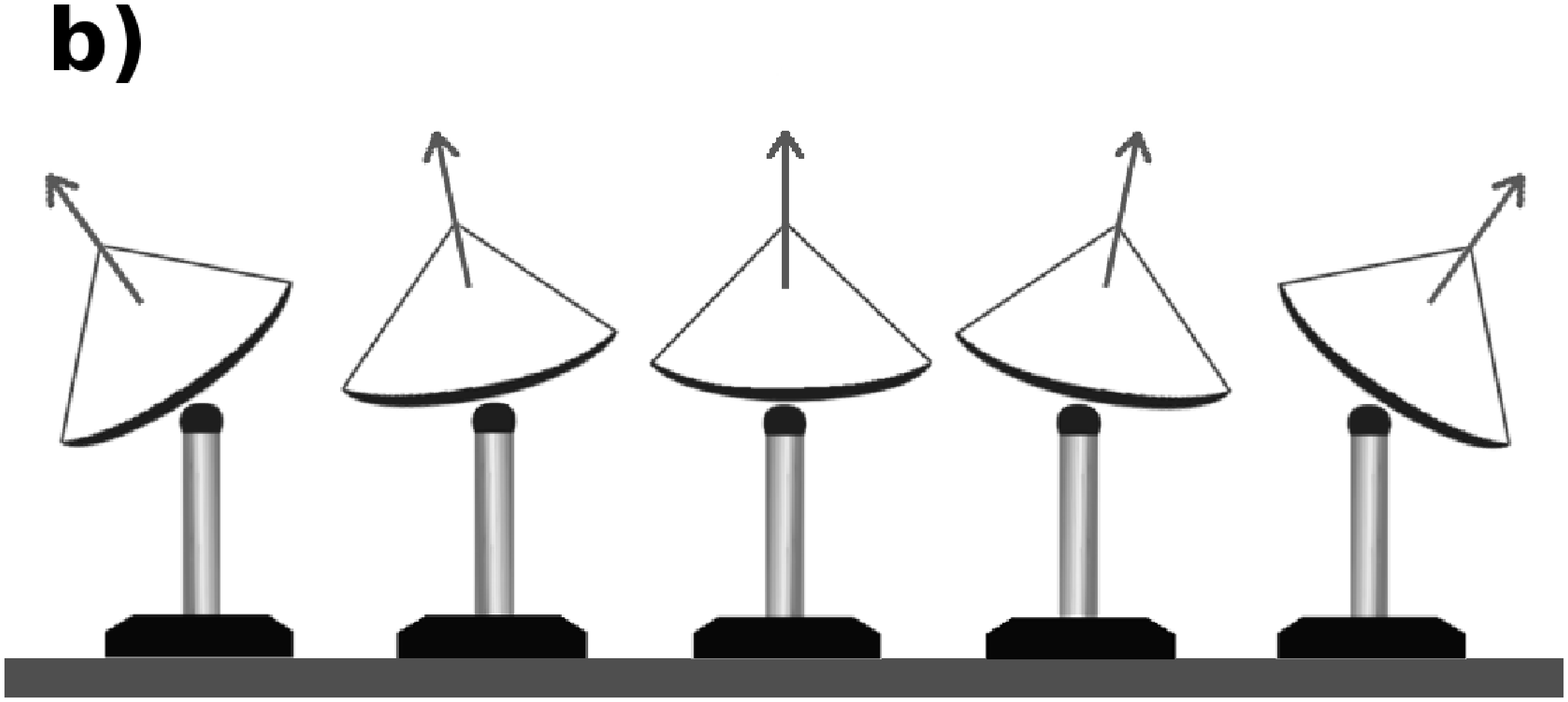}
    \includegraphics[trim = 3mm 0mm 0mm 70mm, clip,scale=0.25]{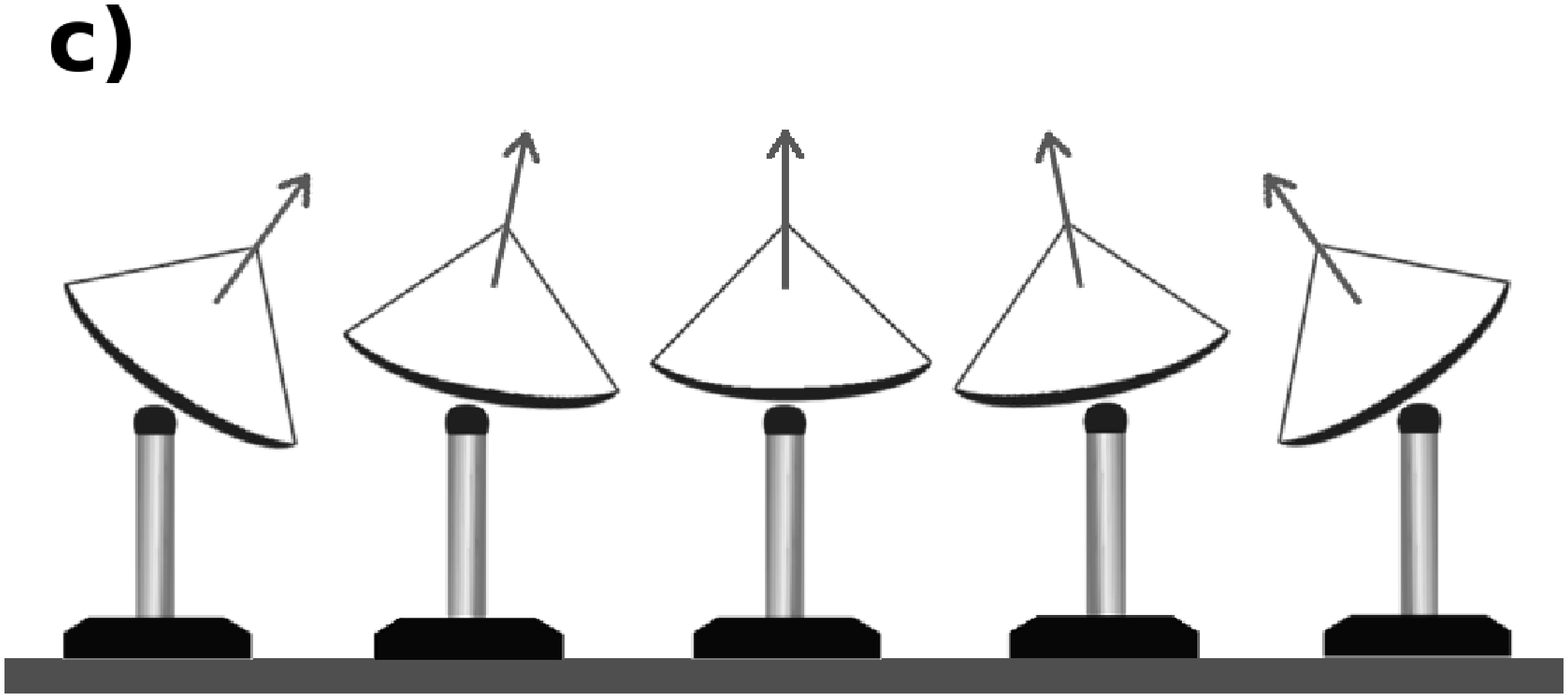}
    
    \caption{Three modes of configuration of the telescope system used in the sky-survey scans: a) normal (parallel) mode; b) divergent mode; c) convergent mode.}
    \label{fig:modes}
  \end{center}
\end{figure}

Fig.~\ref{fig:modes} illustrates possible modes for a large telescope array used for sky surveys. The parallel and divergent configurations were considered before in \citep{Dubus2013}; below  we introduce also a novel, convergent mode (note the difference between our terminology and that of \citep{Dubus2013}, were the parallel mode is referred to as convergent).

The performance of a telescope system operating in the sky survey mode depends on the FOV of the system and the time of observation needed to achieve 
a given significance level, i.e. its sensitivity.

In the simplest approach, sky surveys may be performed with telescopes pointed parallely into the same direction of the sky (Fig.~\ref{fig:modes}a), however, in such a case the FOV of the telescope system is highly limited by the FOVs of individual telescopes.
The FOV of a telescope array can be significantly enlarged by slightly deviating the pointing direction of each telescope. In the divergent mode, telescopes are inclined into the outward direction, see Fig.~\ref{fig:modes}b, by an angle increasing with the telescope distance from the array center. As explained below, a performance improvement for such a configuration can be expected primarily at high energies of primary photons.

For the divergent configuration, images of gamma rays impinging close the array centre are shifted toward the camera edge, which leads to a leakage\footnote{The effect of cutting off an image at the camera edge.} or complete loss of an event.  While the larger loss of events is mostly pronounced for the lower-energy gamma rays, the leakage effect concerns mainly events with higher energies. As a result even if an event is registered it is  poorly reconstructed. On the other hand, orientation of telescopes in the divergent mode is suitable for efficient detection of events with large impact parameter and/or  arriving from directions further from the FOV center  (in both cases mainly with high energies).

Qualitatively, one can expect that those negative effects can be reduced for the opposite orientation, i.e.\ with outer telescopes inclined  toward the array center, see Fig.~\ref{fig:modes}c. 
A quantitative comparison of the performance of the three modes and a related issue, i.e.\ an optimal value of the offset angle (giving the amount of the difference of the pointing directions, as defined below), appears crucial for planning the most efficient survey strategy.

\section{MC simulations}
\label{sec:mc}
For all three modes, we simulate the response of the telescope array to the Extensive Air Showers (EAS) induced by gamma rays and proton background. To simulate the development of EAS we use {\fontfamily{pcr}\selectfont CORSIKA 6.99} code \citep{Heck98,Bernlohr08}, used as a standard in CTA. We simulated $2.1\times 10^{7}$ gamma rays and $3.8\times 10^{8}$ proton events\footnote{including the number of re-used showers} - both with energies between 30 GeV and 10 TeV generated from differential spectra with the spectral index $\Gamma = -2.0$. However, in our analysis, we use event weights corresponding to spectra with $\Gamma = -2.57$ for gamma rays and $\Gamma = -2.73$ for protons. Gamma rays are simulated from a point-like test source with the direction defined by the Zenith angle Za=$20^{\circ}$  and the Azimuth Az=$180^{\circ}$ measured with respect to the magnetic North. The  background proton showers are simulated isotropically with directions within a $10^{\circ}$ half-angle cone (larger than the FOV of all considered modes) centered on the direction of the gamma-ray source. We set the maximum impact parameter for gamma rays to 1000 m and for protons to 1500 m. The detector array is assumed to be located at the Namibian (H.E.S.S.) site at the altitude of 1800 m a.s.l.
\begin{figure}[t]
\begin{center}
  \includegraphics[trim = 10mm 17mm 10mm 15mm, clip,scale=0.4]{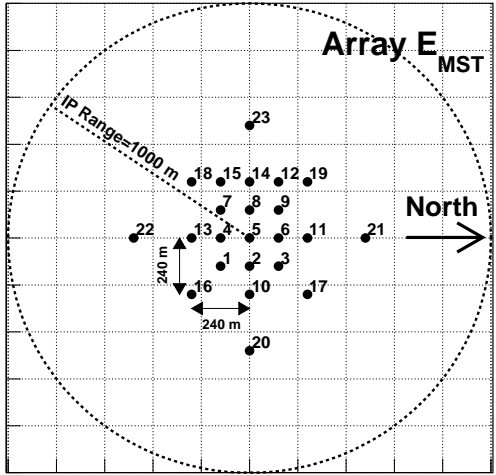}

  \caption{Geometric layout of telescopes used in our simulations. The circle shows the simulated area defined by the maximum impact parameter for gamma rays.}
\label{fig:array}
\end{center}
\end{figure}

\begin{figure}[t]
\begin{center}
  \scalebox{1.75}{%
    \includegraphics{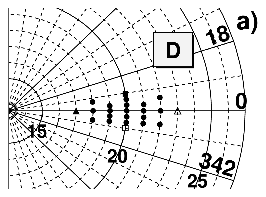}}%
  \scalebox{1.75}{%
    \includegraphics{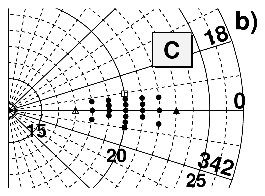}}
  \scalebox{1.75}{%
    \includegraphics{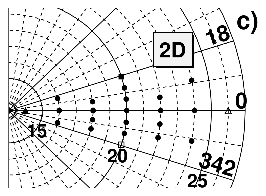}}%
  \scalebox{1.75}{%
    \includegraphics{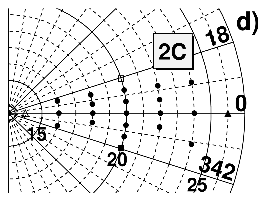}}
\caption{Telescope pointing directions shown in the Zenith (radial) and Azimuth (transversal) polar coordinates for mode D (a), C (b), 2D (c) and 2C (d). For each mode, telescopes no.\ 20 to 23 are indicated with different markers for better illustration of the difference between the divergent and convergent modes.}
\label{fig:array_angle}
\end{center}
\end{figure}

The response of the telescope array is simulated with the CTA {\fontfamily{pcr}\selectfont sim\_telarray} code \citep{Bernlohr08, BernlohrCTA08}. We use the MST subarray of the CTA array E from the so-called {\em production-1}; the subarray includes 23 telescopes with positions shown in Fig.~\ref{fig:array}. The direction of the central telescope no.~5 is always approximately in the center of the FOV of the array (a slight displacement may occur due to the presence of telescopes no.~12 and 15, which break the symmetry); then, this direction is used to define various configurations and we refer to it as the axis of the telescope system. The non-parallel modes are defined by altering the pointing-directions of all remaining telescopes  with respect to this axis (outwards for divergent mode, inwards for convergent mode) by an offset angle, $\alpha$, given by\footnote{except for the most distant telescopes 20-23, for which we use $\alpha = 3 k\cdot d_{\rm tel}\cdot r_{\rm tel} / 8r_{\rm max}$.} $\alpha = k\cdot d_{\rm tel}\cdot r_{\rm tel} / 2r_{\rm max}$, where $k$ is the scaling parameter (=1 or 2, see below), $d_{\rm tel}=8^{\circ}$ is the FOV diameter of a single telescope, $r_{\rm tel}$ is the distance of the telescope from the array center and $r_{\rm max}=480$ m is the largest distance (i.e.\ the one of telescopes no.~20--23). For both non-parallel modes we consider two scales of the angular spread between telescopes, with $k=1$ and $k=2$, referred to as single and double scaled modes, respectively. Then,  in total we consider five configurations: the normal mode ({\bf N}), the single-scaled divergent mode ({\bf D}), the double-scaled divergent mode ({\bf 2D}), the single-scaled convergent mode ({\bf C}) and the double-scaled convergent mode ({\bf 2C}). For example, the telescopes no.~$6$ (with $r_{\rm tel} = 120\;{\rm m}$) and no.~$9$ (with $r_{\rm tel} =170\;{\rm m}$) are offset by $\alpha = 1^{\circ}$ and $1.4^{\circ}$, respectively, for $k=1$ and by $2^{\circ}$ and $2.8^{\circ}$ for $k=2$.  The orientation of telescopes in non-parallel modes is shown in polar coordinates in Fig.~\ref{fig:array_angle}.

We emphasize that the term ''convergent'' used throughout this work refers to pointing scheme where the individual telescope axes cross in an arbitrary point of a shower core. Note that in the C mode the telescope axes converge approximately at a distance of about $7\;{\rm km}$, close to the shower maximum for TeV showers while in the 2C mode the point of convergence is well below the maximum of all showers. 

\begin{table}[t]
\begin{center}
 \begin{tabular}{|c||c||}
  \hline
  {\fontfamily{pcr}\selectfont Sim\_telarray} input & Input value\\
  \cline{1-2}
  \hline \hline
  Telescope type & MST (Type 2)\\
  \hline
  Dish diameter & ${\rm d}=12 \;{\rm m}$\\
  \hline
  Focal length/diameter & ${\rm f/d}=1.3$\\
 \hline
  Camera  Field of view &  ${\rm FoV}=8^{\circ}$\\
  \hline
  Pixel size & $0.18^{\circ}$\\
  \hline
  Photomultipliers & bi-alkali\\
  quantum efficiency &  ${\rm QE}_{\rm peak}=25.7\%$\\
  \hline
  Telescope trigger & Min.\ 4 pe in each of\\
  threshold level & 3 neighboring pixels\\
 \hline
  Min.\ trigger multiplicity & 2 telescopes \\
  \hline
   \end{tabular}
\end{center}
\caption{Parameters assumed in {\fontfamily{pcr}\selectfont sim\_telarray} for our simulations.}
 \label{tab:mc_tel}
\end{table}

The basic technical parameters of the telescopes used in our simulations follow the standard settings of the {\em production-1} (the first CTA MC mass production; see Fig.~18 and Chapt.~8 of \citep{cta} or Chapt.\ 6 of \cite{Bernlohr12}) and are given in Table \ref{tab:mc_tel}. We note that the currently developed MC mass production {\it prod-2} introduces several technical modifications (e.g.~longer read-out intervals and trace integration following the time gradients) which may improve the efficiency of  detection of events with impact parameter above $\sim 400\;{\rm m}$.

\begin{figure*}[t]
\begin{center}
 \scalebox{0.97}{%
  \includegraphics[width=5.5cm,height=5.5cm]{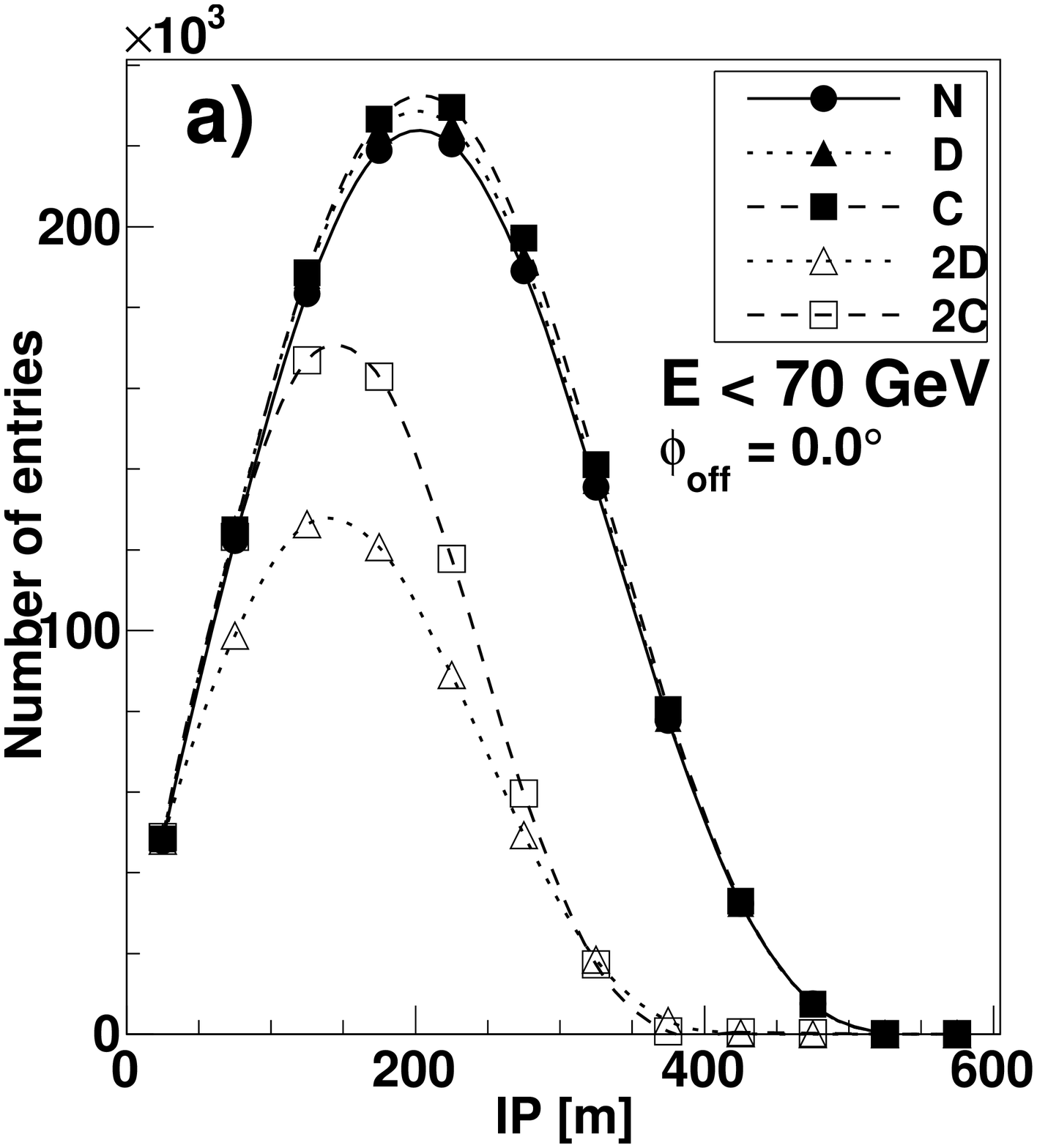}}%
 \scalebox{0.97}{%
  \includegraphics[width=5.5cm,height=5.5cm]{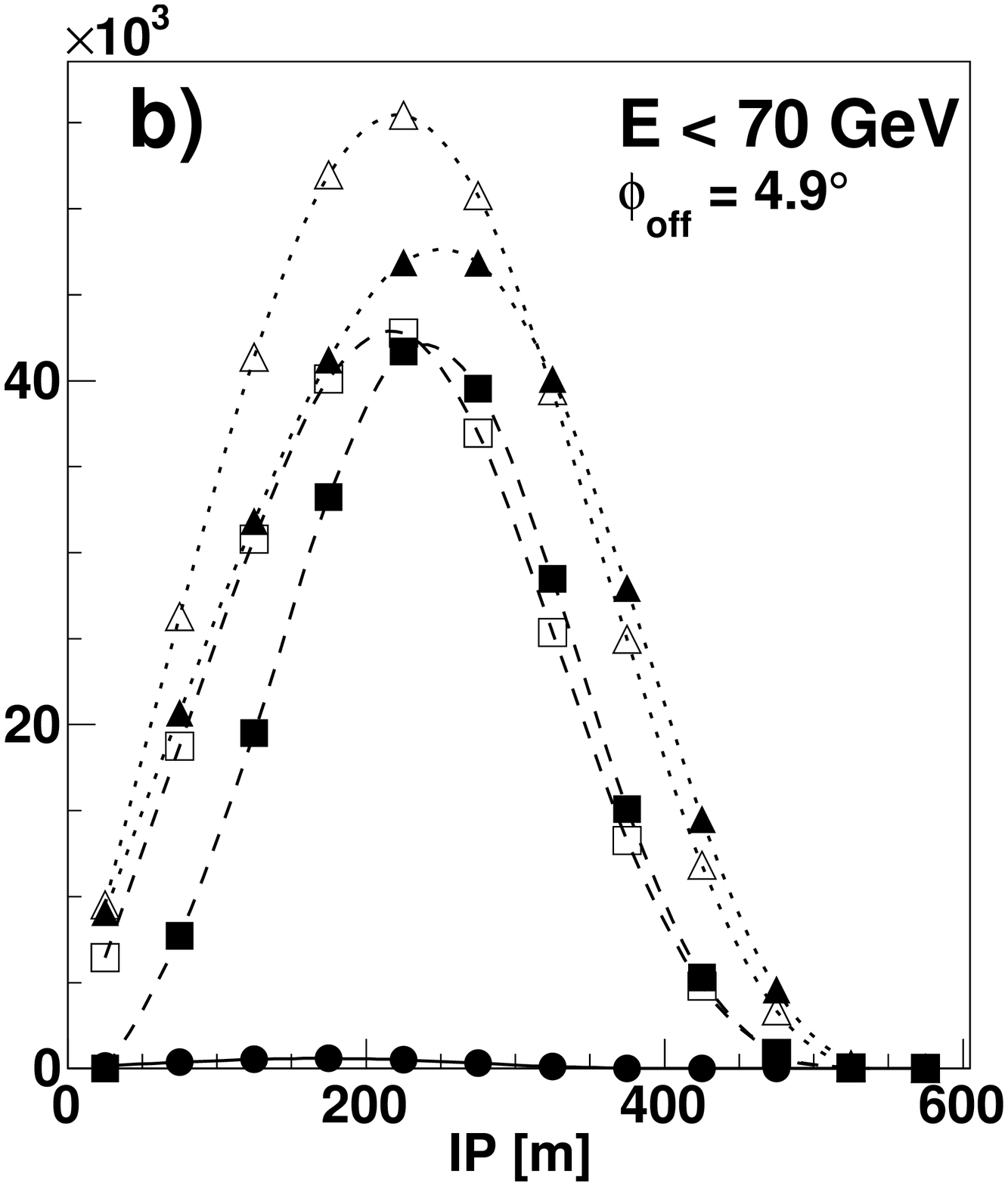}}
 \scalebox{0.97}{%
  \includegraphics[width=5.5cm,height=5.5cm]{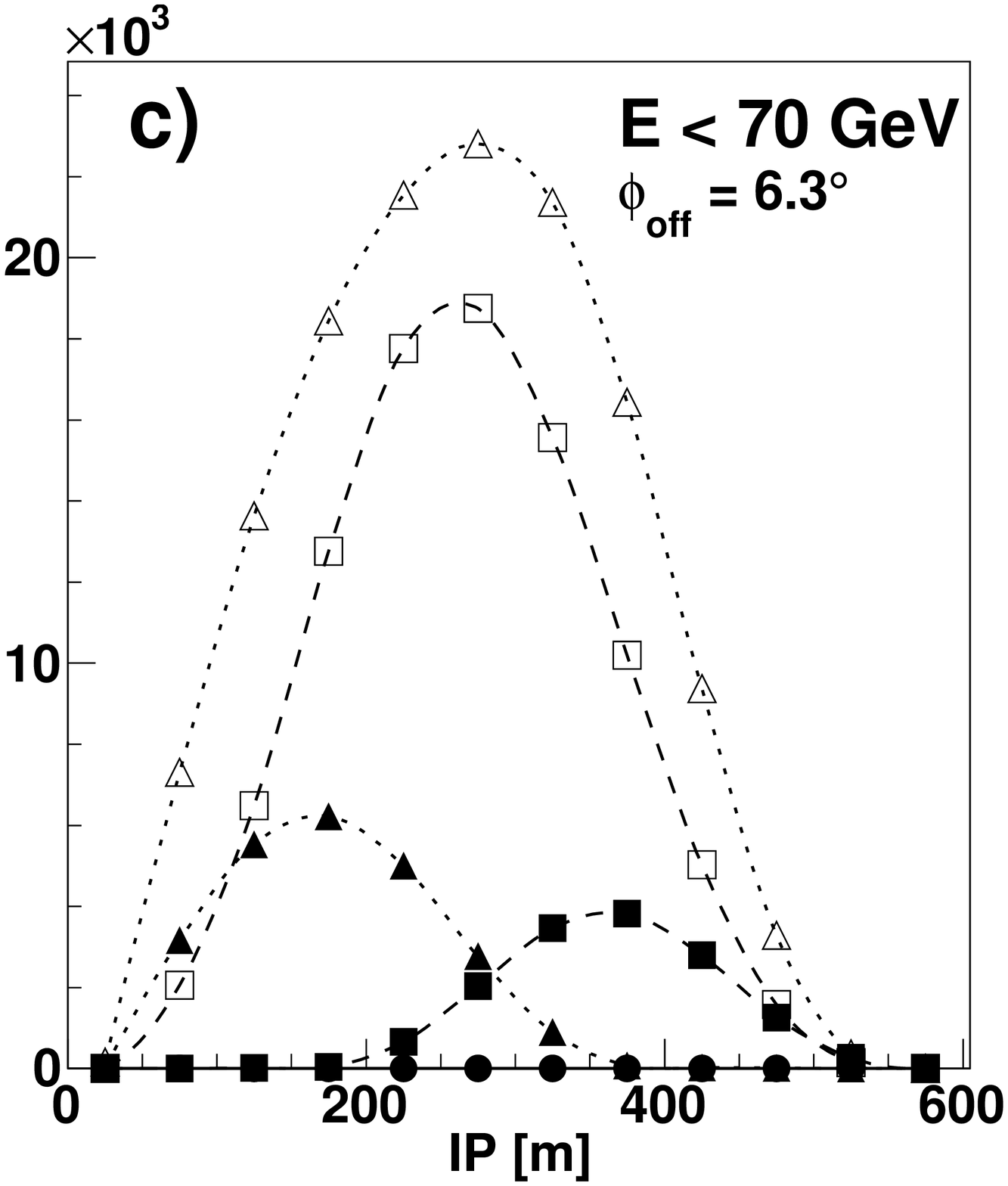}}
 \scalebox{0.97}{%
  \includegraphics[width=5.5cm,height=5.5cm]{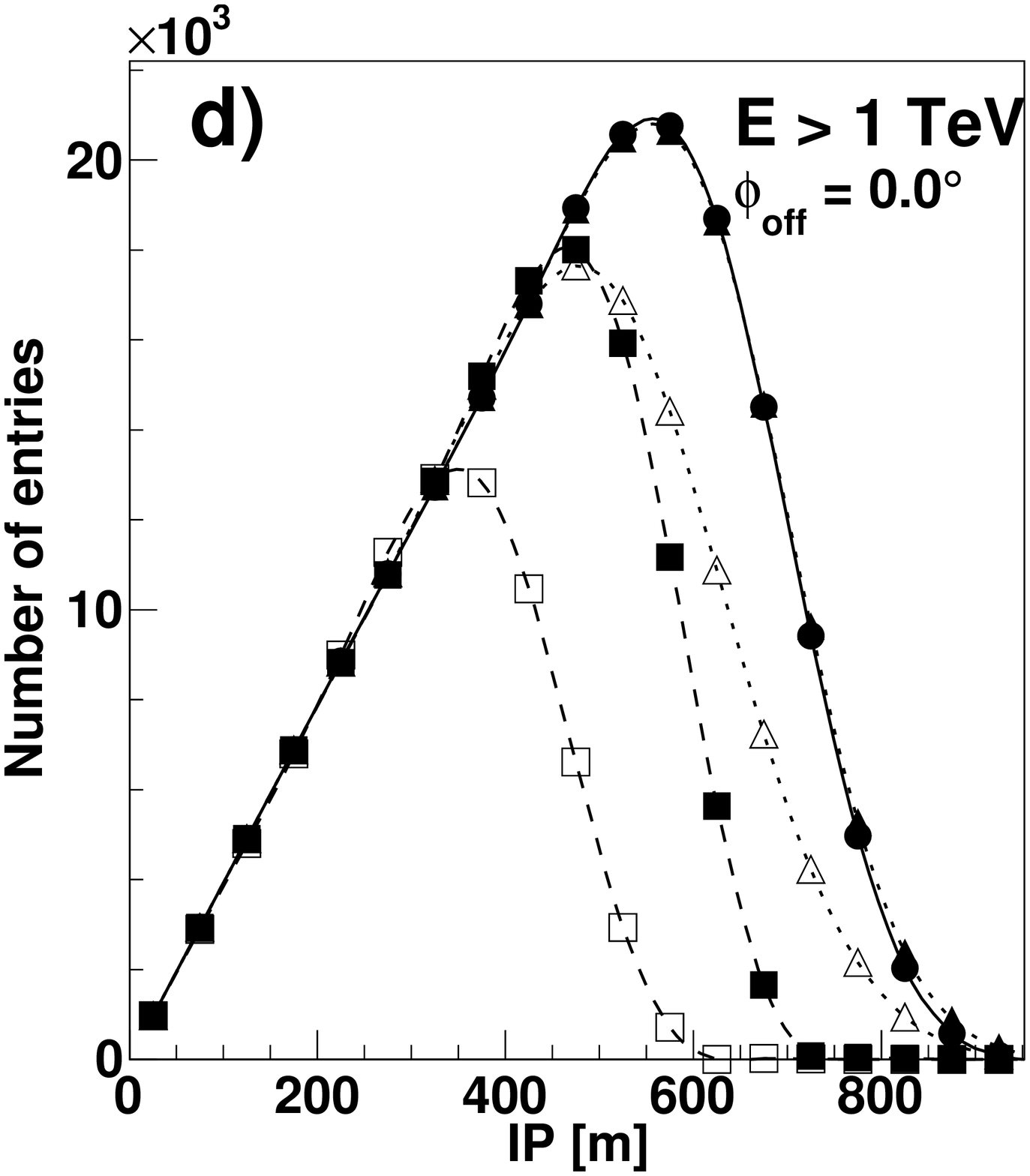}}%
 \scalebox{0.97}{%
  \includegraphics[width=5.5cm,height=5.5cm]{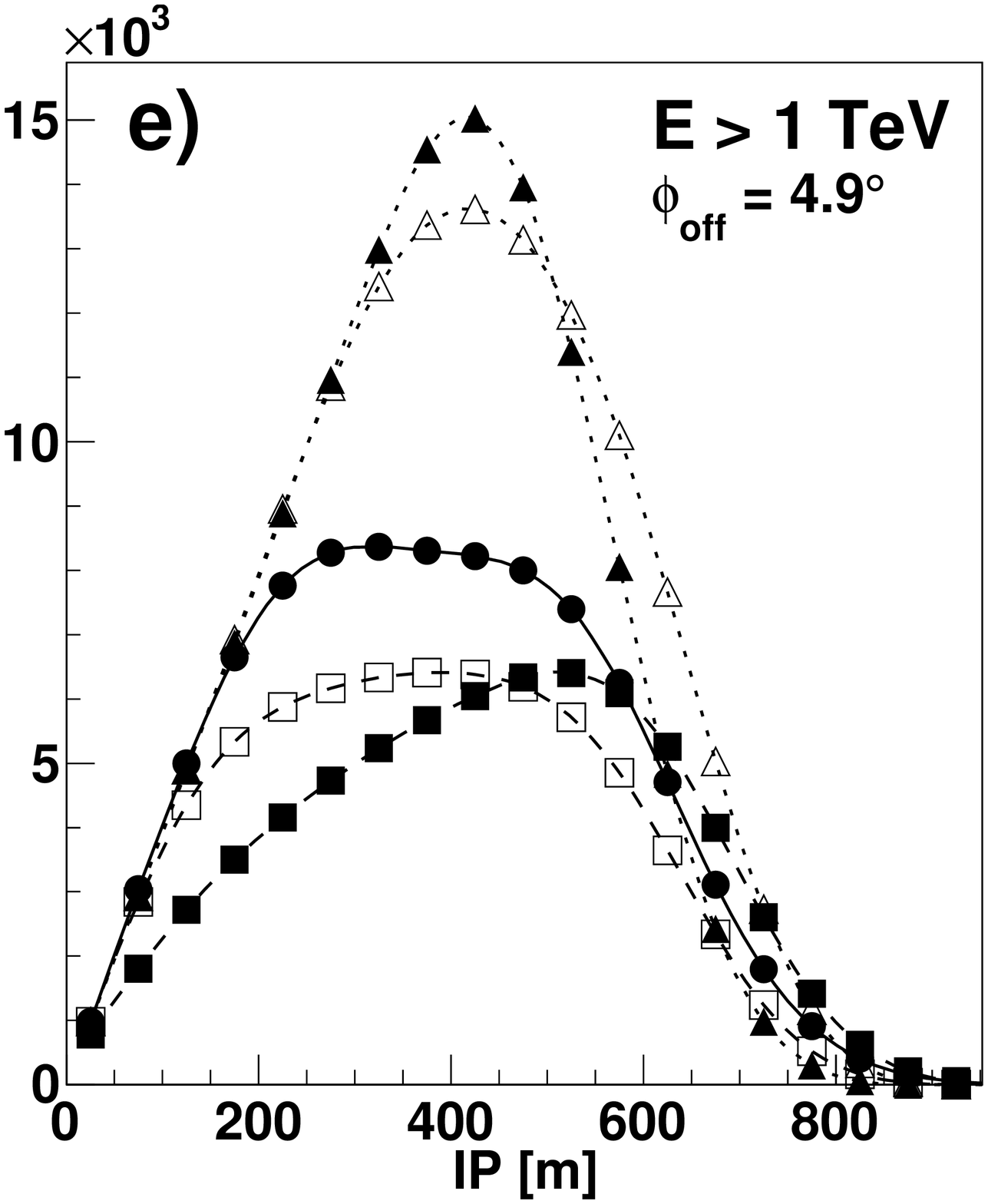}}
 \scalebox{0.97}{%
  \includegraphics[width=5.5cm,height=5.5cm]{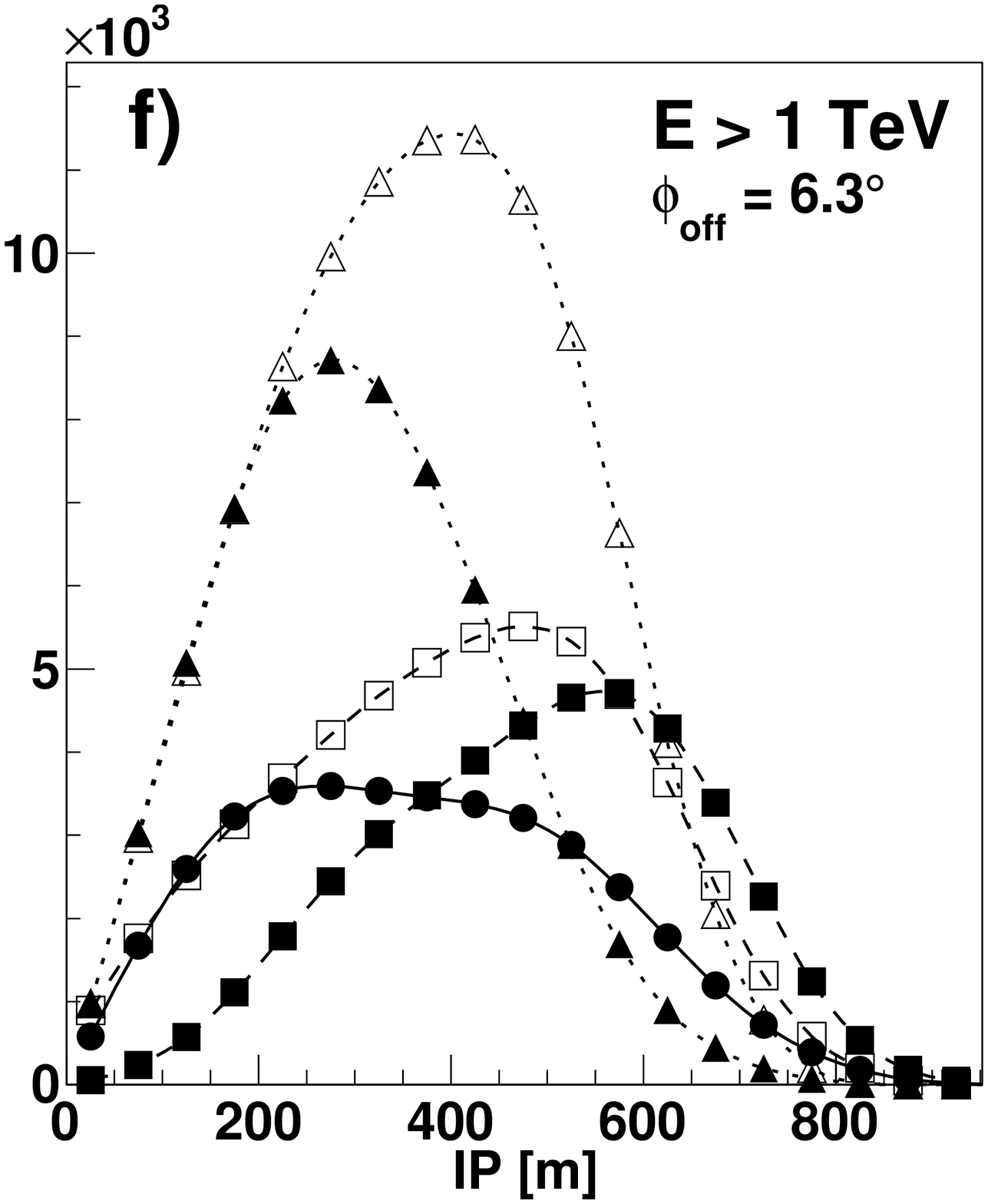}}

\caption{The number of registered gamma-ray events as a function of the impact parameter measured from the array center to the shower core. The top panels are for $E \le 70\;{\rm GeV}$ (the array threshold) and the bottom panels are for $E \ge 1\;{\rm TeV}$. For both energy ranges, $\phi_{\rm off} = 0^{\circ}$ (on-axis),  $4.9^{\circ}$ and $6.3^{\circ}$ from left to right. In each panel the modes are represented by: N - solid lines with filled circles, D - dotted lines with filled triangles, 2D - dotted lines with empty triangles, C - dashed lines with filled squares and 2C - dashed lines with empty squares; the same styles of markers and lines are used for the modes throughout all figures in this paper. In all cases the same number of simulated events were used, so the differences in the number of entries (y-axis) between the modes in each panel correspond to the expected difference of trigger rates.}
\label{fig:impact_energy}
\end{center}
\end{figure*}

Obviously, sky scanning surveys are aimed at the detection of unexpected and non-resolved sources, which may be located anywhere in the scan area. Then, we take into account source locations in various regions of the FOV of the array. To this end we simulate detection of source with different angular distances, $\phi_{\rm off}={\rm f}({\rm Za},{\rm Az})$, from the array axis by moving the FOV of an array with respect to the simulated source direction.
Assuming the rotational symmetry of an array FOV in $({\rm Za},{\rm Az})$ plane, we set Za to zenith angle of the array axis direction and we change only the Az component.
Specifically, we consider test sources with different $\phi_{\rm off}$, with $\phi_{\rm off}$ between $0.0^{\circ}$ and $7.0^{\circ}$ and we use the step of $\Delta \phi_{\rm off} = 0.7^{\circ}$. The case of the source with $\phi_{\rm off} = 0.0^{\circ}$ is referred to as the on-axis source and all cases with $\phi_{\rm off} \ne 0.0^{\circ}$ as off-axis.

Our analysis is based on Hillas image parameters \citep{Hillas85} obtained from {\fontfamily{pcr}\selectfont Sim\_telarray} simulations using the {\fontfamily{pcr}\selectfont read\_cta} code, being a part of the CTA simulation package. Images are cleaned using the tail-cuts of 5.5/11 photo-electrons (pe).

The shower direction and energy reconstruction used in this work follows the CTA standard baseline methods as described in Sect.~3 of \citep{Bernlohr12}. The shower direction is determined with classical stereo reconstruction which uses Hillas image parameters obtained from individual telescopes and then transformed into the common reference frame (CRF) of coordinates. In our approach CRF is defined by the plane perpendicular to the array axis. The use of CRF is a general method yielding proper reconstruction for both the parallel and the non-parallel pointing schemes. The shower parameters, which result from a stereo reconstruction carried out in CRF, then are used in the energy reconstruction. The energy reconstruction, in turn, uses the correlation matrices between energy, image SIZE and reconstructed impact parameter. Parameters of correlation matrices are stored in the so-called lookup tables (LUTs). In this work, to avoid the possible biasing of the results, the energy reconstruction is done in two steps using two separate event samples. First, with the gamma-ray training sample, we create LUTs. Then, the LUTs are used to determine the energy of both gamma rays and protons in the independent test samples. As the stereo parameters stored in LUTs depend on the observing mode, we use LUTs specific for each studied mode.

\section{{\bf Results}}
\label{sec:results}

We compare the basic parameters used to describe the performance of ground-based gamma ray detectors, cf.\ \citep{hinton09}, separately at the trigger and analysis levels, i.e.\ before and after gamma/hadron separation cuts. The analysis level parameters allow for a straightforward comparison of the overall performance of the modes, however, they depend on the chosen gamma/hadron separation method and then they can  be improved if optimized procedures are developed.
In turn, the trigger level parameters provide information independent of the chosen analysis method and then they 
 are directly related to the detector physics.

\subsection{Trigger level parameters}
\label{sec:trigger level parameters}

\subsubsection{Impact parameter distributions}
\label{sec:rates collection}

\begin{figure}[t]
\begin{center}
  \scalebox{0.95}{%
    \includegraphics[width=4.6cm,height=4.6cm]{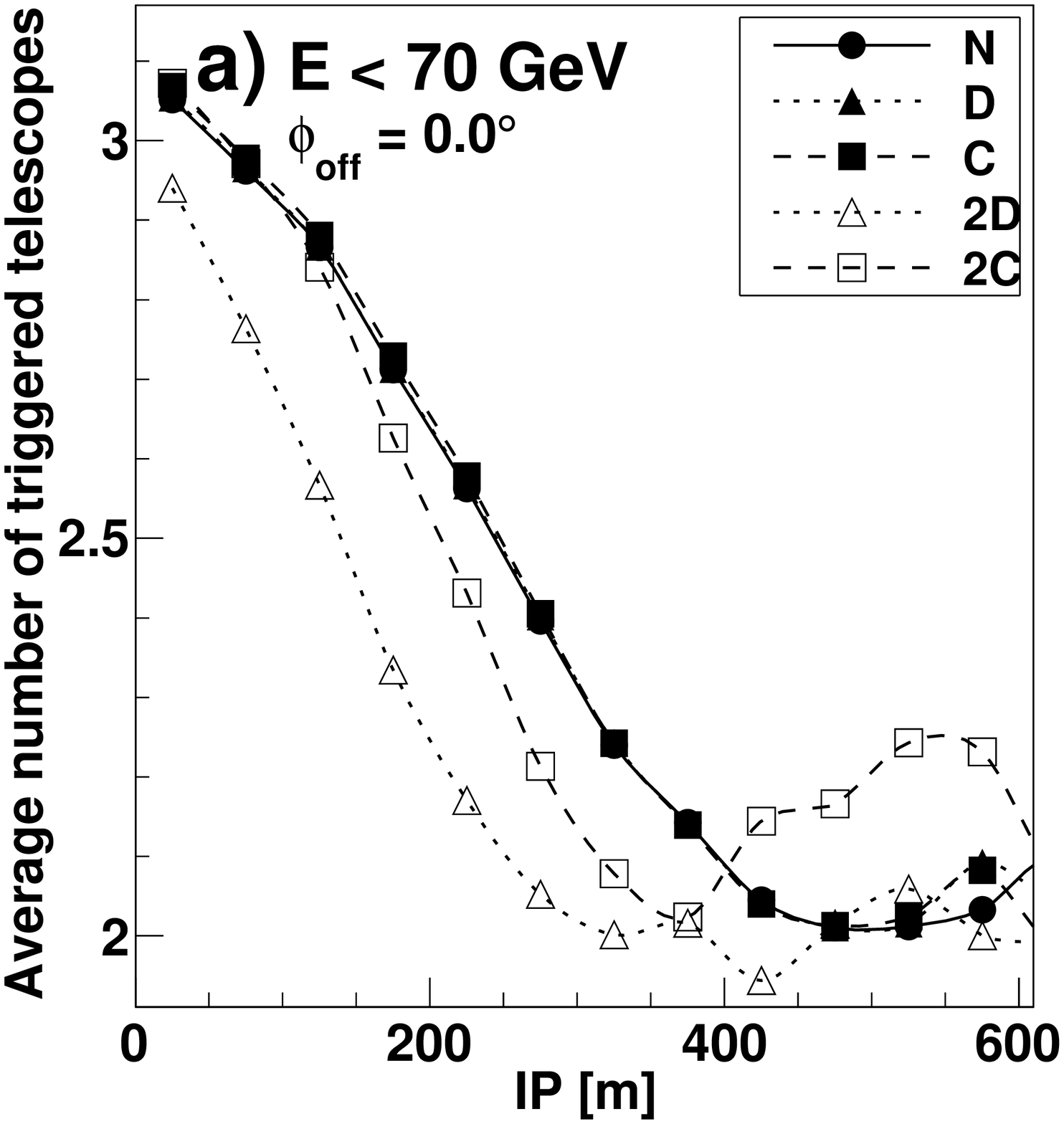}}%
  \scalebox{0.95}{%
    \includegraphics[width=4.6cm,height=4.6cm]{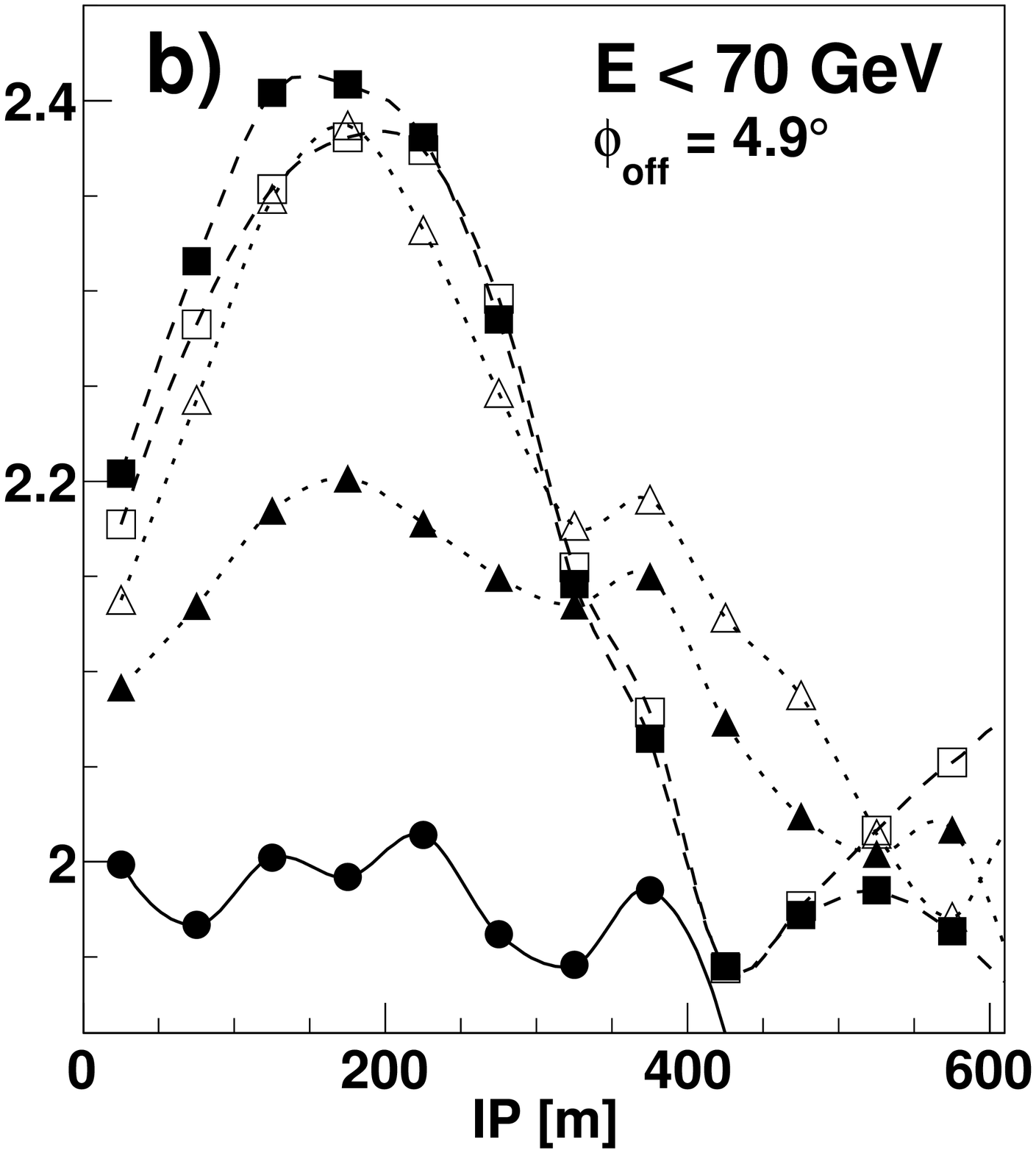}}
  \scalebox{0.95}{%
    \includegraphics[width=4.6cm,height=4.6cm]{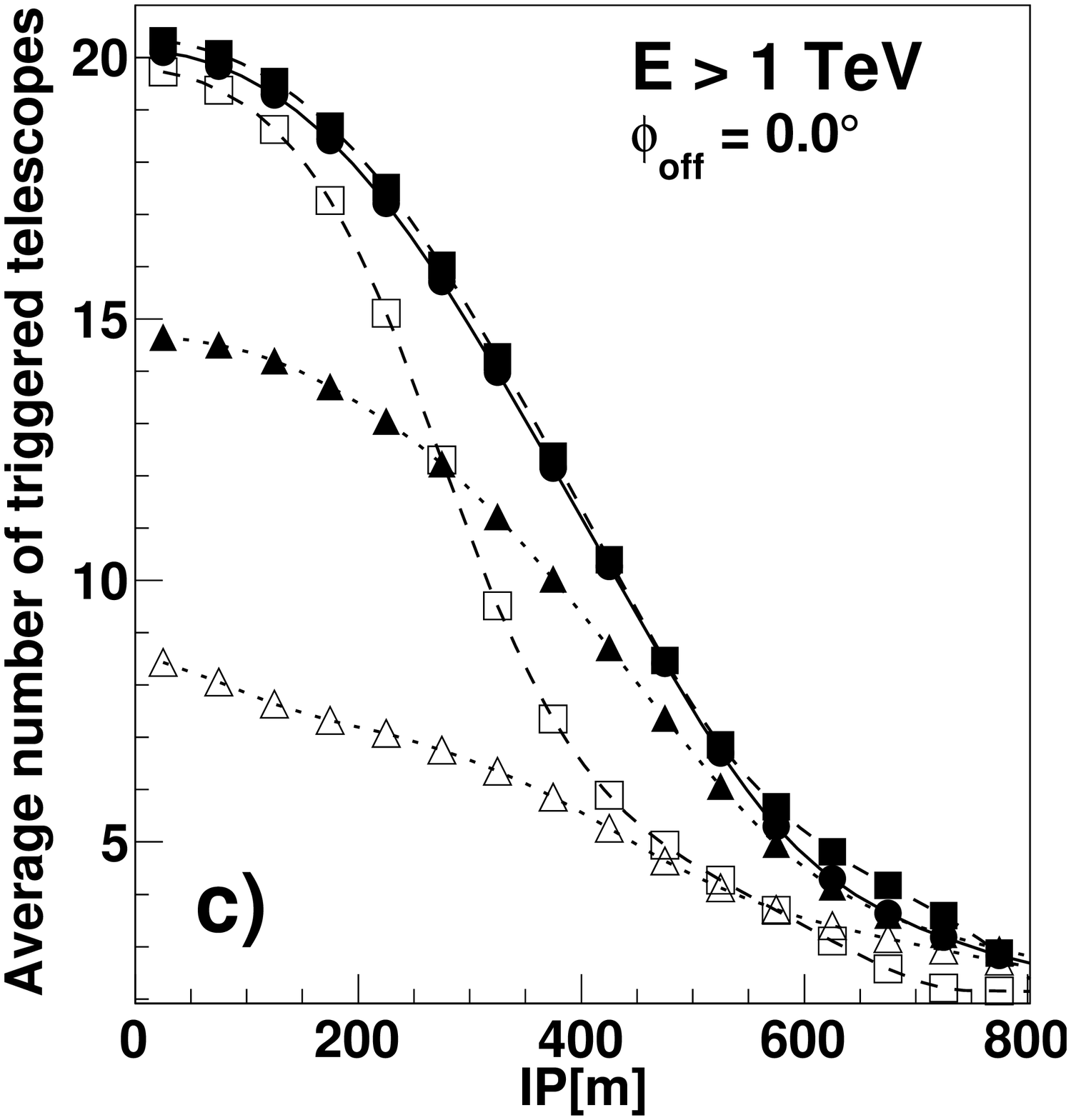}}%
  \scalebox{0.95}{%
    \includegraphics[width=4.6cm,height=4.6cm]{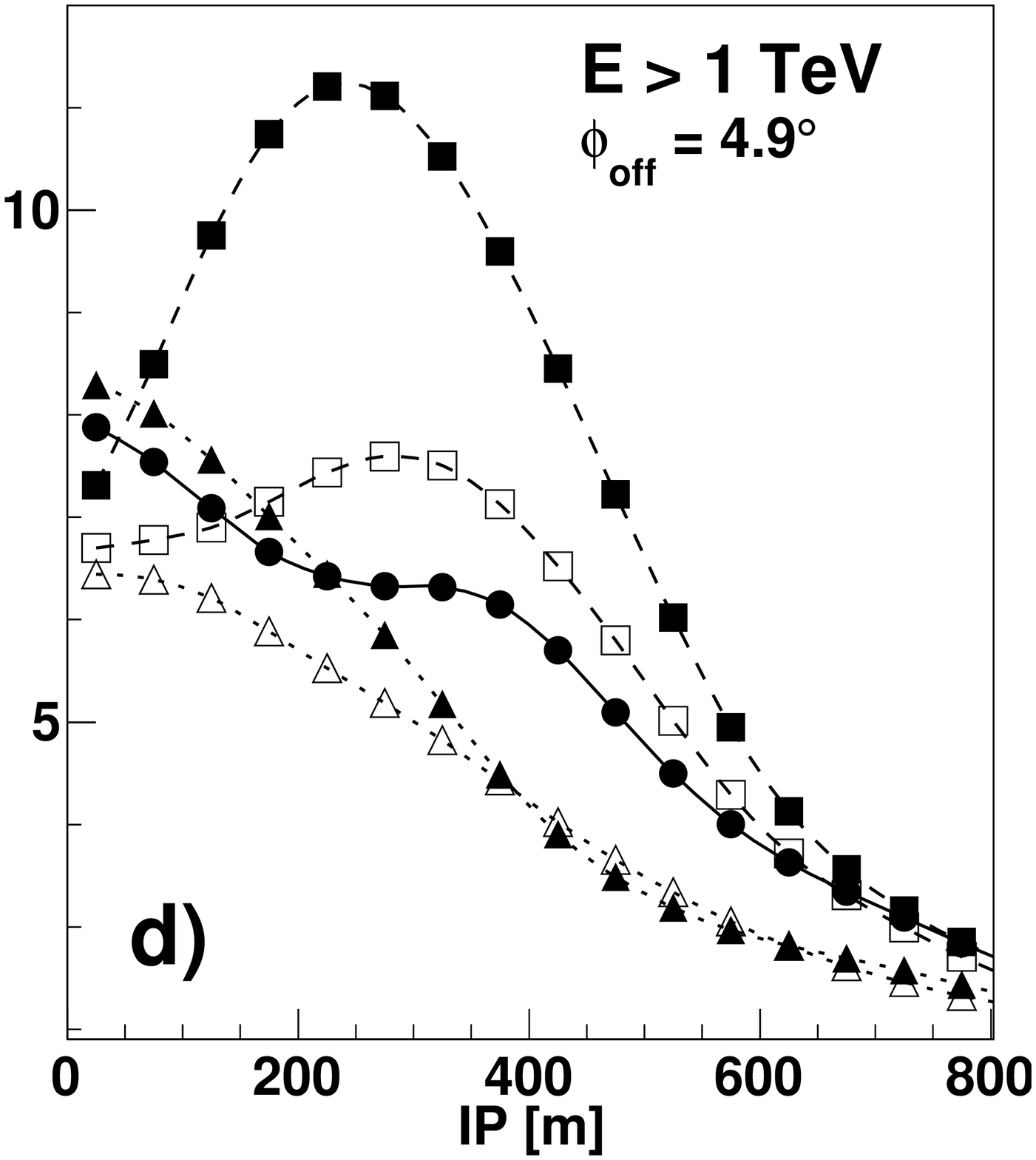}}

\caption{The average number of telescopes triggered by gamma-ray events as a function of the impact parameter (measured from the array center) for the five modes, represented by the same line and marker styles as in Fig.~\ref{fig:impact_energy}.
Top panels: $E \le 70\;{\rm GeV}$, bottom panels: $E \ge 1\;{\rm TeV}$, left: $\phi_{\rm off} = 0^{\circ}$, right:  $\phi_{\rm off} =4.9^{\circ}$.}
\label{fig:Ntrg_impacts}
\end{center}
\end{figure}

We first illustrate basic effects underlying the difference of performance discussed in following sections.
The optimization of the performance should maximize the number of gamma-ray events which (1) are detected, and then (2) have properly reconstructed arrival direction and energy. While (1), i.e.\ efficient detection, is mainly determined by the single telescope response, (2) improves with a larger number of triggered telescopes (regardless of the analysis method). 
In Figs.~\ref{fig:impact_energy} and \ref{fig:Ntrg_impacts} we show the relevant distributions of the number of registered events, $n_{\rm tr}$,
and the number of triggered telescopes, $N_{\rm tel}$, as functions of the impact parameter, IP, measured from the {\it array center} to the shower core.

Low energy events  with energies around the energy threshold of the MST array, $E \sim 70\; {\rm GeV}$, are typically registered by a single telescope at distances not exceeding the radius of the Cherenkov light pool on the ground, $\sim 120\; {\rm m}$, which then for a whole telescope array corresponds to the IP $\sim 200$ m. As seen in Figs.~\ref{fig:impact_energy}(a--c), for a large $\phi_{\rm off} \gtrsim 5^{\circ}$ (i.e.\ larger than the radius of FOV of a single MST) the use of non-parallel modes has a huge effect in the low energy range; all non-parallel modes give $n_{\rm tr}$ by 2 orders of magnitude larger than the N mode. Furthermore, in this regime (i.e.\ with low $E$ and large $\phi_{\rm off}$),  the divergent modes are notably better than the convergent modes especially at larger IP. For sources located on-axis, $n_{\rm tr}({\rm IP})$ is almost the same for the N, D and C modes, whereas for the 2D and 2C modes $n_{\rm tr}$ is much smaller, which shows that in this regime (low $E$ and small $\phi_{\rm off}$) the leakage effect is important only for larger telescope  offset angles ($\alpha$).

High energy events,  in the TeV range, produce a  large density of  Cherenkov photons, which is sufficient to trigger a telescope at distances significantly exceeding the light-pool radius, hence on average ${\rm IP}  \gtrsim  r_{\rm max}$. At high energies, both divergent modes give much larger $n_{\rm tr}$ than the C, 2C or N mode regardless of $\phi_{\rm off}$, except for small $\phi_{\rm off}$ for which N and D give similar (largest) values of $n_{\rm tr}$, see Figs \ref{fig:impact_energy}(d-f). 

As also seen in  Fig.~\ref{fig:impact_energy}, at all energies the single scaled modes give much larger $n_{\rm tr}$ than the corresponding double scaled modes up to $\phi_{\rm off} \sim 5^{\circ}$. However, at large $\phi_{\rm off} > 5^{\circ}$ the double modes are more efficient and especially at high energies the 2D  mode gives a much better detection efficiency than other modes.

The number of  triggered telescopes is much smaller for low energy events, on average $N_{\rm tel} \sim 2-3$ for all observation modes and any location of the source, than for high energy events. However, for the latter it strongly depends on both the source location and the mode. Namely, for an on-axis source (Fig.~\ref{fig:Ntrg_impacts}c) and for small IP ($< 300$ m), $N_{\rm tel}$ is much smaller for modes D and 2D (for the latter by over a factor of 2) than in other modes, which is an obvious consequence of the loss of the event in the outer, most diverged telescopes. A more complex effect occurs for off-axis sources, for which in some configurations the displacement of the image from the camera center due to the increasing IP balances the (opposite) displacement due to the increasing $\phi_{\rm off}$
in some ranges of IP, however, it results in a significant increase of $N_{\rm tel}$ only in mode C (see  Fig.~~\ref{fig:Ntrg_impacts}d).

Briefly summarizing  the above, we can expect the largest trigger rates in mode D (for smaller $\phi_{\rm off}$) or 2D (for larger $\phi_{\rm off}$)  and the most precise reconstruction in mode C. The overall performance results from the combination of these two features and then the above basic properties do not allow to assess which mode can be most suitable for a specific type of observation. To this end a extensive comparison of the standard IACT performance parameters is needed with use of a possibly most effective analysis methods.

\subsubsection{Collection areas and trigger rates}

\begin{figure}[t!]
\begin{center}
  \scalebox{0.95}{%
    \includegraphics[trim = 3mm 0mm 2mm 0mm, clip, width=4.6cm,height=4.6cm]{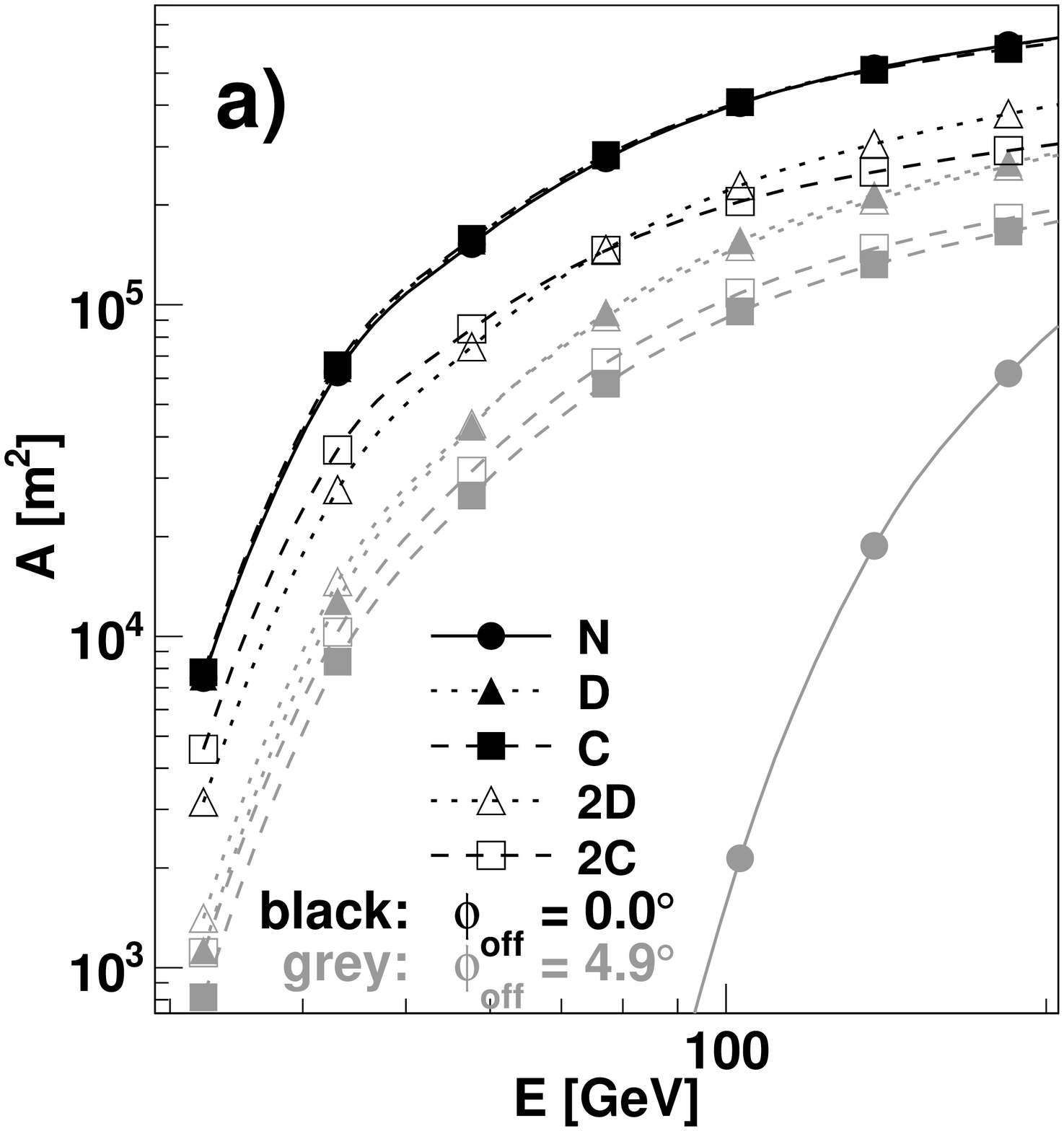}}%
  \scalebox{0.95}{%
    \includegraphics[trim = 3mm 0mm 2mm 0mm, clip, width=4.6cm,height=4.6cm]{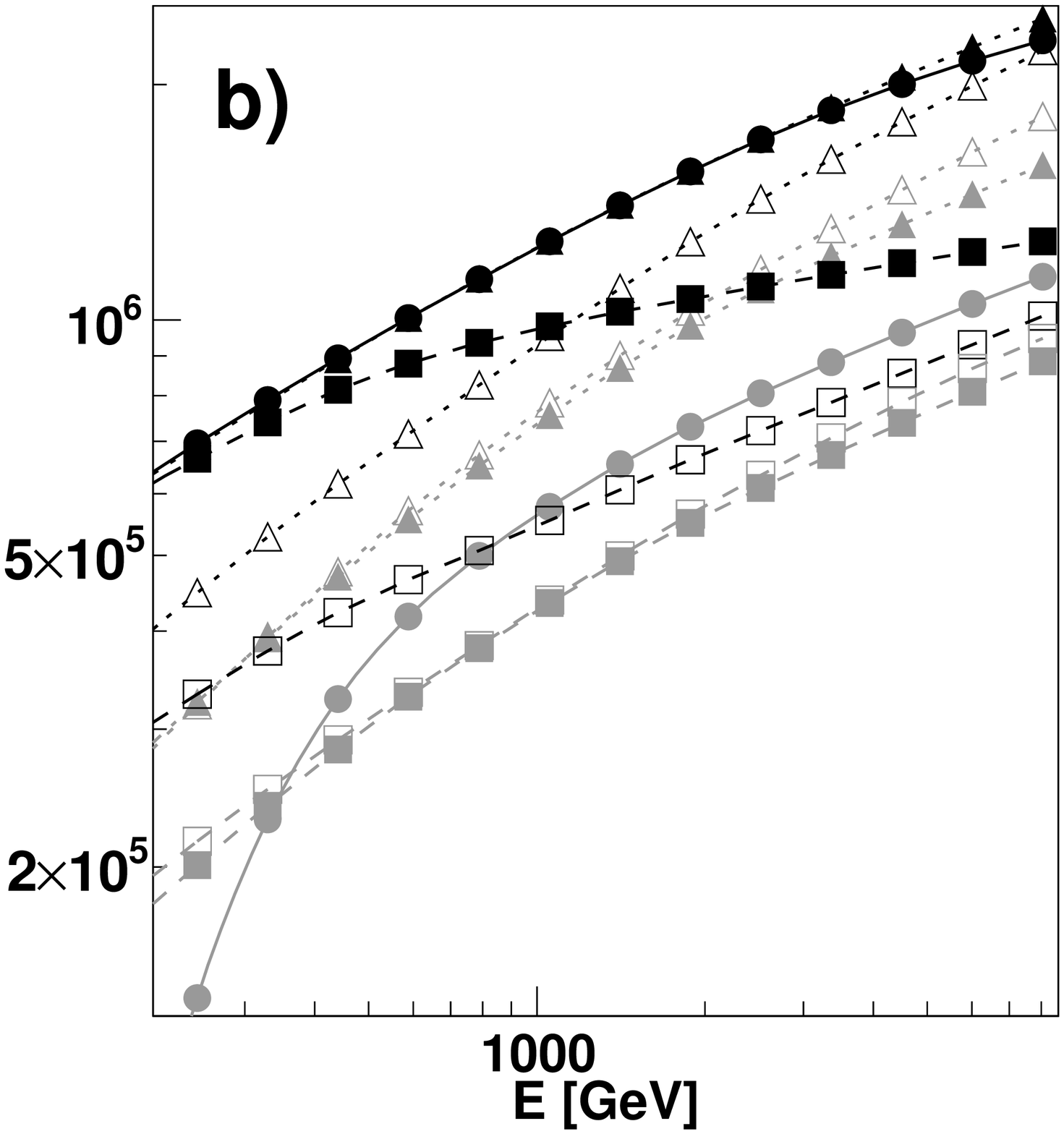}}
  \scalebox{0.95}{%
    \includegraphics[trim = 3mm 0mm 2mm 0mm, clip, width=4.6cm,height=4.6cm]{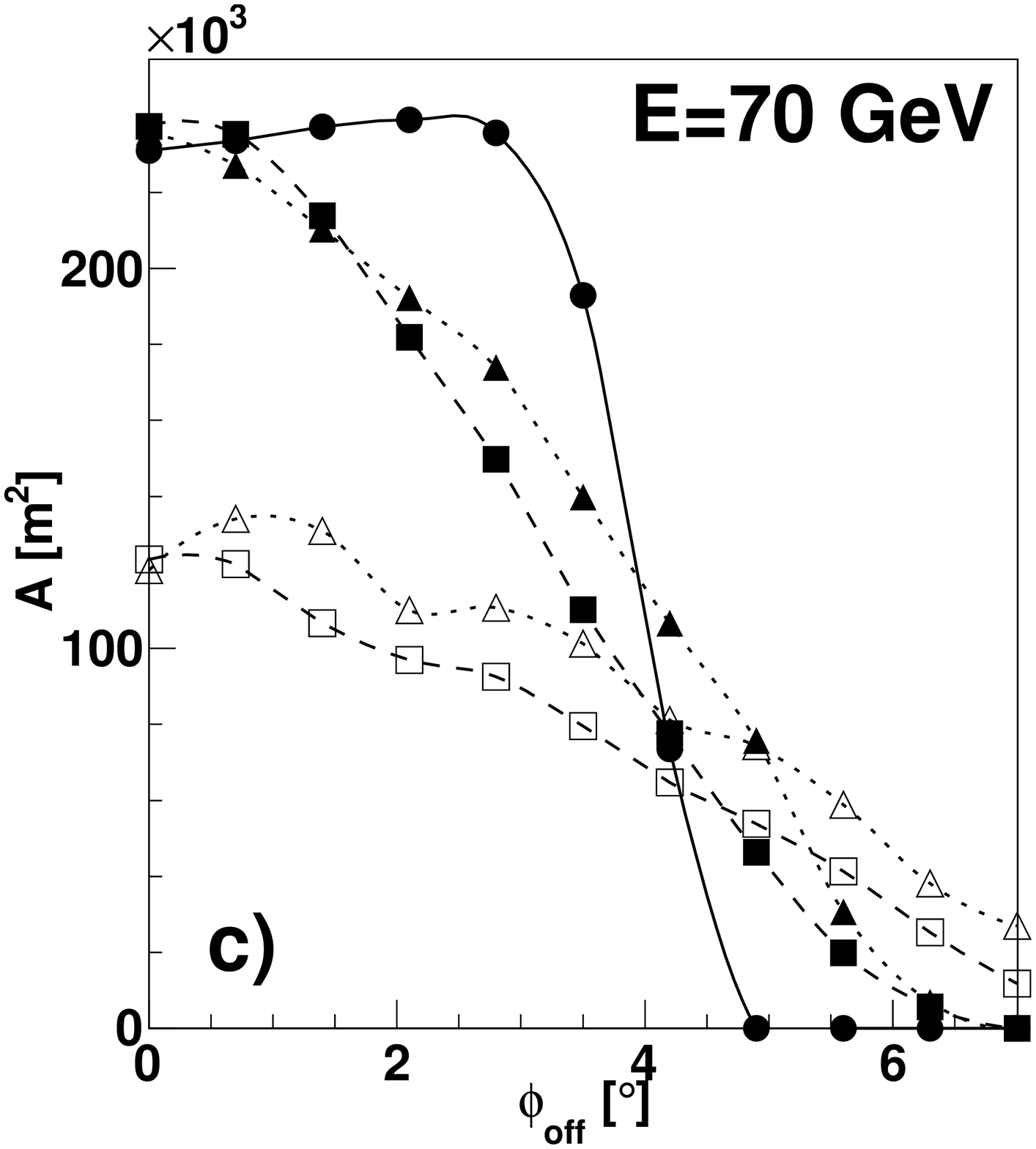}}%
  \scalebox{0.95}{%
    \includegraphics[trim = 3mm 0mm 2mm 0mm, clip, width=4.6cm,height=4.6cm]{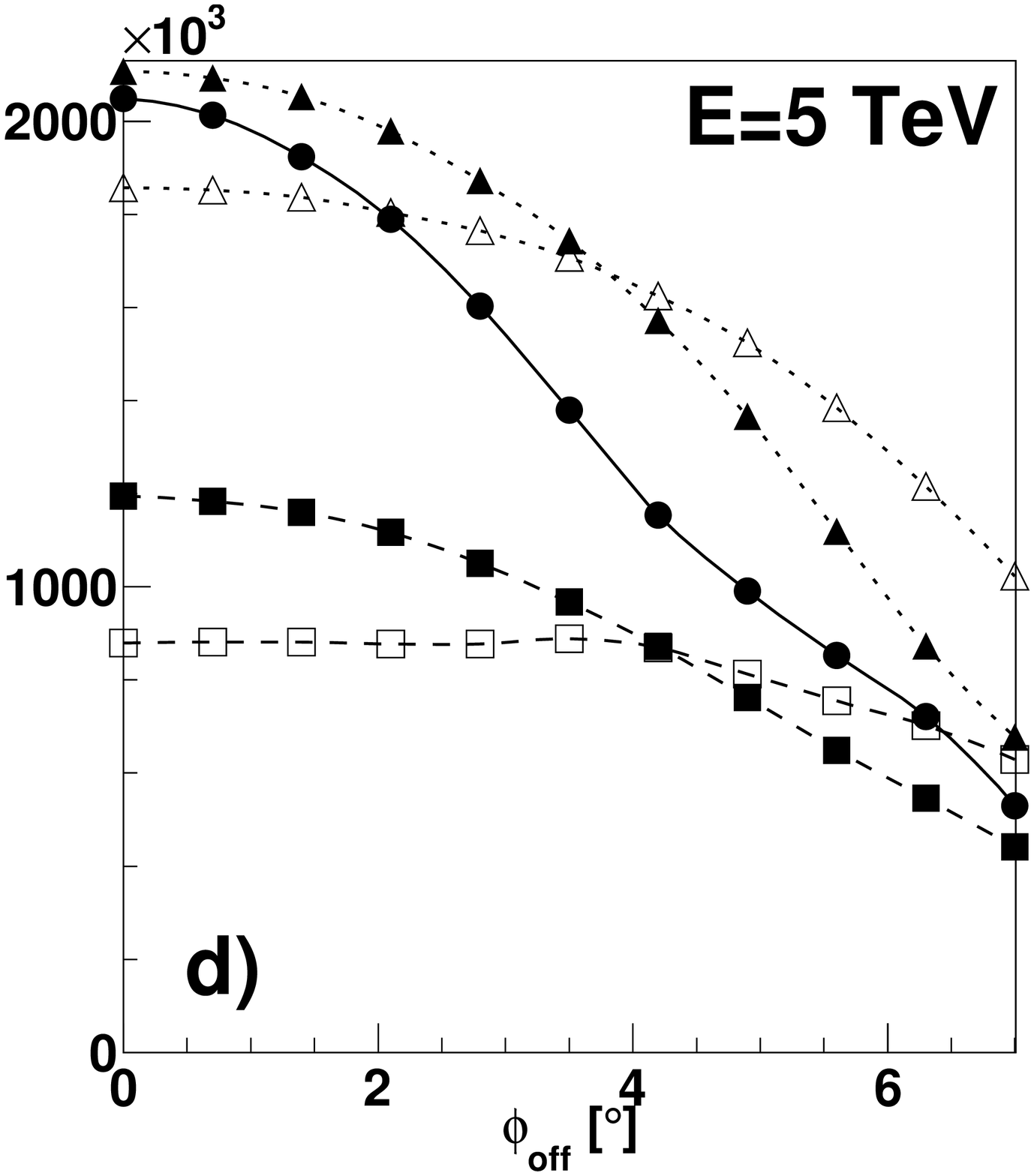}}

\caption{Trigger-level collection area as a function of $E$ (top panels) and $\phi_{\rm off}$ (bottom panels) for the five modes, represented by the same line and marker styles as in Fig.~\ref{fig:impact_energy}. In the top panels the black lines and markers are for $\phi_{\rm off} = 0^{\circ}$ and the grey ones for  $\phi_{\rm off} =4.9^{\circ}$. Panel (c) is for $E = 70\; {\rm GeV}$ and panel (d) for $E = 5\; {\rm TeV}$.}
\label{fig:areas}
\end{center}
\end{figure}

The collection area of an IACT array, $A(E)$, can be obtained from MC simulations from the number of triggered and simulated events, $n_{\rm tr}$ and $n_{\rm sim}$, and the simulated area, $S$, as $A(E) = [n_{\rm tr}(E)/n_{\rm sim}(E)] \times S$. 
Fig.\ \ref{fig:areas}  presents the $E$-- and $\phi_{\rm off}$--dependence of $A$. Essentially it reflects the dependence of $n_{\rm tr}$ on $E$ and $\phi_{\rm off}$ discussed above. In particular, 

\noindent
(i)
the largest values of $A$ correspond to $\phi_{\rm off}=0^{\circ}$; then, $A$ decreases with increasing $\phi_{\rm off}$, but the shape of $A(\phi_{\rm off})$ significantly differs between the modes. The most rapid decrease occurs in mode N, the flattest $A(\phi_{\rm off})$ correspond to double scaled modes, for which $A$ is larger than for single scaled modes at $\phi_{\rm off} \gtrsim (4-5)^{\circ}$. 

\noindent
(ii)
for small $\phi_{\rm off}$ and low $E$ ($\sim 70\; {\rm GeV}$),   $A$ in double scaled modes is roughly half that of single-scaled modes.

\noindent
(iii) for small $\phi_{\rm off}$ and high $E$ ($\sim 5\; {\rm TeV}$), $A$ is approx.~by a factor of 2 larger in divergent modes than in convergent modes. 

\begin{figure}[t]
\begin{center}
  \includegraphics[trim = 6mm 1mm 0mm 1mm, clip, scale=0.458]{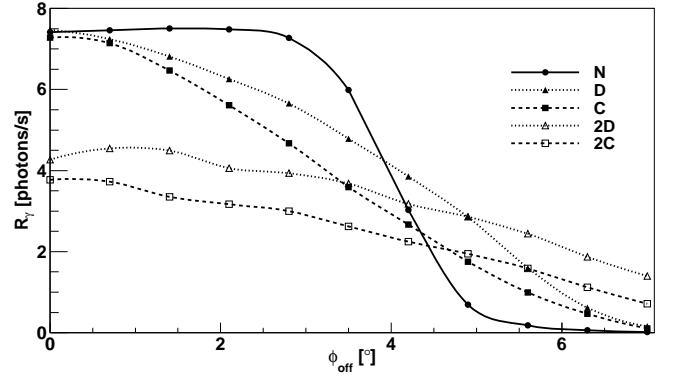}
\caption{The total trigger rates for gamma rays, $R_{\gamma}$, for a Crab-like source (see text) as a function of $\phi_{\rm off}$ for the five modes, represented by the same line and marker styles as in Fig.~\ref{fig:impact_energy}.}
\label{fig:gamma_total_rates}
\end{center}
\end{figure}

The collection area provides information independent of the energy spectrum of a source. To illustrate efficiency of the modes in a realistic data-taking, we use as a reference the spectrum measured from the Crab nebula around 1 TeV, $F_{\gamma}(E)=2.79 \times 10^{-11} (E/1 {\rm TeV})^{-2.57}\; {\rm cm^{-2}s^{-1}TeV^{-1}}$ (see Chapt.~8 in \citep{cta}), and we calculate the total  trigger rates, $R_{\gamma}$, by integrating over $E$ the differential trigger rates,  $R_{\gamma}(E) = F_{\gamma}(E)\times A(E)$. The resulting $R_{\gamma}(\phi_{\rm off})$ is shown in Fig.~\ref{fig:gamma_total_rates}. The major contribution to  $R_{\gamma}$ comes from low $E$ events, then the dependence on $\phi_{\rm off}$ is close to that of $A(\phi_{\rm off})$ at low $E$ (shown in Fig.~\ref{fig:areas}c).

We also find the total trigger rate for protons, $R_{p}$, by integrating over $E$ the differential trigger rate, $R_{p}(E) = \int {\rm d}\Omega(\omega) F_{p}(E)\times A_{p}(E, \omega)$, where $F_{p}(E)=8.72 \times 10^{-6} (E/1 {\rm TeV})^{-2.732}\; {\rm cm^{-2}s^{-1}sr^{-1}TeV^{-1}}$ (i.e., at $1\;{\rm TeV}$ comparable with proton flux obtained with BESS-TeV experiment \citep{Bess2007}), $A_p$ is the collection area for protons and $\Omega$ is the simulated proton cone angle related with the half opening angle, $\omega$, by $\Omega(\omega)=2\pi(1-\cos\omega)$.

Comparing the values of $R_{p}$ given in Table \ref{tab:bkg_total_rates} with $R_{\gamma}$ in Fig.~\ref{fig:gamma_total_rates}, we see that for an on-axis source $R_{p}$ are larger by three orders of magnitude even for the most efficient (in gamma events) modes N, D and C, and the ratio of $R_{\gamma}$ to $R_{p}$ decreases with increasing $\phi_{\rm off}$. Notably, $R_{p}$ in the convergent modes is  by  $\sim 15\; \%$ larger than in the divergent modes\footnote{note that apart from a potential advantage in a better significance, arrays triggering less background events are preferred for easier acquisition of data}.

 We emphasize that in our background simulations we do not take into account other components, e.g.\ helium, electrons etc., and thus some performance parameters, e.g.~absolute values of sensitivities, could differ from more realistic estimations for CTA. On the other hand, it should not affect significantly the relative comparison of different modes. We briefly discuss expected changes in sensitivities due to contribution of other background components at the end of Sect.~\ref{sec:summary}.

\begin{table}[t]
\begin{center}
 \begin{tabular}{|c|c|c|c|c|}
   \hline
   \multicolumn{5}{|c|}{\bf Total background proton rates [protons/s]}\\
   \hline
   {\bf N} & {\bf D} & {\bf C} & {\bf 2D} & {\bf 2C} \\
   \hline
  \; 7888\; & \;7446\; & \;8224\; & \;6633\; &\; 8032\; \\
   \hline
 \end{tabular}
\end{center}
\caption{Total background proton rates, $R_{p}$, for the studied modes.}
 \label{tab:bkg_total_rates}
\end{table}

\subsubsection{Acceptance}
\label{sec:field_of_view_acc}

As noted above, the $A(\phi_{\rm off})$ dependence is different in the considered  modes, in particular, it is much flatter in modes 2C and 2D, whereas  in modes N, D and C it is strongly enhanced at  $\phi_{\rm off} \lesssim 5^{\circ}$. To compare the detection efficiency in the total observed area, we define the 
acceptance, ${\rm Acc}(E)=\int A(E,\phi_{\rm off}){\rm d}\Omega$, where ${\rm d}\Omega=2\pi\sin\phi_{\rm off}\times{\rm d}\phi_{\rm off}$ (assuming the variation of $A(E,\phi_{\rm off})$ across the FOV depends exclusively on $\phi_{\rm off}$). As ${\rm d}\Omega$ increases with the offset angle $\phi_{\rm off}$, the relatively stronger contribution to ${\rm Acc}(E)$ comes from the regions with larger $\phi_{\rm off}$. In other words ${\rm Acc}(E)$ can be used as a measure of the performance of the array for an isotropic population of sources within the FOV.

\begin{figure}[t]
\begin{center}
  \includegraphics[trim = 6mm 1mm 0mm 5mm, clip, scale=0.458]{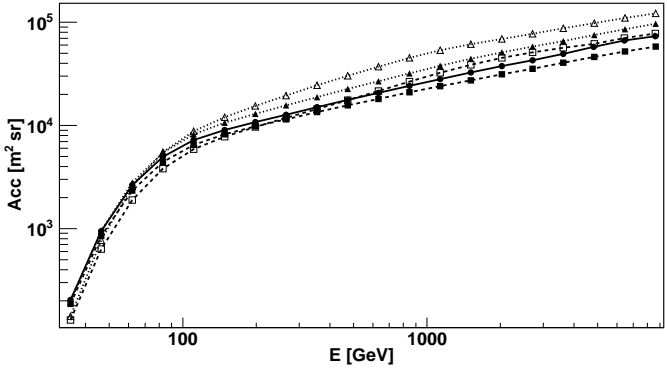}

\caption{Acceptance at the trigger level as a function of energy for the five modes, represented by the same line and marker styles as in Fig.~\ref{fig:impact_energy}.
}
\label{fig:FOV_ACC}
\end{center}
\end{figure}

As we see in Fig.~\ref{fig:FOV_ACC}, the difference between ${\rm Acc}$ in different modes is larger at high energies, where also larger ${\rm Acc}$ correspond to the divergent modes and the largest (for mode 2D) and smallest (for mode C) acceptances differ by over a factor of 2. At low energies the differences are smaller, but also here the largest ${\rm Acc}$ corresponds to modes D and 2D and the smallest to mode 2C.

\subsubsection{Reconstruction}
\label{sec:reconstruction}
In this section we compare the accuracy of the direction and energy reconstruction  in different modes.
The geometric reconstruction method used for IACT relies on the Hillas image parameters, describing geometrical properties of an image in a telescope camera plane. The arrival direction of a primary particle is estimated from the intersection of the major axes of Hillas ellipses of images from each pair of triggered telescopes (averaged over all pairs if more than two telescopes triggered). The quality of the direction reconstruction is given by the angular resolution, defined by the containment radius, $r_{f}$, of the area in the camera plane enclosing a fixed fraction $f$ of the reconstructed directions (for details see Sect.~4.5 in \citep{Szanecki2013}).

\begin{figure}[t]
\begin{center}
  \includegraphics[trim = 6mm 1mm 0mm 4mm, clip, scale=0.458]{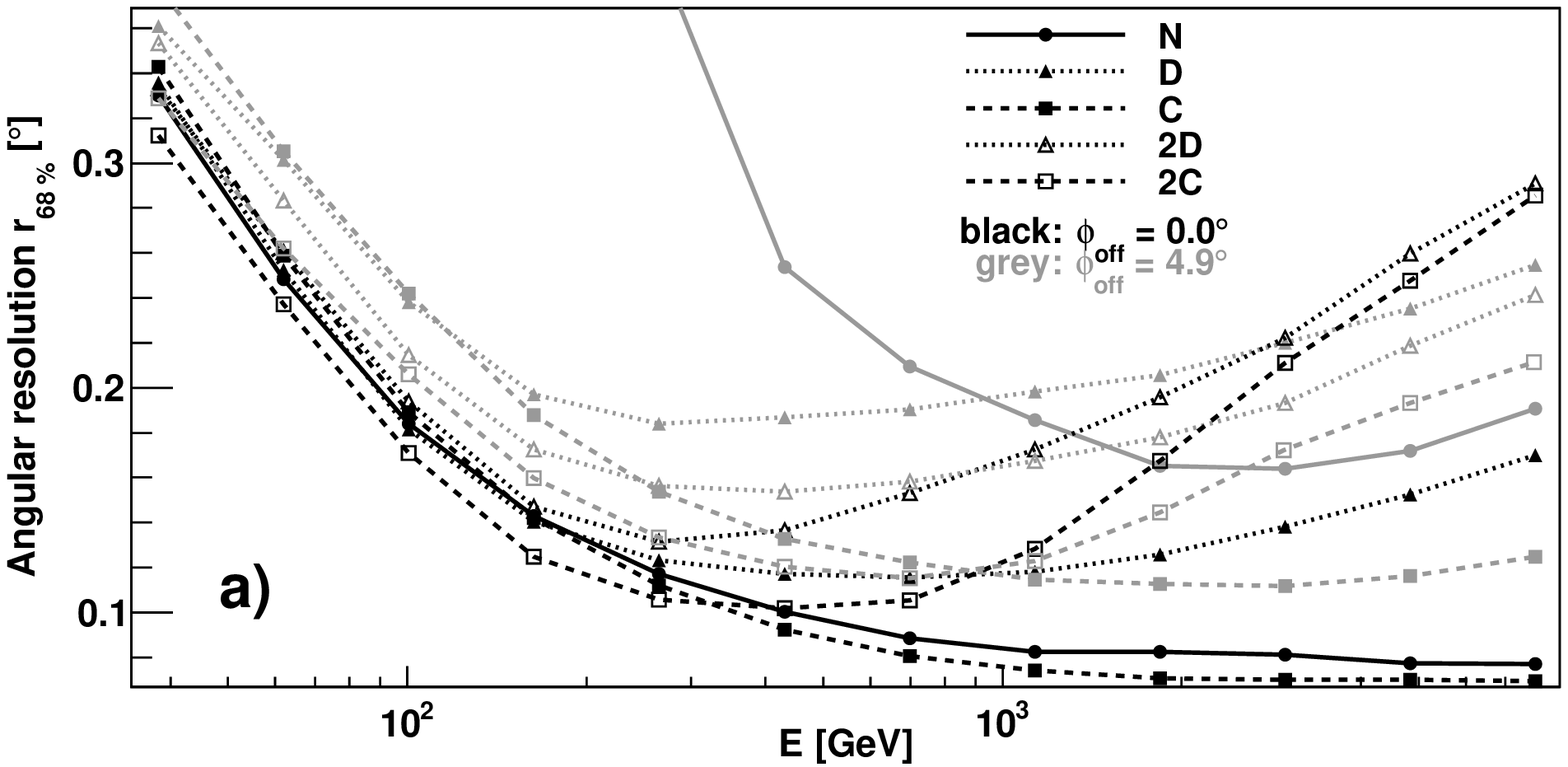}
  \includegraphics[trim = 6mm 1mm 0mm 1mm, clip, scale=0.458]{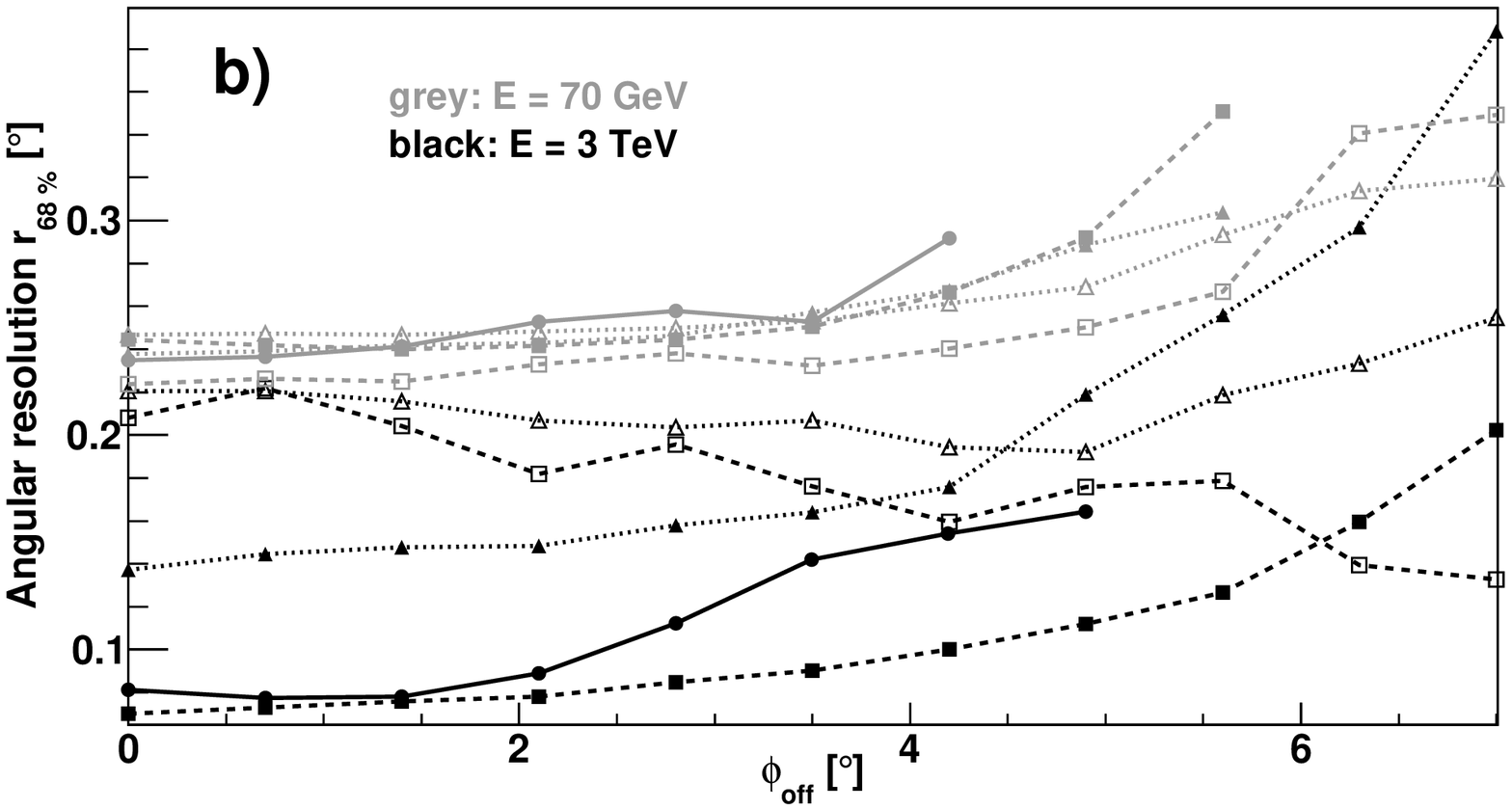}
\caption{Angular resolution before background suppression cuts for the $68\;\%$ containment radius as a function of $E$ (a) and $\phi_{\rm off}$ (b), for the five modes represented by the same line and marker styles as in Fig.~\ref{fig:impact_energy}. In panel (a)
the black lines and markers are for $\phi_{\rm off} = 0^{\circ}$ and the grey ones for  $\phi_{\rm off} =4.9^{\circ}$. In panel (b)
the black lines and markers are for $E = 3\; {\rm TeV}$ and the grey ones for  $E = 70\; {\rm GeV}$.
The lack of several points in panel (b) is due to an  insufficient statistics to compute the angular resolution.}
\label{fig:ang_res}
\end{center}
\end{figure}

The angular resolution for $f=68\%$ is shown in Fig.~\ref{fig:ang_res}. The difference between the modes results mostly from two effects. First, the quality of the direction reconstruction improves with the increasing number of telescopes triggered by an event, as a larger number of major axis crossing points tends to reduce the reconstruction inaccuracies. Secondly, the angular resolution is spoiled by the leakage effect (the truncated images typically do not contribute to improvement of reconstruction). We note that both effects add up and yield a much better angular resolution in the convergent than in the divergent modes; in the former $N_{\rm tel}$ is typically much larger (see Fig.\ \ref{fig:Ntrg_impacts}) and in the latter the leakage effect is typically stronger, especially at high energies. An additional, but weaker effect involves different shapes of images, with more elongated images allowing for a more precise determination of the image major axis and, then, a better estimation of the particle arrival direction. This effect results in a slightly better angular resolution in mode C than in D at small $E$ and large $\phi_{\rm off}$, for which mode C has a larger average IP (implying more elongated images), see Fig.\ \ref{fig:Ntrg_impacts}.

In standard analysis of parallel mode configuration the angular resolution typically improves with increasing $E$, for which Hillas ellipses are better constrained. Such an improvement indeed occurs in mode N and C for
an on-axis source (Fig.~\ref{fig:ang_res}a). However, in other modes the angular resolution worsens with increasing $E$, above $\sim 500$ GeV, and we note that this is mostly due to the leakage effect, by which modes D, 2D and 2C are strongly affected at TeV energies. As a result, large differences occur in this large energy range, in particular at $E \gtrsim 3$ TeV the best resolution of $\simeq 0.07^{\circ}$ in mode C is by over a factor of 3 better than the resolutions in modes 2D and 2C and twice better than the resolution in mode D. Notably,
the resolution in modes N, D and C significantly worsens with the increase of $\phi_{\rm off}$,
whereas for 2D and 2C it is less sensitive to $\phi_{\rm off}$ and at $\phi_{\rm off} \gtrsim 5^{\circ}$ the double scaled modes have a better resolution than single scaled ones.
At low energies, $\sim 100$ GeV, all modes have a poor resolution, $> 0.2^{\circ}$, and it is less dependent on both $\phi_{\rm off}$ and the mode.

The estimation of the primary particle energy relies on  its correlation with the SIZE of an image (the total number of photo-electrons in image) and the impact parameter. SIZE is obtained by integrating the light collected in each pixel belonging to an image and the impact parameter is estimated with use of DIST Hillas parameter (distance between the reconstructed source position and the center of gravity of an image), to which it is proportional. The energy reconstruction quality of IACT telescope systems is conventionally described by two quantities, the energy resolution and the energy bias. The first one is defined by the value of the standard deviation of a Gaussian fit of the  distribution of $(E_{\rm rec}-E)/E$, where $E_{\rm rec}$ and $E$ are the reconstructed and true (i.e.\ simulated) energy, respectively.

The bias is the mean of the fitted Gaussian. The differences in energy biases - the most apparent at energies close to the array threshold - reflect the possible changes of the effective energy threshold among the studied modes. The value of the effective energy threshold, being usually taken as the first point of the reconstructed energy spectrum, is relevant to properly identify the spectrum of observed source in the possibly largest energy range.

\begin{figure}[t]
\begin{center}
  \scalebox{0.95}{%
    \includegraphics[trim = 3mm 1mm 2mm 0mm, clip, width=4.6cm,height=4.6cm]{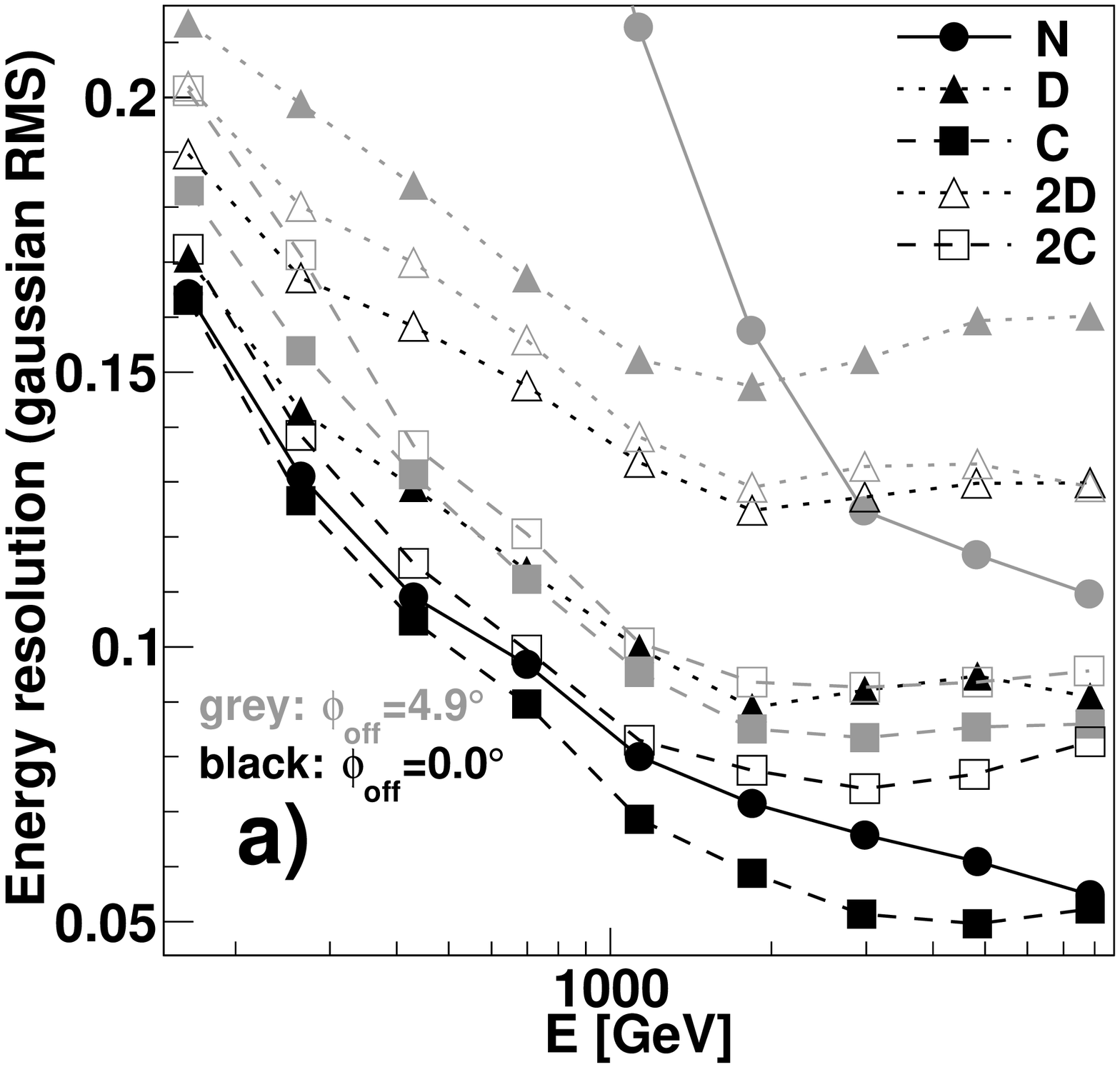}}%
  \scalebox{0.95}{%
    \includegraphics[trim = 3mm 1mm 2mm 0mm, clip, width=4.6cm,height=4.6cm]{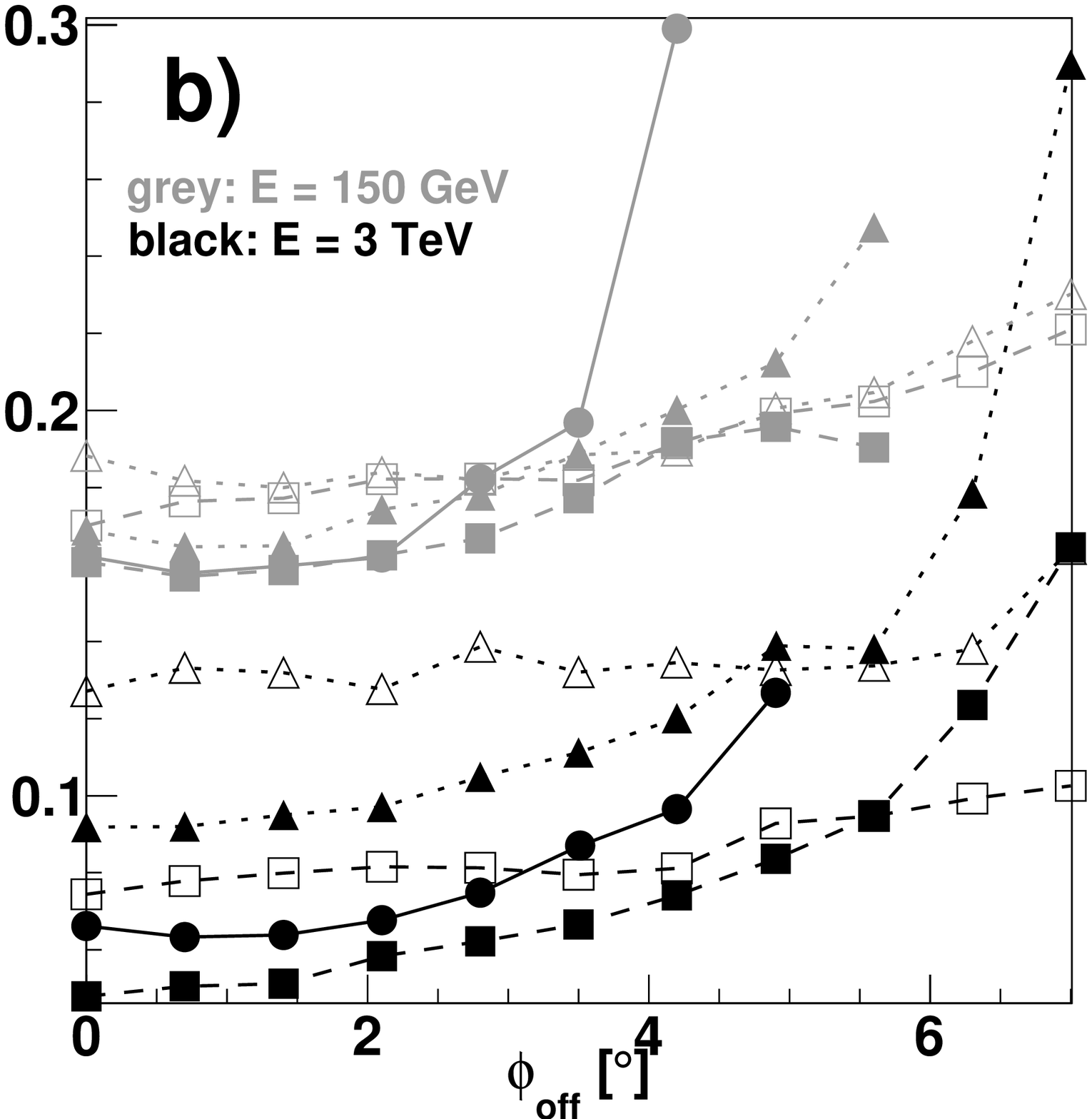}}
  \includegraphics[trim = 6mm 1mm 5mm 1mm, clip, scale=0.458]{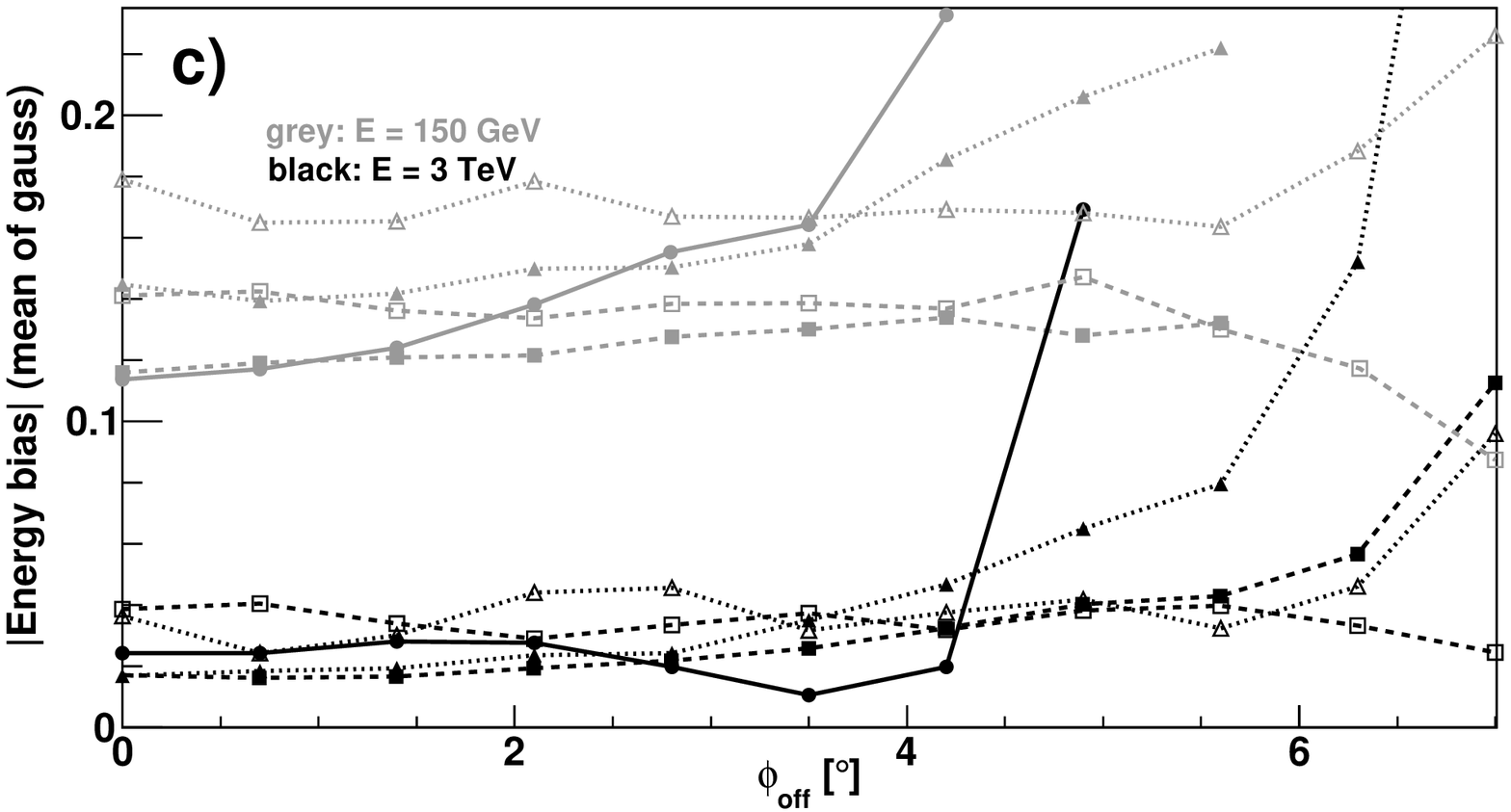}

\caption{Top panels show the trigger-level energy resolution as a function of $E$ (a) and  $\phi_{\rm off}$ (b) and bottom panel (c) shows the trigger-level  energy bias as a function of $\phi_{\rm off}$, for the five modes represented by the same line and marker styles as in Fig.~\ref{fig:impact_energy}.
In panel (a)
the black lines and markers are for $\phi_{\rm off} = 0^{\circ}$ and the grey ones for  $\phi_{\rm off} =4.9^{\circ}$.
In panels (b) and (c) the black lines and markers are for $E = 3\:{\rm TeV}$ and the grey ones for    $E = 150\:{\rm GeV}$.}
\label{fig:energy_res}
\end{center}
\end{figure}

As seen in Fig.~\ref{fig:energy_res}, qualitatively, the energy resolution shows the same dependence on the mode, $E$ and $\phi_{\rm off}$ as the angular resolution and this results from the same effects, involving the number of telescopes which are  triggered and survive the image reconstruction (which improves the geometrical reconstruction)  and the leakage effect which spoils the energy reconstruction especially in non-parallel modes (except for mode C) at high energies. 
Apart from this, the standard effects affect the energy dependence, namely,
the Hillas ellipses are better defined for higher $E$ allowing for a more precise estimation of the center of gravity of an image, and hence a 
more accurate reconstruction of IP,
furthermore, higher energy gamma rays give larger SIZES, less sensitive to fluctuations.
In all modes the best energy resolution is obtained at several TeV and the overall best value of $\Delta E /E \sim 0.05$ of the mode C is by a factor of $\sim 3$ better than that in mode 2D.

Also the energy bias is the lowest for mode C, except for large $\phi_{\rm off}$ (see Fig.~\ref{fig:energy_res}c). In all modes, the bias is negative at high $E$ and positive at low $E$. It has the lowest  absolute value, $\lesssim 0.05$, at high $E$ and low $\phi_{\rm off}$, and increases with the decrease of $E$ and increase of $\phi_{\rm off}$. The most biased energy reconstruction occurs for D mode. At energies of $\approx 150\;{\rm GeV}$ and at offsets of $>4^{\circ}$ the expected bias for D mode is by $\approx 60\%$ larger in comparison with bias for both convergent modes.

\subsection{Analysis level parameters}
\label{sec:analysis level parameters}

\subsubsection{Gamma/hadron separation and sensitivity}
\label{sec:sensitivity}

To separate the gamma-ray signal from the dominating hadronic background we apply the standard gamma/hadron separation method with shape and direction cuts. Specifically, we use cuts on the mean scaled WIDTH (MSW), mean scaled LENGTH (MSL) and $\theta^{2}$ (see \citep{Daum97} for definitions),  which are the most effective separators. To improve the background rejection we also dynamically optimize the cut on the minimal number of telescopes used for reconstruction of an event. We do not apply any nominal distance cuts (in e.g.~the DIST parameter).

 Cuts are optimized to provide the best sensitivity. To avoid biasing the results with optimization on statistical fluctuations, we first tune the cuts on separated MC sub-samples (train samples) of gamma and background events. Using such optimized cuts, we calculate sensitivities on independent MC test samples.
An optimization of cuts with train samples is performed independently for each observational mode, each energy bin and for each value of $\phi_{\rm off}$. Note that  we calculate the point source sensitivity, however, the term sensitivity is used  throughout the paper for brevity..

Sensitivity is the key parameter  describing the  performance of a telescope array. It gives the magnitude of the flux observed from a source in a given time at a particular significance level; we assume the usual level defined by the gamma-ray signal exceeding the background estimate by 5 standard deviations ($\sigma$)
and we find sensitivities for an observation time of 50 h.

In our analysis we use the definition of differential sensitivity which follows from the
Li \& Ma significance formula (see eq.~(17) in \citep{LiMa1983}), commonly applied in IACT experiments, including CTA.  
We also apply the standard CTA conditions for a significant detection \citep{cta}, i.e.\
 gamma-ray signal exceeding $5\%$ of the remaining background and at least 10 events in excess, $N_{\rm ex}$.
To calculate the differential sensitivity we divide the reconstructed energy range into five bins per decade and we integrate gamma and proton differential rates (see Sect.~\ref{sec:rates collection}) to get ON-axis counts, $N_{\rm \textlcsc{on}}$, and OFF-axis counts, $N_{\rm \textlcsc{off}}$, for each simulated $\phi_{\rm off}$.
In our analysis, the off-axis events are binned in 11 rings around the array FOV centre. Specifically, rings (angular bins) are defined by 11 consecutive ranges of offsets: $\left[0.0^{\circ}-0.5^{\circ}\right]$, $\left[0.5^{\circ}-1.0^{\circ}\right]$, $\left[1.1^{\circ}-1.7^{\circ}\right]$, $\left[1.8^{\circ}-2.4^{\circ}\right]$, $\left[2.5^{\circ}-3.1^{\circ}\right]$, $\ldots$ , $\left[6.7^{\circ}-7.3^{\circ}\right]$, i.e.~off-axis angular bins are centered around each $\phi_{\rm off}$. We checked that the contribution of background events with the real offset $>10.0^{\circ}$ and the reconstructed offset $<7.0^{\circ}$ is negligible even for modes 2C and 2D.

\begin{figure*}[t]
\begin{center}
  \scalebox{0.95}{%
    \includegraphics[trim = 10mm 5mm 1.9mm 0mm, clip, width=6.0cm,height=6.0cm]{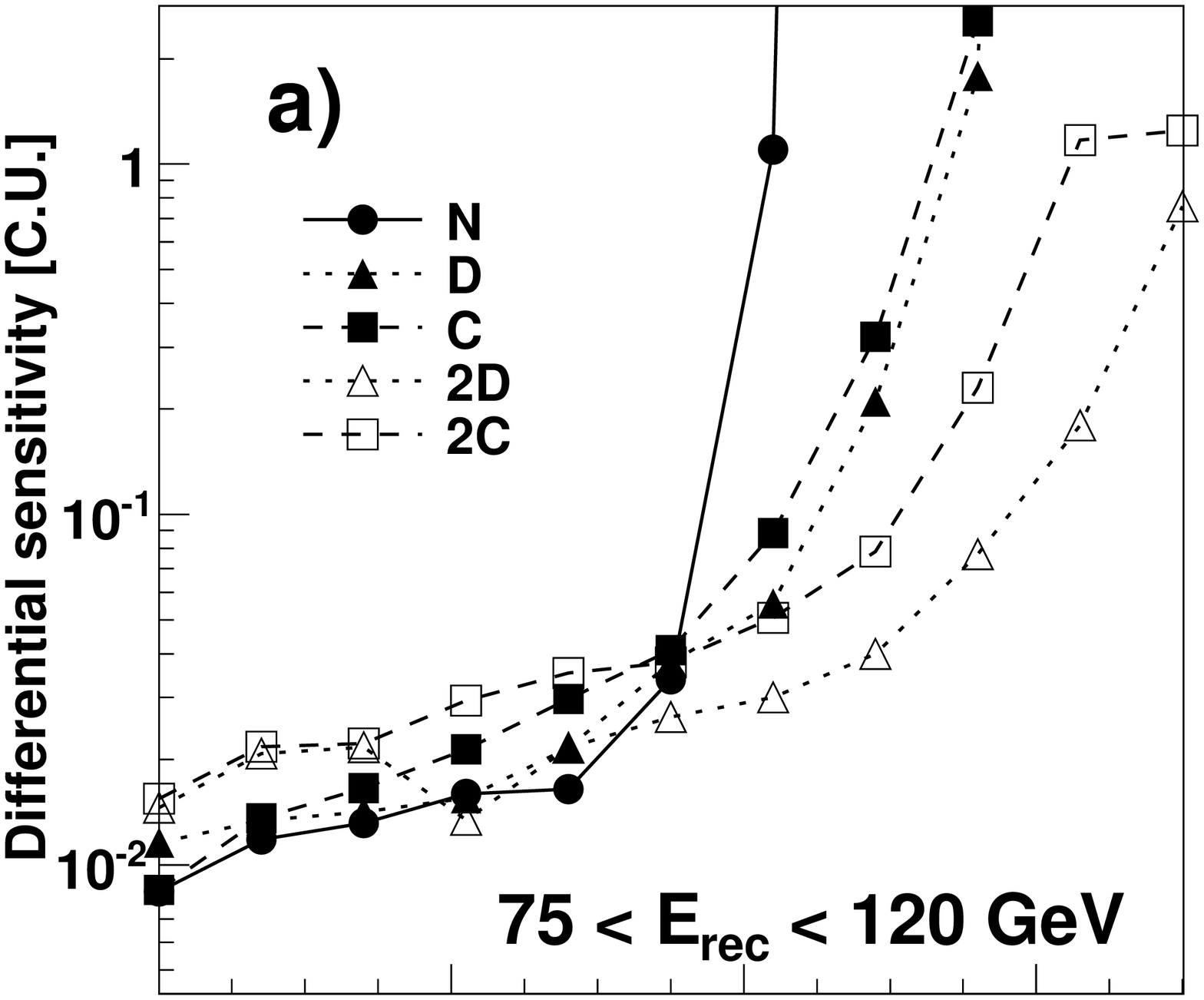}}%
  \scalebox{0.95}{%
    \includegraphics[trim = 10mm 5mm 1.9mm 0mm, clip, width=6.0cm,height=6.0cm]{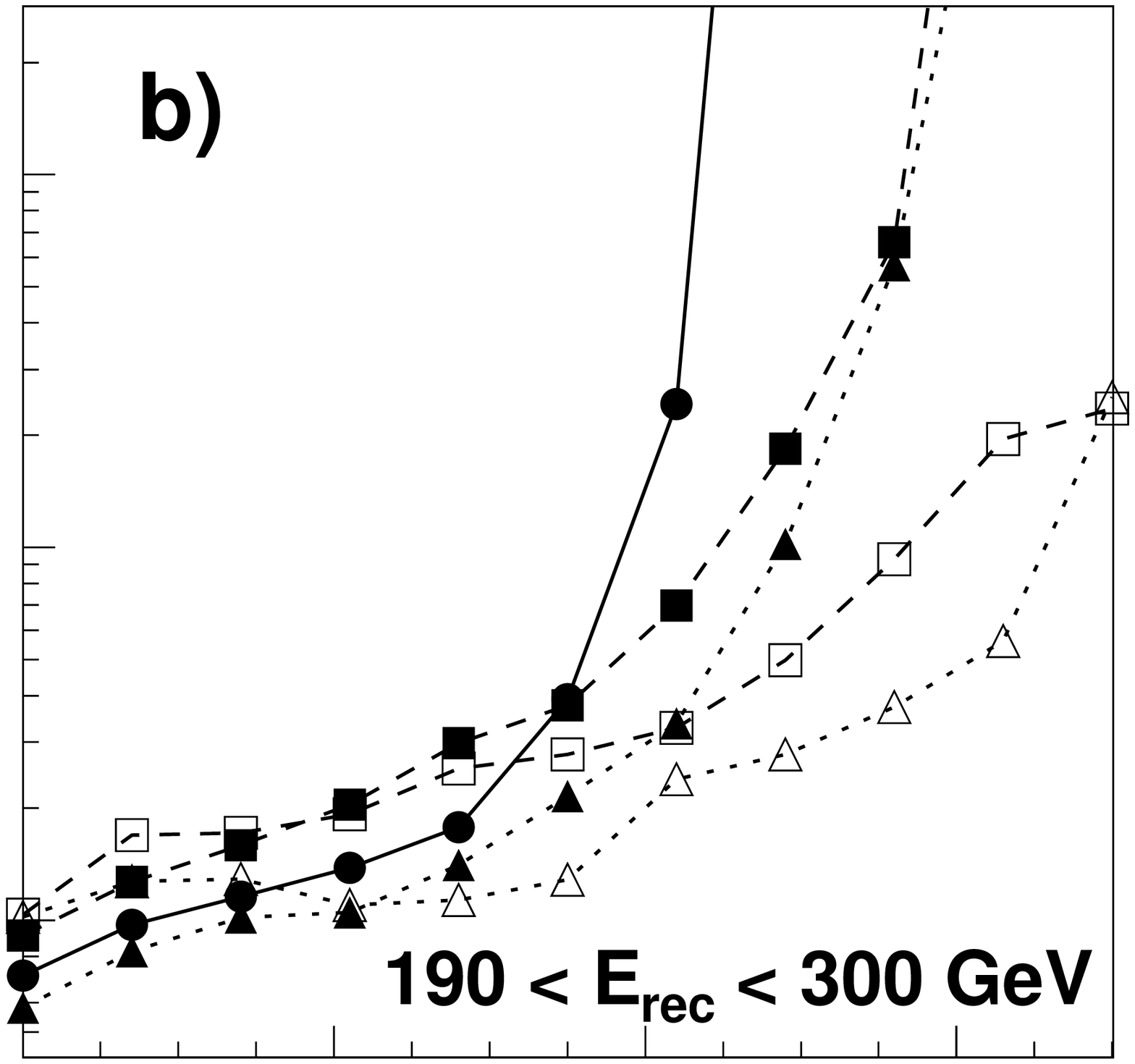}}\\
  \scalebox{0.95}{%
    \includegraphics[trim = 10mm 5mm 1.9mm 0mm, clip, width=6.0cm,height=6.0cm]{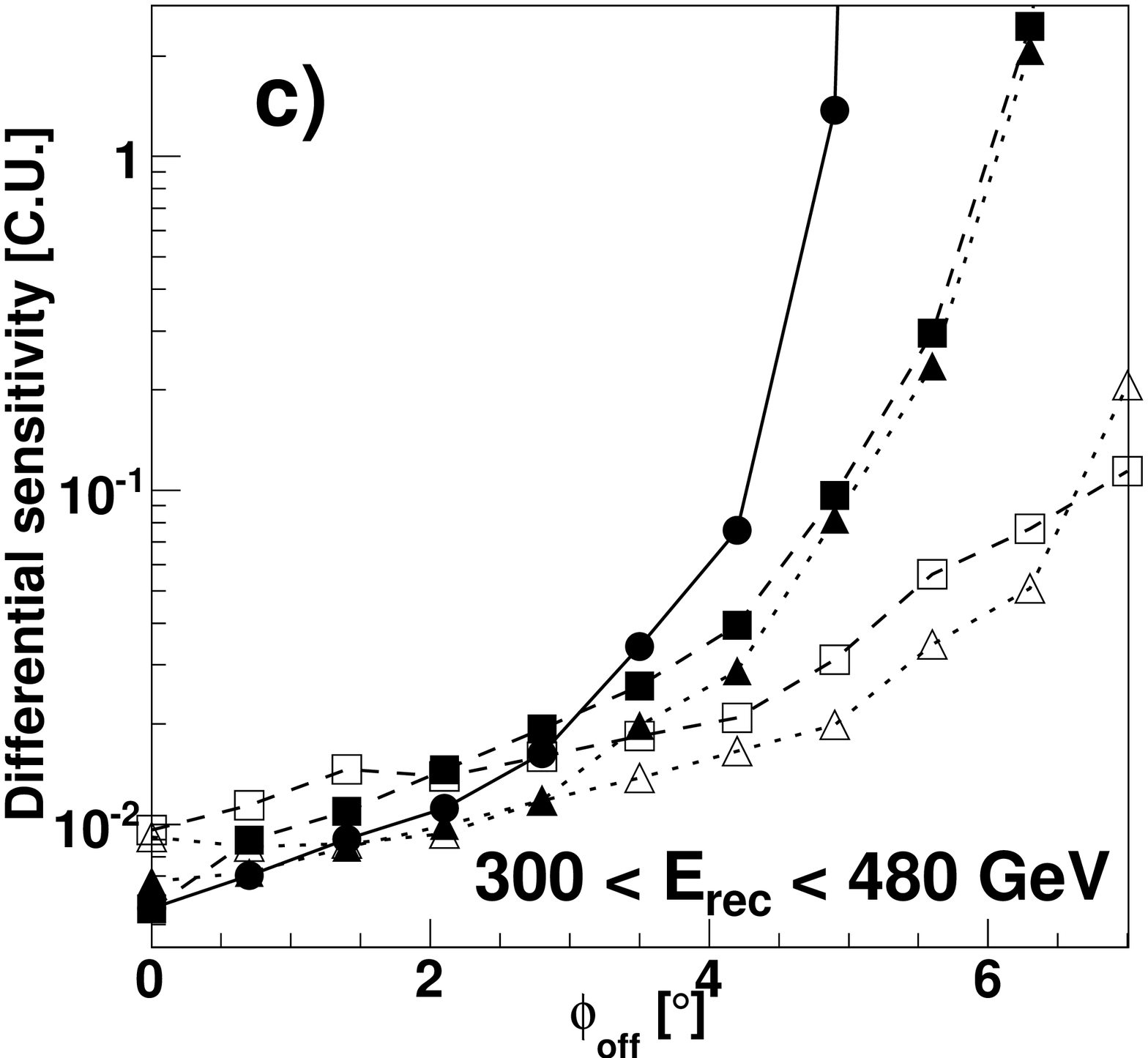}}%
  \scalebox{0.95}{%
    \includegraphics[trim = 10mm 5mm 1.9mm 0mm, clip, width=6.0cm,height=6.0cm]{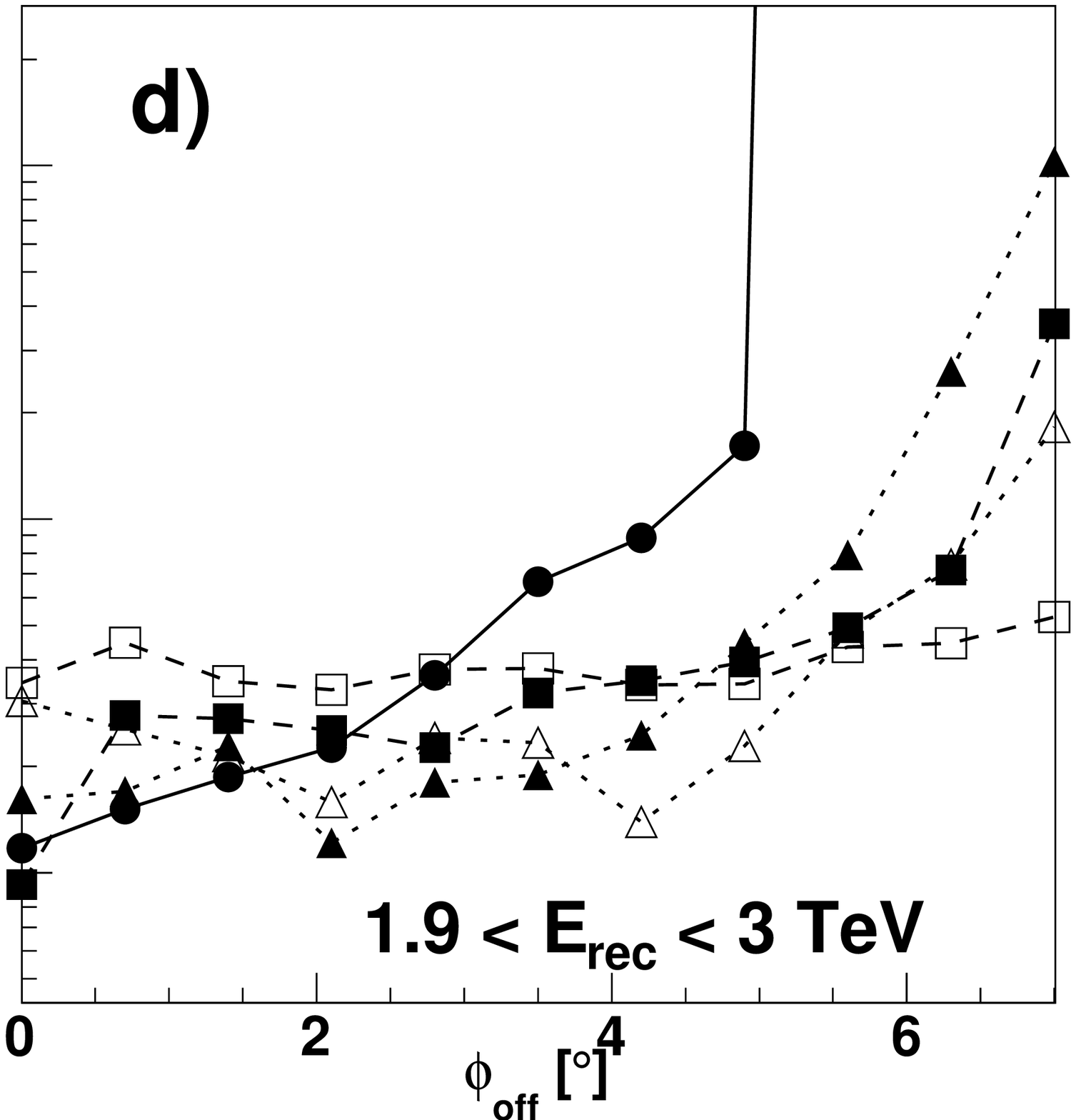}}
\caption{Differential sensitivity  as a function of $\phi_{\rm off}$  in four bins of the reconstructed energy bins, given in the panels, for the five modes represented by the same line and marker styles as in Fig.~\ref{fig:impact_energy}.
Standard CTA conditions for a significant detection are used
   ($5\sigma$ significance in $50\; {\rm h}$ with gamma excess at a level of at least 5\% of the remaining background). Sensitivity is given in standard HEGRA Crab Units, $1{\rm C.U.}\equiv 2.79\times10^{-11}E^{-2.57}\;{\rm cm^{-2}s^{-1}TeV^{-1}}$.}
\label{fig:sensitivity}
\end{center}
\end{figure*}

In Fig.\ref{fig:sensitivity} we present the sensitivity as a function of $\phi_{\rm off}$  in several energy bins.
Note that for mode N the number of events registered at   $\phi_{\rm off} \gtrsim 4^{\circ}$ is too small to satisfy the detection conditions specified above, so in this mode we formally obtain an infinite required flux outside of the nominal FOV of an individual MST.

To increase the statistic of proton events, all the events from each ring are selected. The total number of events selected in a ring is scaled to account for the difference of area between the ring and standard circular OFF regions. In each ring, the ratio, $\eta$, of the ON-source to the OFF-source proton events used in background estimation, is set to 1/5. Then, the detection conditions are given by $N_{\rm ex}\equiv N_{\rm \textlcsc{on}}-\eta N_{\rm \textlcsc{off}} >10$ and $N_{\rm ex} \geq 5\%N_{\rm bkg}$, where $N_{\rm bkg}\equiv \eta N_{\rm \textlcsc{off}}$.

\begin{figure}[t]
\begin{center}
 \scalebox{0.95}{%
   \includegraphics[trim = 3mm 5mm 2mm 5mm, clip, width=4.6cm,height=4.6cm]{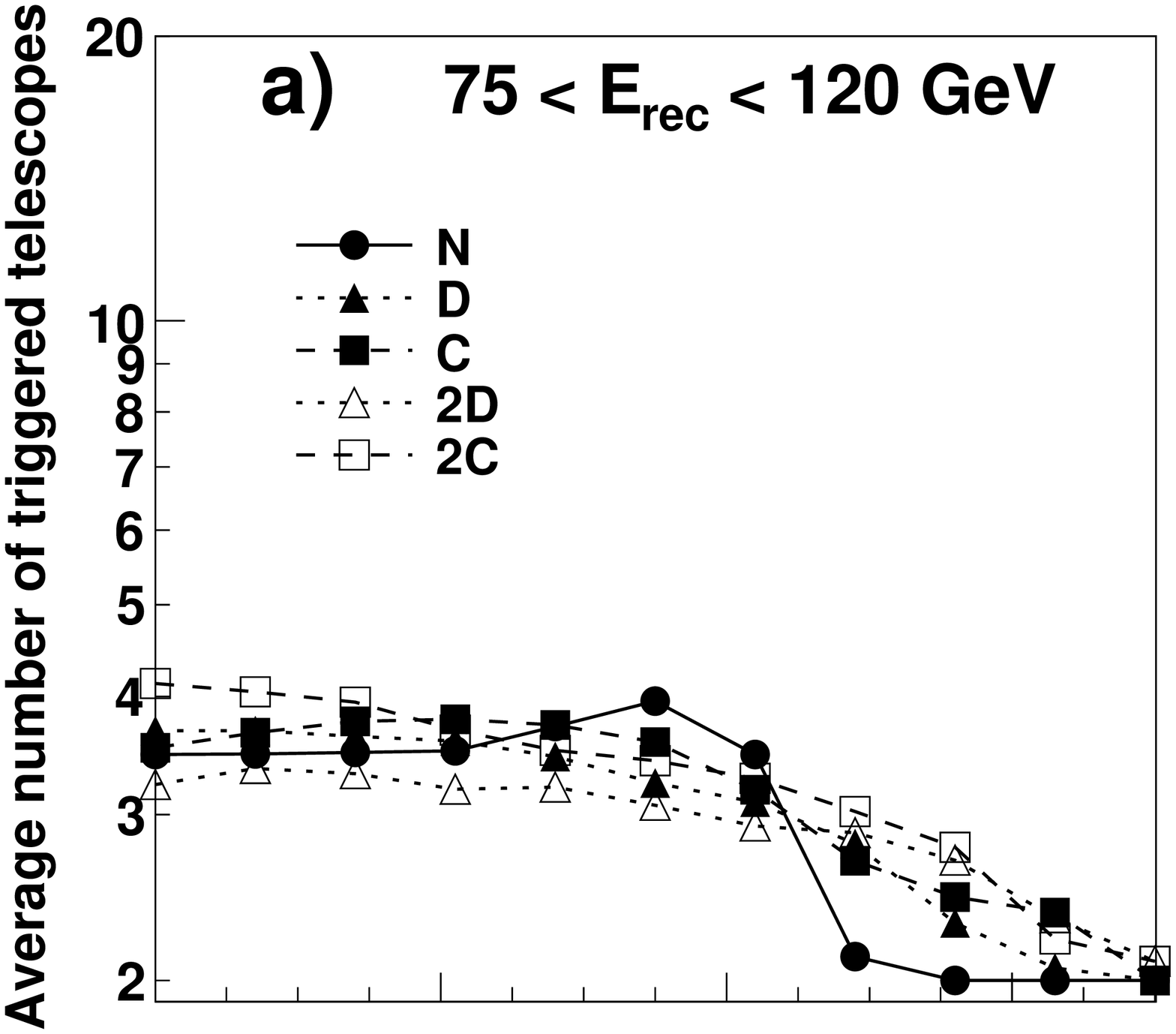}}%
 \scalebox{0.95}{%
   \includegraphics[trim = 3mm 5mm 2mm 5mm, clip, width=4.6cm,height=4.6cm]{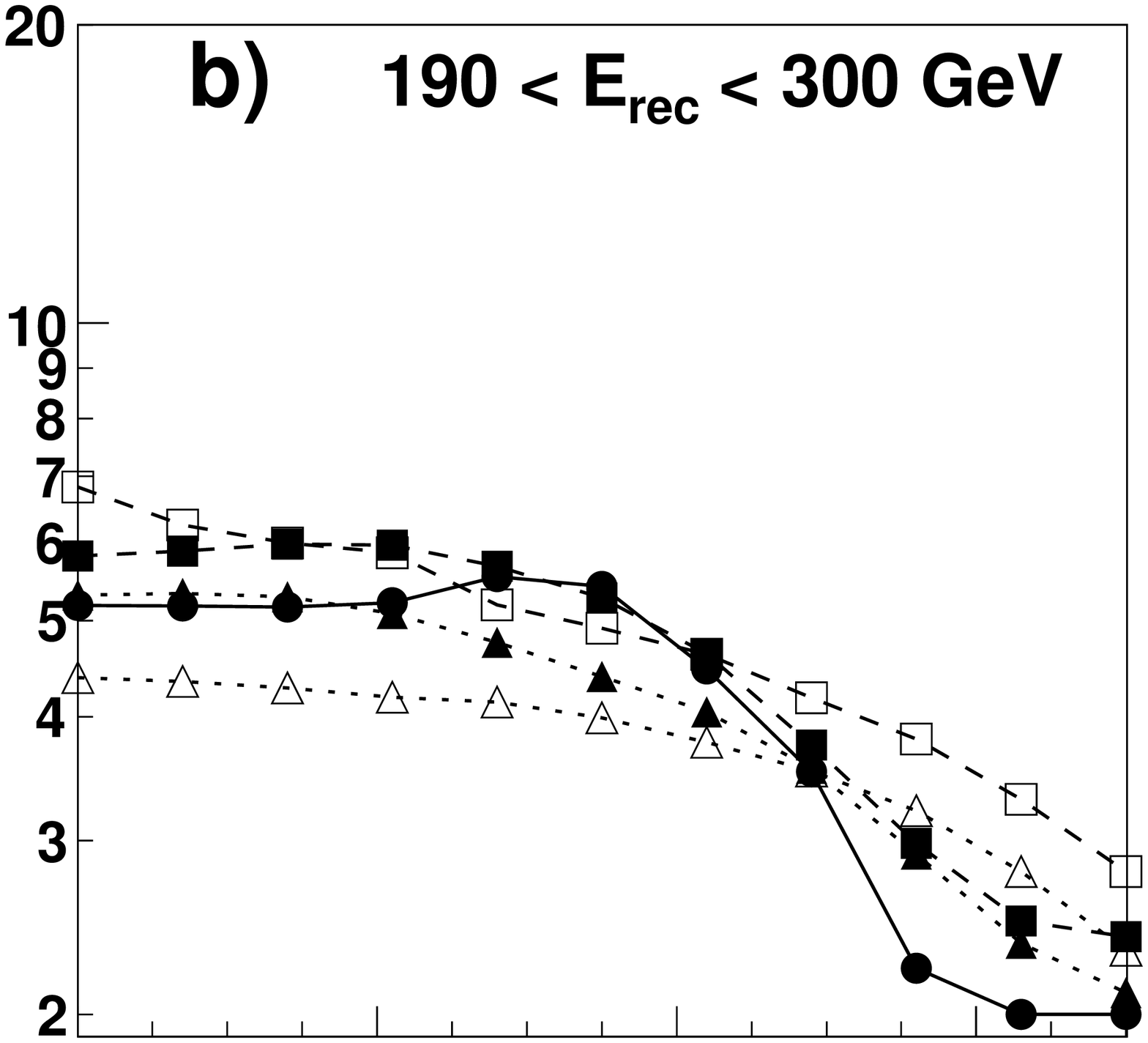}}
 \scalebox{0.95}{%
   \includegraphics[trim = 3mm 2mm 2mm 5mm, clip, width=4.6cm,height=4.6cm]{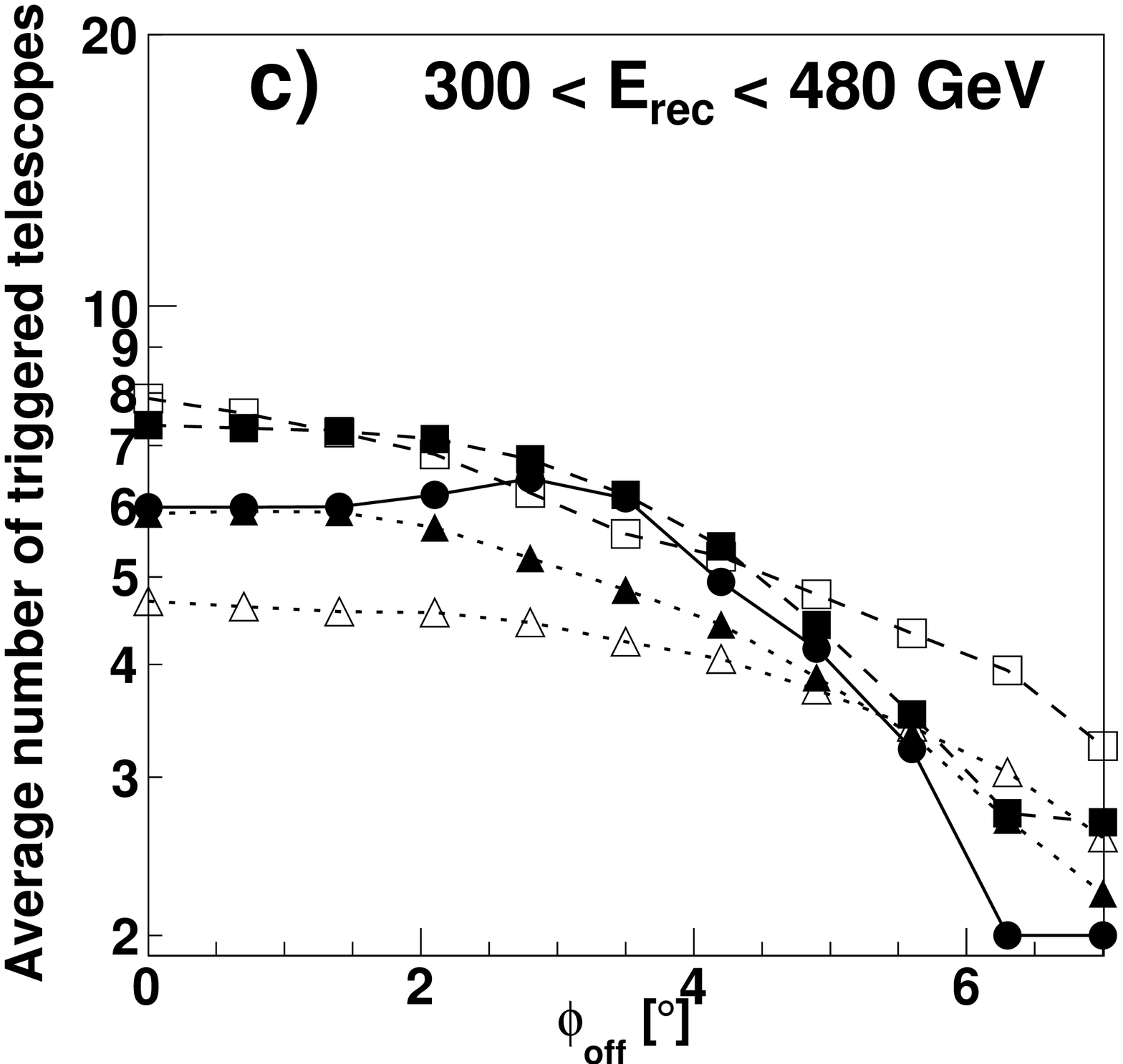}}%
 \scalebox{0.95}{%
   \includegraphics[trim = 3mm 2mm 2mm 5mm, clip, width=4.6cm,height=4.6cm]{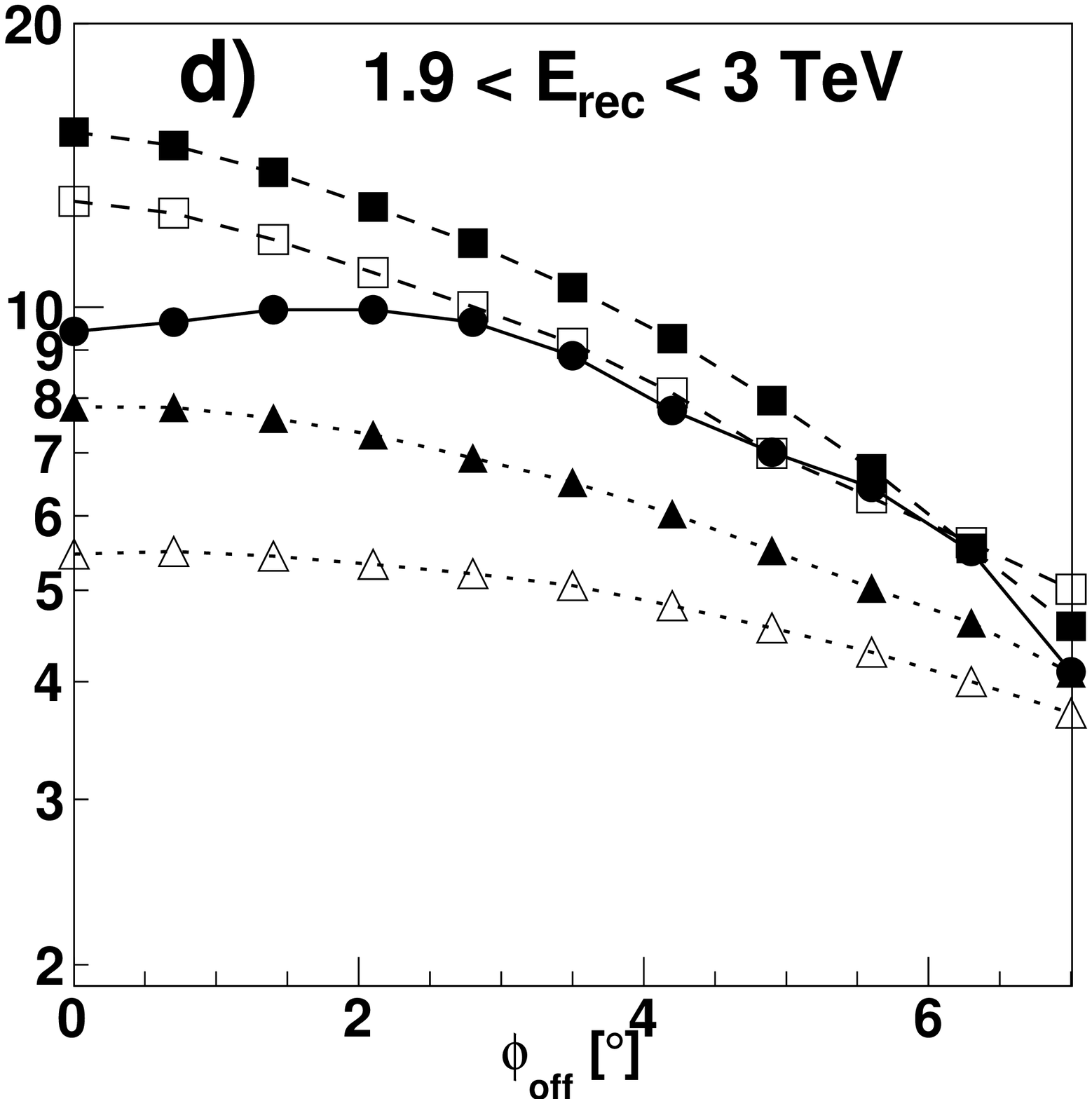}}

\caption{The average number of telescopes triggered by gamma-ray events as a function of $\phi_{\rm off}$ for the five modes, represented by the same line and marker styles as in Fig.~\ref{fig:sensitivity}. Panels (a)--(d) correspond to energy ranges used in panels (a)--(d) in Fig.~\ref{fig:sensitivity}.}
\label{fig:N_trg_off}
\end{center}
\end{figure}

The sensitivity of a telescope system is mainly determined by the effectiveness of gamma/hadron separation cuts and the (trigger) efficiency of gamma-events detection;
obviously, the differential sensitivity depends also on the accuracy of the energy reconstruction. We note that the efficiency of separation cuts at given $E$ and $\phi_{\rm off}$  differs 
significantly  between the modes, with differences of the fraction of rejected proton events exceeding an order of magnitude.
The efficiency of the $\theta^{2}$ cut is directly related with the angular resolution and 
the efficiency of the cut on the number of triggered telescopes depends on $N_{\rm tel}$ (see Fig.~\ref{fig:Ntrg_impacts} and Fig.~\ref{fig:N_trg_off}) and, then, these separation cuts are particularly efficient in both C and 2C modes. Since for the convergent modes $N_{\rm tel}$ rapidly increases with the energy (see Fig.~\ref{fig:N_trg_off}),  the cut on the number of triggered telescopes in convergent modes becomes particularly efficient above $\approx 300\; {\rm GeV}$. For those energies and $\phi_{\rm off} \gtrsim 4^{\circ}$, the sensitivities of convergent modes are comparable or even better than those of the corresponding divergent modes, see Figs.~\ref{fig:sensitivity}(c--d). In mode 2C we additionally note highly efficient MSW cuts  at large $\phi_{\rm off}$. 

The separation cuts in the divergent modes are typically less efficient,
nevertheless, the sensitivity of mode 2D is much better than those of other modes for large $\phi_{\rm off}$, while mode D has a good sensitivity at small $\phi_{\rm off}$, in both cases due to high trigger efficiencies of gamma events. 

Summarizing this section, we emphasize two properties crucial for our final conclusions:

\noindent
(i) a high sensitivity of mode C for $\phi_{\rm off}=0$, resulting from highly efficient cuts in both the direction and the number of telescopes (see Fig.~\ref{fig:N_trg_off}),

\noindent
(ii) a high sensitivity of mode 2D at large $\phi_{\rm off}$, resulting from the high trigger efficiency.

As a final remark we emphasize that the sensitivities reported in this section serve only the purpose of comparing the different pointing configurations. The more advanced gamma/hadron separation methods are currently deployed in CTA on a more realistic background profiles. The sensitivities provided with those methods give a more  adequate expectations of the absolute sensitivities of the CTA instrument.

\subsubsection{Collection areas and acceptance}
\label{sec:rates_collection_analysis}

\begin{figure}[t]
\begin{center}
  \includegraphics[trim = 6mm 1mm 0mm 5mm, clip,scale=0.459]{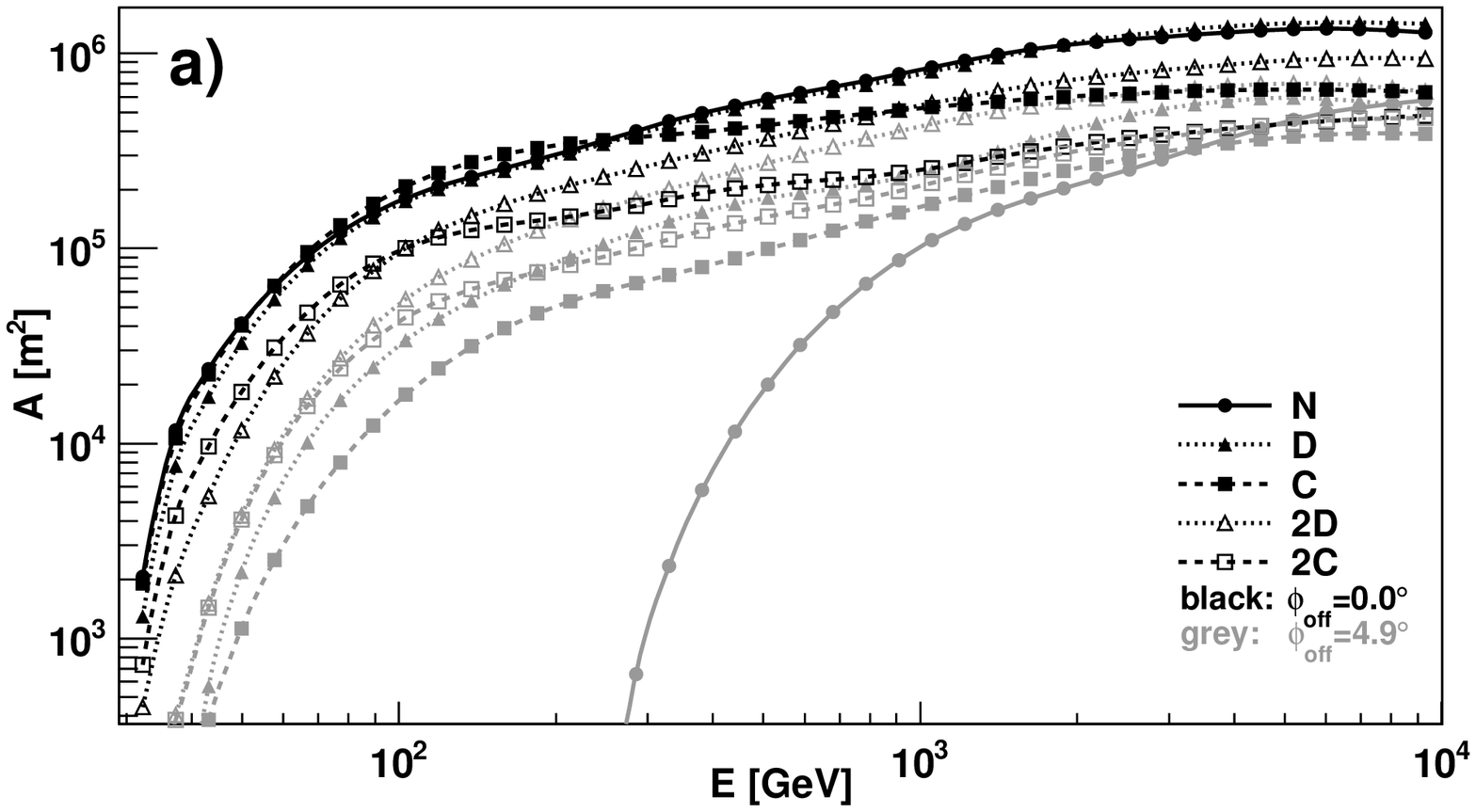}
  \scalebox{0.95}{%
    \includegraphics[trim = 0mm 0mm 0mm 0mm, clip, width=4.6cm,height=4.6cm]{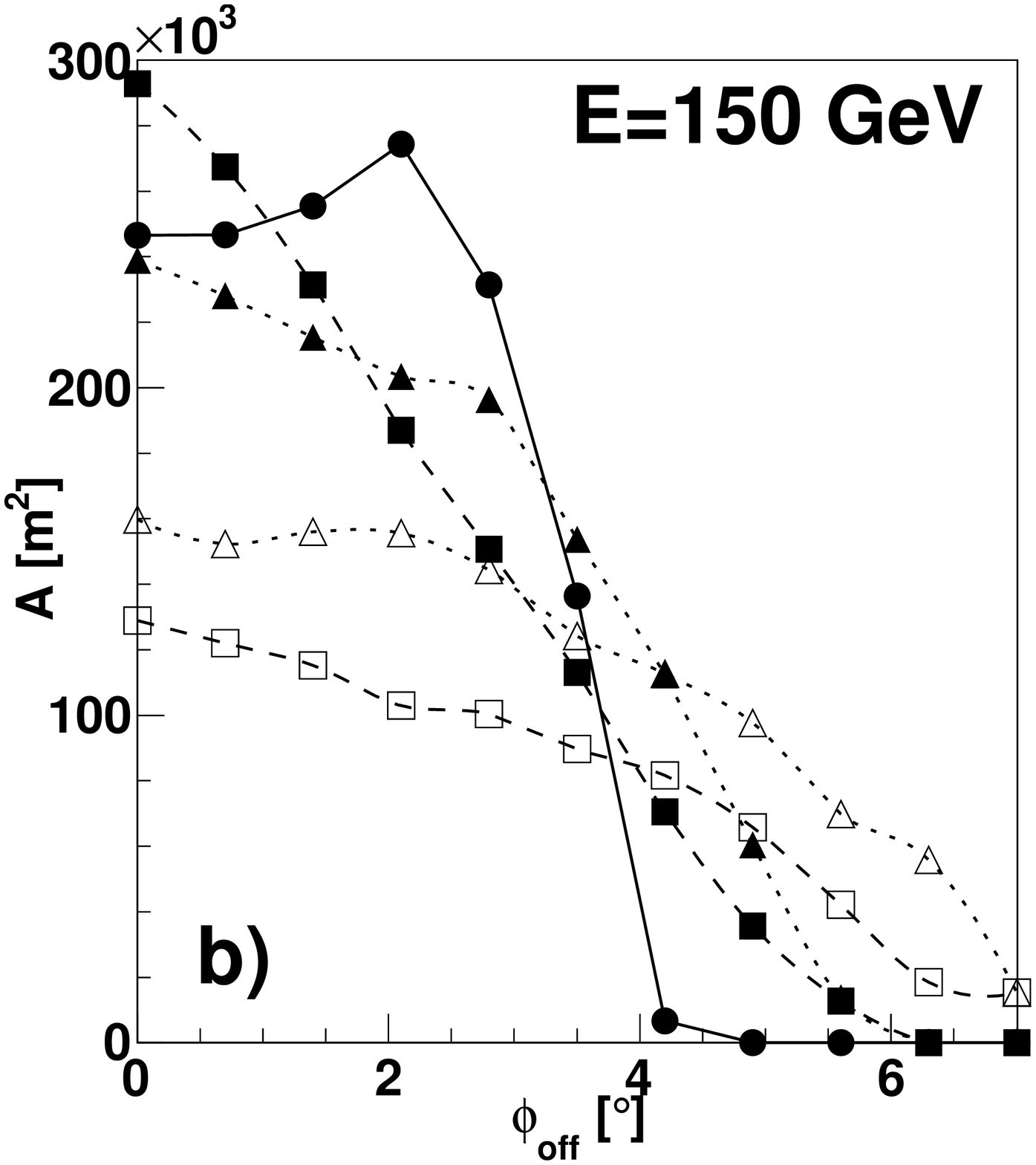}}%
  \scalebox{0.95}{%
    \includegraphics[trim = 0mm 0mm 0mm 0mm, clip, width=4.6cm,height=4.6cm]{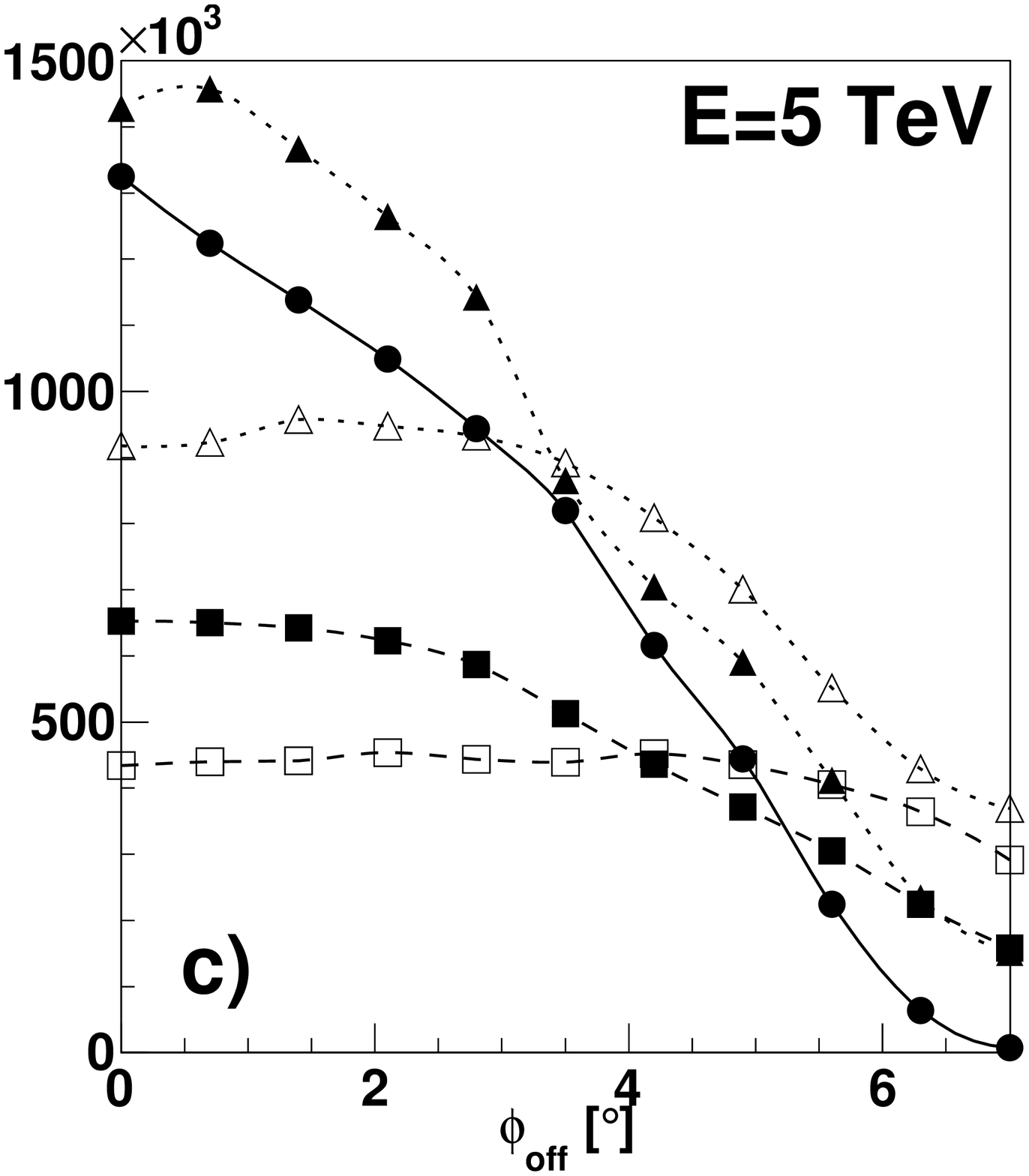}}

\caption{Analysis-level collection area as a function of $E$ (top panel a) and $\phi_{\rm off}$ (bottom panels) for the five modes, represented by the same line and marker styles as in Fig.~\ref{fig:impact_energy}. In panel (a) the black lines and markers are for $\phi_{\rm off} = 0^{\circ}$ and the grey ones for  $\phi_{\rm off} =4.9^{\circ}$. Panel (b) is for $E = 150\; {\rm GeV}$ and panel (c) for $E = 5\; {\rm TeV}$.}
\label{fig:collection_areas_analysis}
\end{center}
\end{figure}

\begin{figure}[t]
\begin{center}
  \includegraphics[trim = 6mm 1mm 0mm 5mm, clip,scale=0.458]{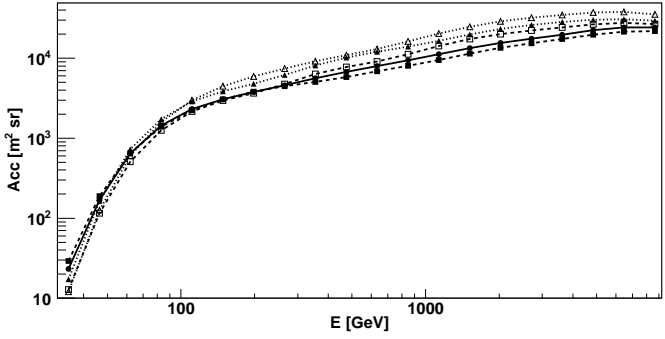}

\caption{Acceptance at the analysis level as a function of simulated energy for the five modes, represented by the same line and marker styles as in Fig.~\ref{fig:impact_energy}.}
\label{fig:FoV_cuts}
\end{center}
\end{figure}

Rejection of gamma-ray events in gamma/hadron separation reduces the collection areas at the analysis level with respect to the trigger level by a factor of $\sim 2$ for $\phi_{\rm off} = 0^{\circ}$ and by a factor of several for $\phi_{\rm off} >4^{\circ}$, see Figs.~\ref{fig:areas} and \ref{fig:collection_areas_analysis}. This implies a relatively weaker contribution from large $\phi_{\rm off}$ to Acc at the analysis level and, furthermore, the decrease of Acc by a factor of several in all modes, see Figs.~\ref{fig:FOV_ACC} and \ref{fig:FoV_cuts}.   However, the 2D mode has the largest  collection areas at large   $\phi_{\rm off}$ and also the largest Acc in the whole energy range, similarly as at the trigger level. We also note that the fraction of rejected events is different in various modes at low energies, however, at   $E \gtrsim 1\; {\rm TeV}$
it is similar in all modes and hence at high energies the ratio of their collection areas or Acc values is roughly the same at the analysis and trigger levels.

\subsubsection{Quality of reconstruction}
\label{sec:reconstruction_analysis}

The background rejection is accompanied with a large rejection, exceeding a factor of 2 in magnitude, of gamma-ray events (c.f.~Fig.~\ref{fig:areas} and Fig.~\ref{fig:collection_areas_analysis}). Major part of excluded shower images are of low quality, thus their rejection improves the reconstruction efficiency. Consequently, the reconstruction performance of an array at the analysis level is in principle at least as good as before introduction of the background suppression cuts.

In this section we briefly consider changes in reconstruction efficiency of studied modes at the analysis level. We do not discuss differences in energy bias as they are comparable with those before introduction of gamma/hadron separation cuts (for details see Sect.~\ref{sec:reconstruction}). 

\begin{figure}[t!]
\begin{center}
  \includegraphics[trim = 3mm 0mm 0mm 0mm, clip,scale=0.458]{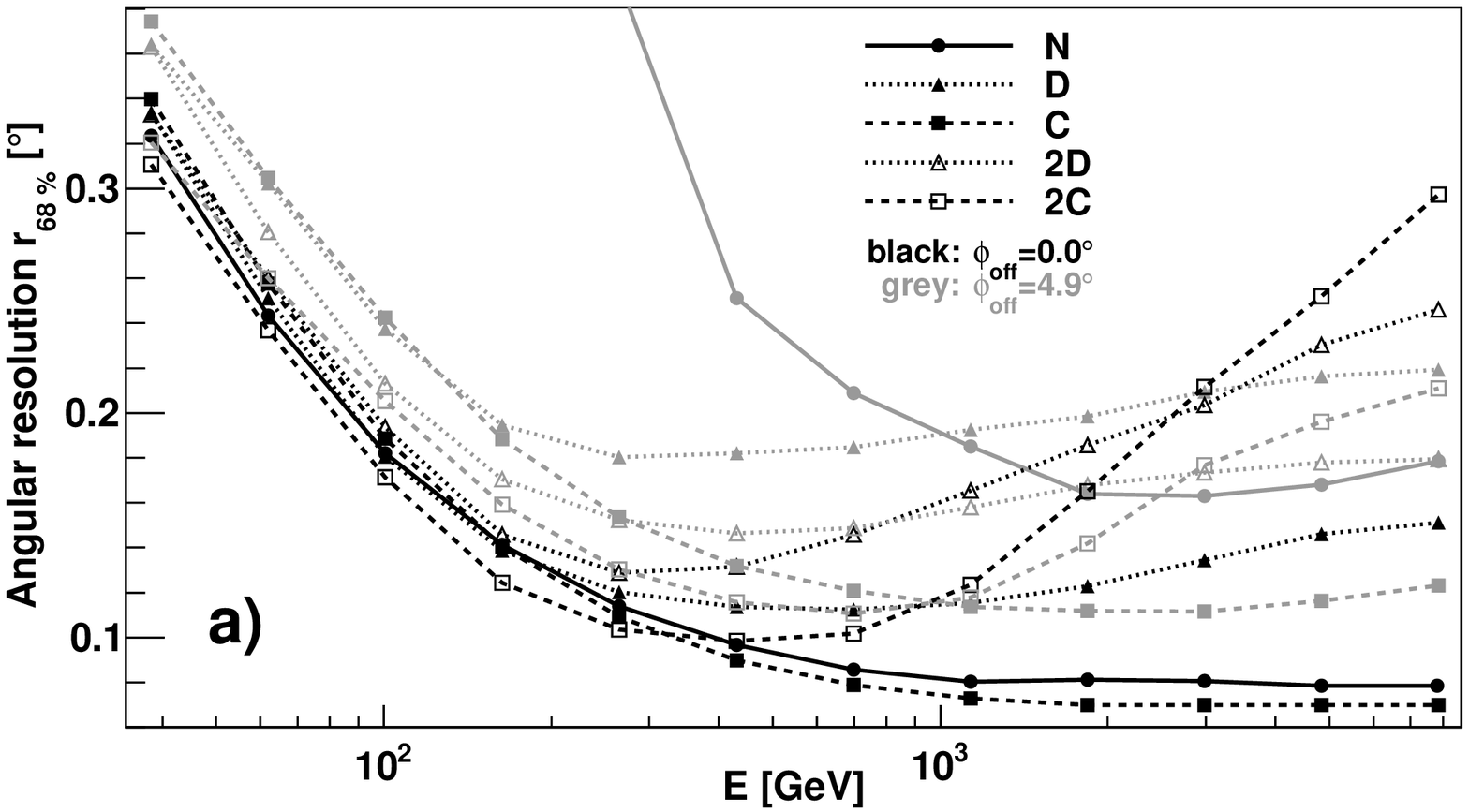}
  \scalebox{0.95}{%
    \includegraphics[trim = 3mm 2mm 2mm 0mm, clip, width=4.6cm,height=4.6cm]{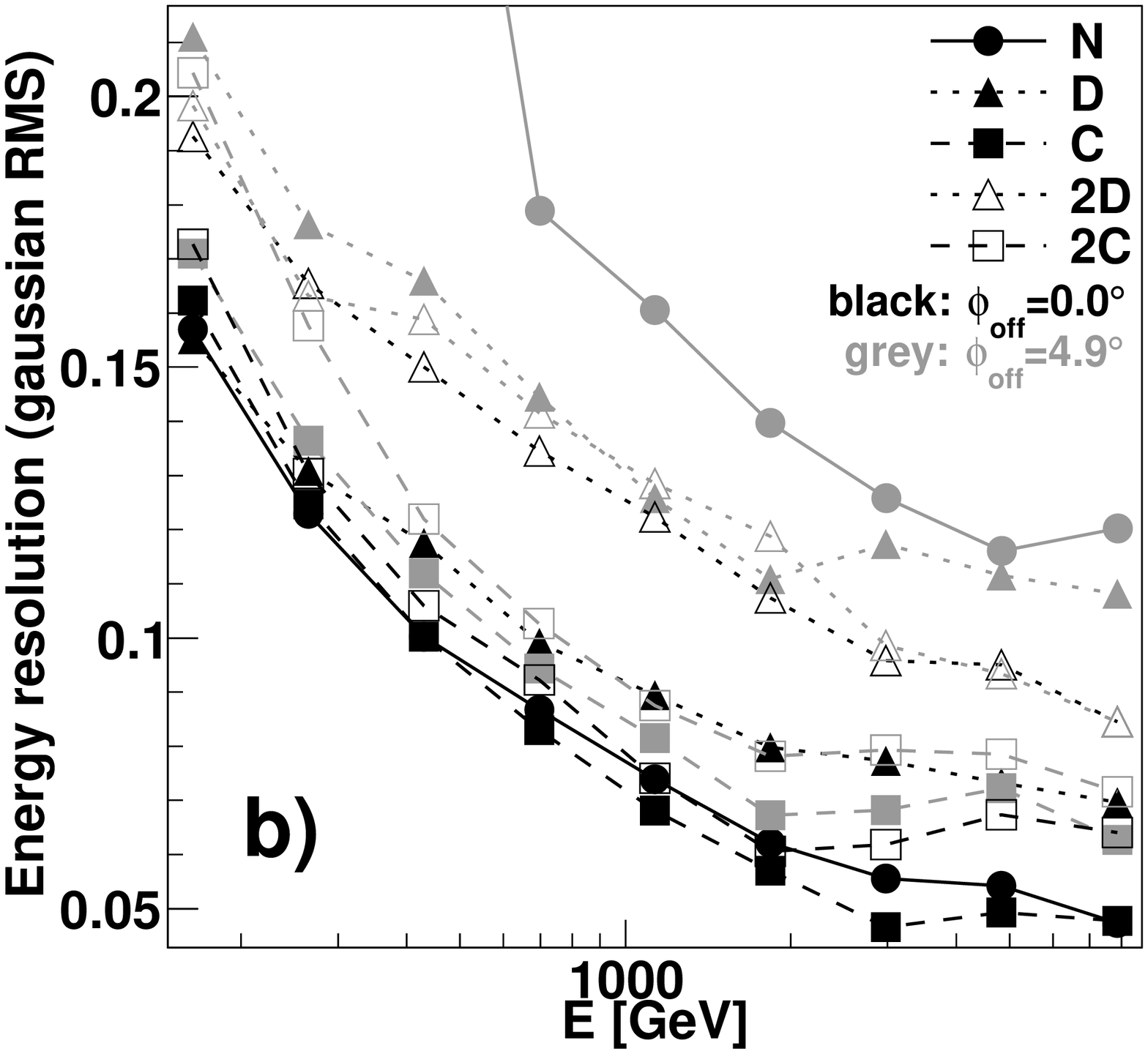}}%
  \scalebox{0.95}{%
    \includegraphics[trim = 3mm 2mm 2mm 0mm, clip, width=4.6cm,height=4.6cm]{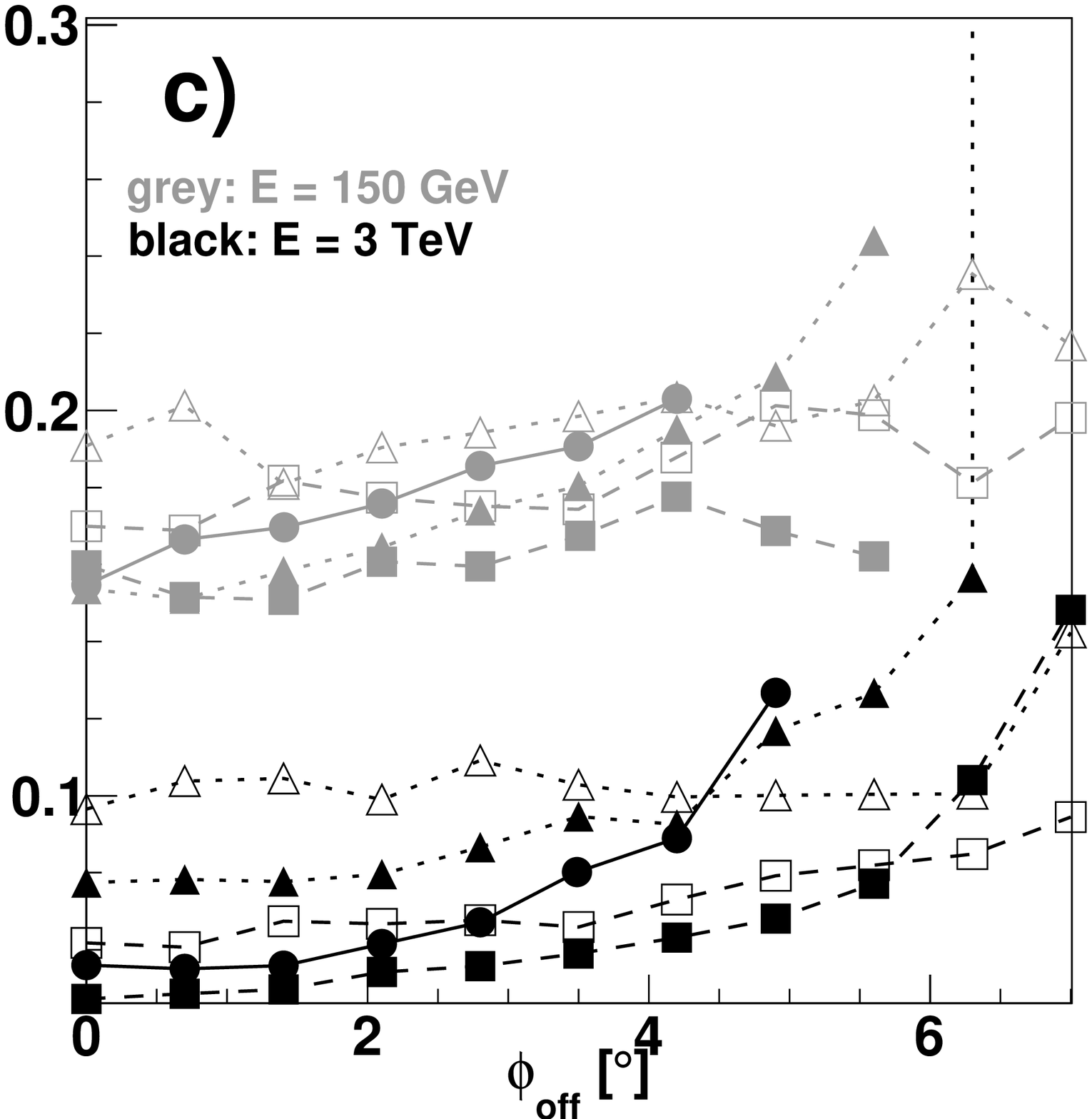}}

\caption{Analysis-level angular and energy resolutions, for the five modes represented by the same line and marker styles as in Fig.~\ref{fig:impact_energy}. Panel (a) shows the angular resolution for the $68\;\%$ containment radius as a function of the simulated energy.
Bottom panels show the energy resolution as a function of the simulated energy (a) and  $\phi_{\rm off}$ (b).
The black lines and markers in panels (a,b) are for $\phi_{\rm off} = 0^{\circ}$ and the grey ones for  $\phi_{\rm off} =4.9^{\circ}$. 
In panel (c)
the black lines and markers are for $E = 3\; {\rm TeV}$ and the grey ones for  $E = 150\; {\rm GeV}$.}
\label{fig:en_res_cuts}
\end{center}
\end{figure}

By virtue of used cuts the large leakage effect noticed in Sect.~\ref{sec:reconstruction}, which mainly affected divergent arrays, is significantly reduced. The resulting improvement in angular resolution is the most noticeable in the high energy range $>2\; {\rm TeV}$ for 2D mode where for  $\phi_{\rm off}=4.9^{\circ}$ it is comparable with N and it is even better than 2C (see Fig.~\ref{fig:en_res_cuts}a). Interestingly for both convergent modes the effect of leakage being the lowest before use of analysis cuts, is not further reduced at the analysis level - the angular resolutions corresponding to C and 2C do not significantly change.

Similar improvement is noticed in energy resolution of arrays set in divergent modes, see Fig.~\ref{fig:en_res_cuts} (b and d) (c.f.~with Fig.~\ref{fig:energy_res} (a and b)). For example, at fixed energy of $3\;{\rm TeV}$, the energy resolutions for D and 2D modes are improved by almost $25\; \%$. For other modes, the energy reconstruction after gamma-selection cuts is also more accurate. However the decrease of the corresponding values of energy resolution is less than $10\; \%$. Note that the C mode which provides the best energy resolution before gamma-selection cuts (see Fig.~\ref{fig:energy_res} (a and b)) remains the best also after use of cuts.

\begin{figure*}[t]
  \begin{center}
  \includegraphics[trim = 3mm 0mm 0mm 11mm, clip,scale=0.458]{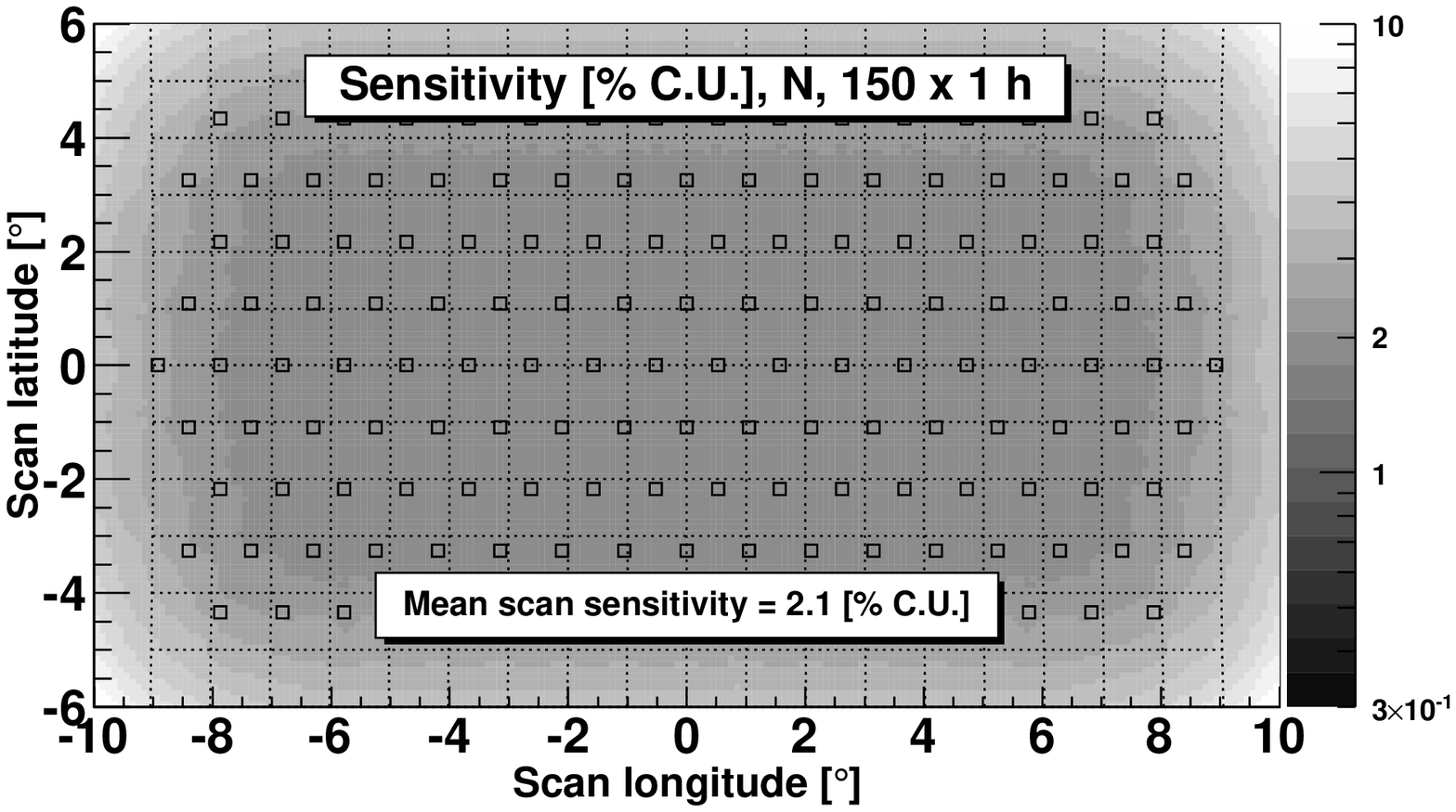}\\
  \includegraphics[trim = 3mm 7mm 0mm 11mm, clip,scale=0.458]{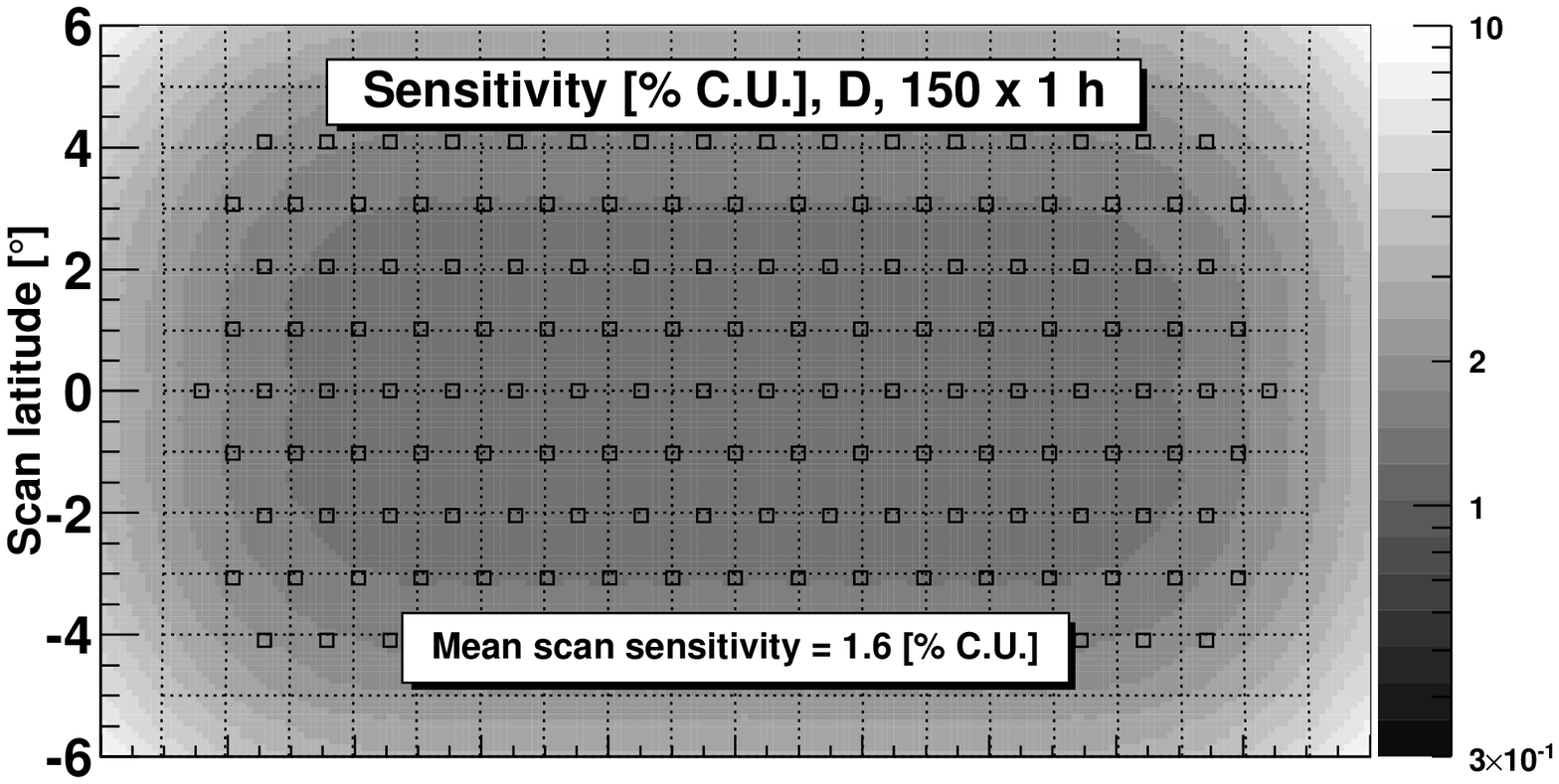}
  \includegraphics[trim = 3mm 7mm 0mm 11mm, clip,scale=0.458]{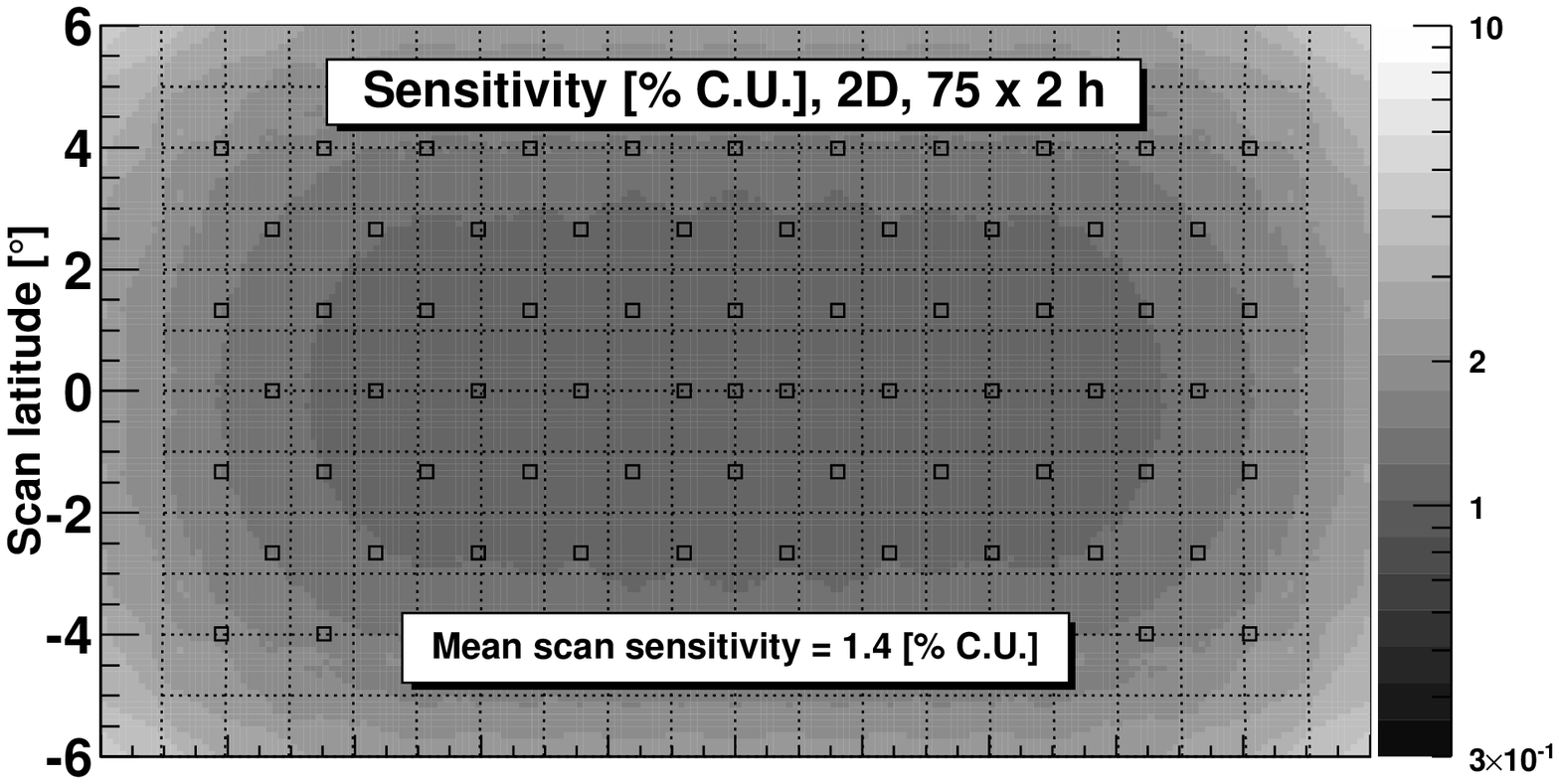}
  \includegraphics[trim = 3mm 7mm 0mm 11mm, clip,scale=0.458]{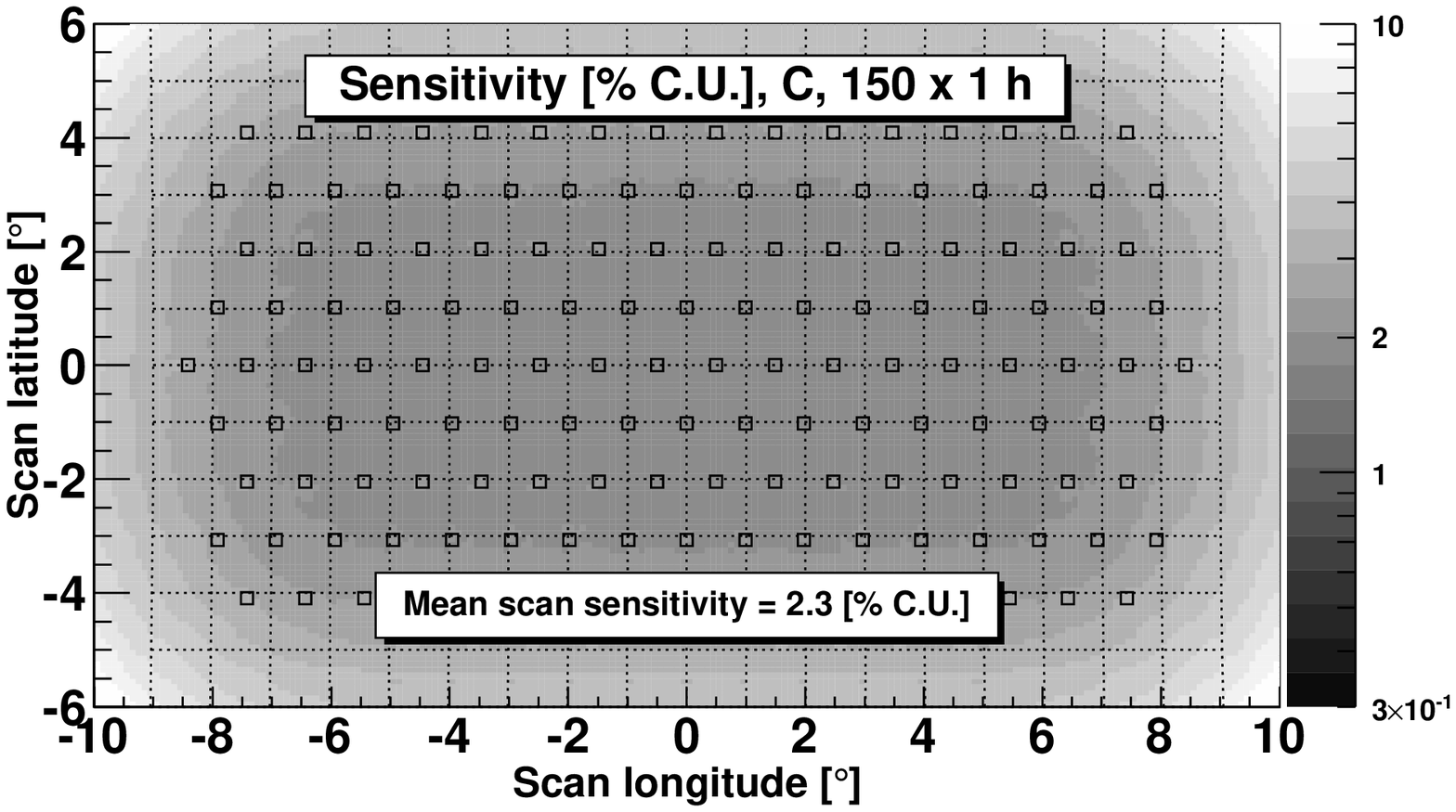}
  \includegraphics[trim = 3mm 7mm 0mm 11mm, clip,scale=0.458]{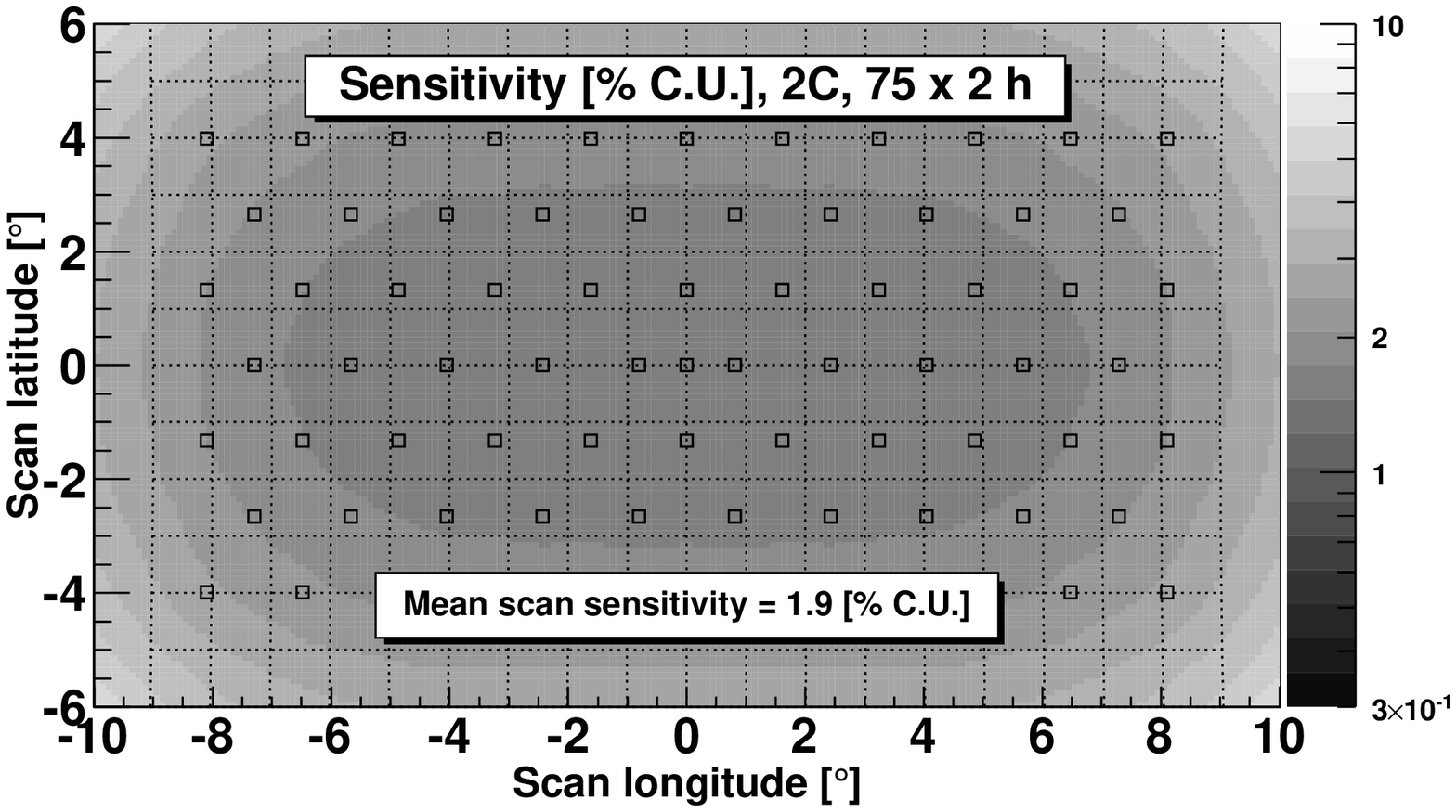}
    \caption{Sensitivity at energies between 300 and $480\; {\rm GeV}$ in sky scans performed with the five modes. The sky area of $20^{\circ}\times 12^{\circ}$ is covered with the same exposure time of 150 h for each mode. We use a simple multiple-row scan strategy to uniformly expose the investigated sky region. Because of the relatively larger effective FOV of modes 2D and 2C, the number of pointings for these modes is reduced by a factor of 2. The mean scan sensitivity achieved after the total exposure time of 150 h is given for each mode for a robust comparison of sky-survey efficiencies.}
  \label{fig:sky_scans_C}
\end{center}
\end{figure*}

In this section we present results of sky scans performed with the five modes on a fixed sky area of $20^{\circ}\times 12^{\circ}$, see Fig.~\ref{fig:sky_scans_C}.
For modes 2D and 2C we use two times fewer pointings, taking into account the large effective FOV and the significantly better sensitivity at large $\phi_{\rm off}$ in these modes. The total exposure/scan time is the same for each mode. To roughly compare the scan efficiency of each studied mode, we calculate the mean sensitivity, $S_{\rm MODE}$,  achieved during the scan time of 150 h.

Technically, to evaluate $S_{\rm MODE}$ we compute the total gamma-ray exposure and the total background rate for each sky coordinate from all the sky pointing positions. Then for each coordinate we calculate the sensitivity (flux fraction) satisfying the conditions described in Sec.~\ref{sec:sensitivity}. $S_{\rm MODE}$ is finally computed by averaging the achieved sensitivities over the whole map of coordinates. To avoid possible large contributions from regions where the flux fractions are very large (to achieve the significance of 5 $\sigma$),  the averaging is carried out with a use of the harmonic mean. The relative statistical uncertainty, $\Delta S_{\rm MODE}/ S_{\rm MODE}$, of $S_{\rm MODE}$, estimated with an RMS of the distribution of significance over a whole sky map, is of order of $0.2$.

\subsubsection{Sky scans}
\label{sec:sky_scans}

In all modes, we find the best mean sensitivities in the energy bin between 300 and $480\; {\rm GeV}$ and Fig.~\ref{fig:sky_scans_C} presents the sensitivity map for this bin; the mean values of $S$ at higher and lower $E$ are given in Table \ref{tab:mean_sens}.
\begin{table}[h]
  \begin{center}
    \begin{tabular}{|c|c|c|c|c|c|}
      \hline
      \multicolumn{6}{|c|}{\bf Mean scan sensitivities [\% C.U.]}\\
      \hline
      Energy [GeV] &  N & D & C & 2D & 2C \\
      \hline\hline
      120-190 & 2.5 & 2.0 & 3.5 & 2.0 & 3.6 \\
      300-480 & 2.1 & 1.6 & 2.3 & 1.4 & 1.9 \\
      1900-3000 &4.2 &2.3 & 3.0 &2.3 &3.2 \\
      \hline
    \end{tabular}
  \end{center}
\caption{Mean sensitivities in sky scans performed with the five modes
 for the pointing layouts presented in Fig.~\ref{fig:sky_scans_C}.}
 \label{tab:mean_sens}
\end{table}

We see that at $E=(300-480)$ GeV the best scan results are obtained with the 2D mode. Its $S_{\rm 2D}$ is by $\approx 33\%$ better than $S_{\rm N}$, which implies that a scan can be performed with mode 2D in a time shorter by a factor $(S_{{\rm N}}/S_{{2\rm D}})^{2}\approx 2.3$ than with mode N. Also for modes D and 2C, $S_{{\rm D}}=1.6 \% \;{\rm C.U.}$ and $S_{{\rm 2C}}=1.9 \%\;{\rm C.U.}$ indicate reduced sky-scan times, by a factor of $\approx 1.7$  and $\approx 1.2$ with respect to mode N. The worst mean scan sensitivity corresponds to mode C.

Properties similar to the above occur also at low and high $E$, although here  $S_{\rm 2D} \simeq S_{\rm D}$. However, we regard the comparison of mean $S$ 
at the intermediate $E$ as the most informative.

We emphasize that the mean scan sensitivities shown above resulting from a simple analysis methods and are used only for relative comparisons of modes. The more advanced analysis techniques with introduction of a more realistic background profiles could lead to a different absolute values of sensitivities. On the other hand, the possible contribution of the other background components is expected to not too significantly change the relative differences between mean sky sensitivities obtained by the modes. By virtue of a very approximate method of using gamma rays with different offsets from the camera center as a proxy for diffuse electrons (being one of the most irreducible background components)  we estimated the possible changes of sensitivities due to electrons contribution. The resulting decrease of relative differences between the mean sky sensitivities does not exceed $\sim 10\%$ and consequently the final results presented in this work, at least qualitatively, hold.

\section{Summary and discussion}
\label{sec:summary}
Using the CTA MST subarray, we studied various modes of observation with a large array of IACTs. 
Our results should help in the development of an optimal observing  strategy, making a full use of the capabilities of CTA.

\begin{figure}[t!]
\begin{center}
  \scalebox{0.95}{%
    \includegraphics[trim = 3mm 2mm 2mm 0mm, clip, width=4.9cm,height=4.9cm]{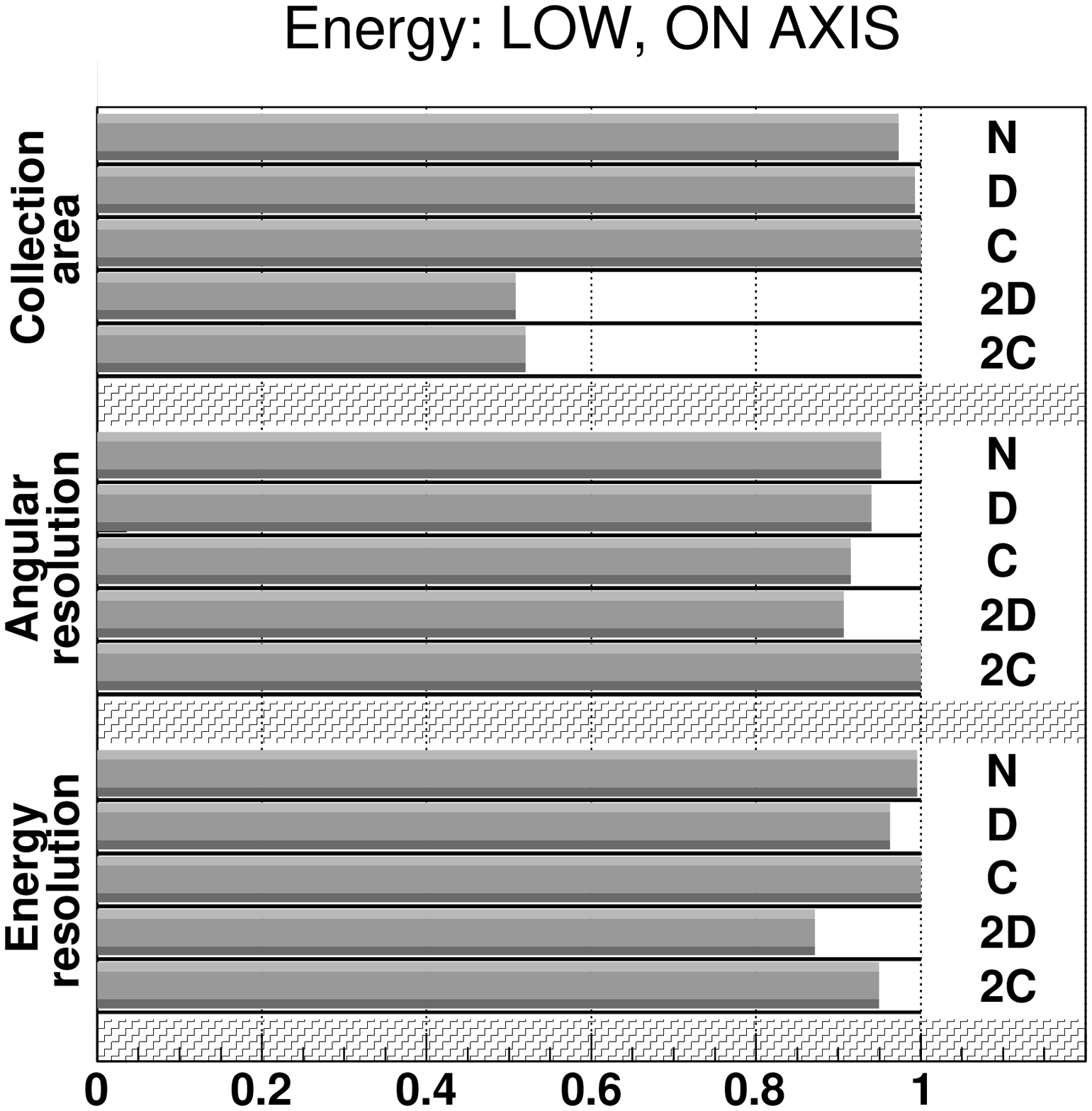}}%
  \scalebox{0.95}{%
    \includegraphics[trim = 3mm 2mm 2mm 0mm, clip, width=4.9cm,height=4.9cm]{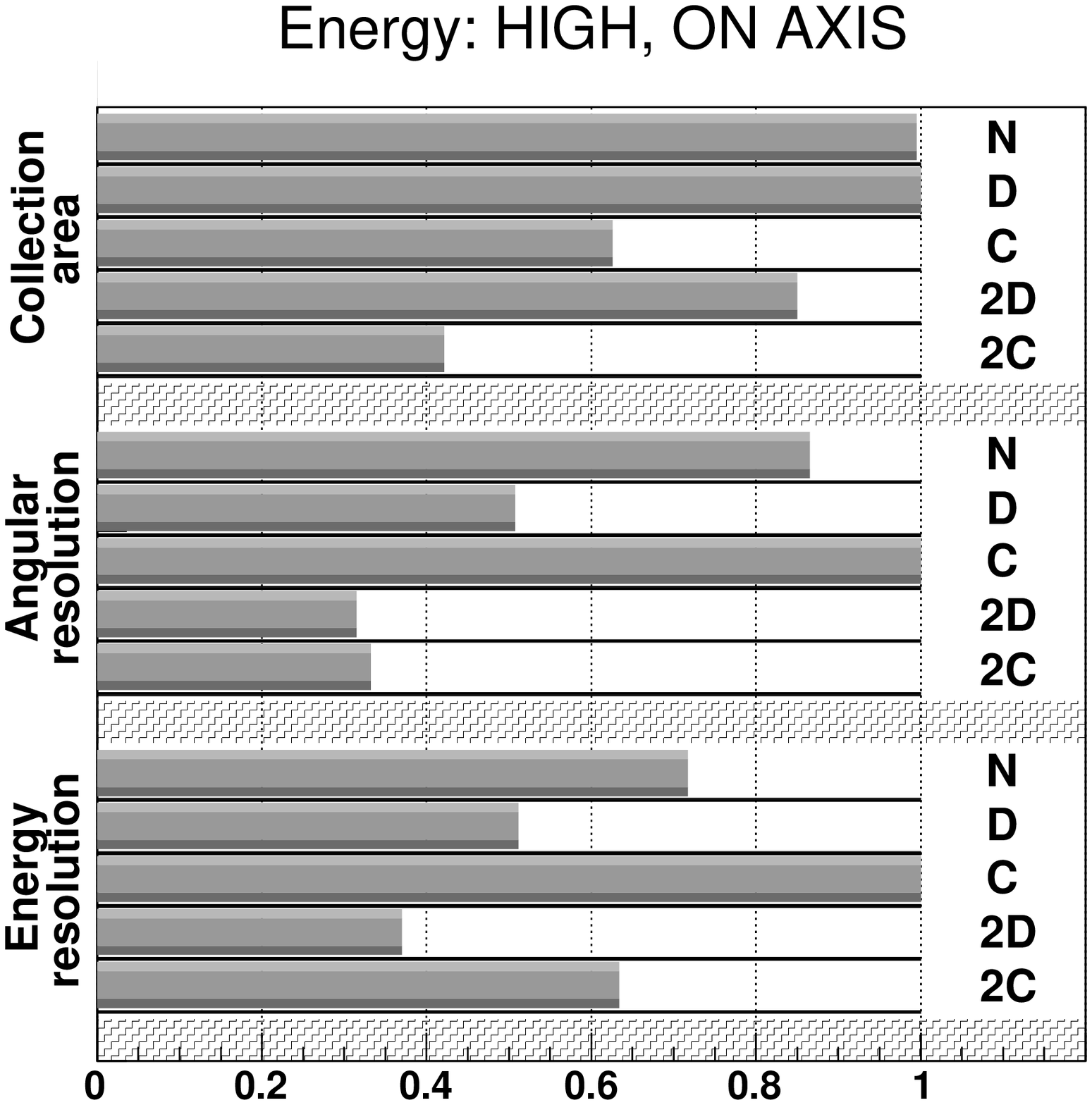}}
  \scalebox{0.95}{%
    \includegraphics[trim = 3mm 2mm 2mm 0mm, clip, width=4.9cm,height=4.9cm]{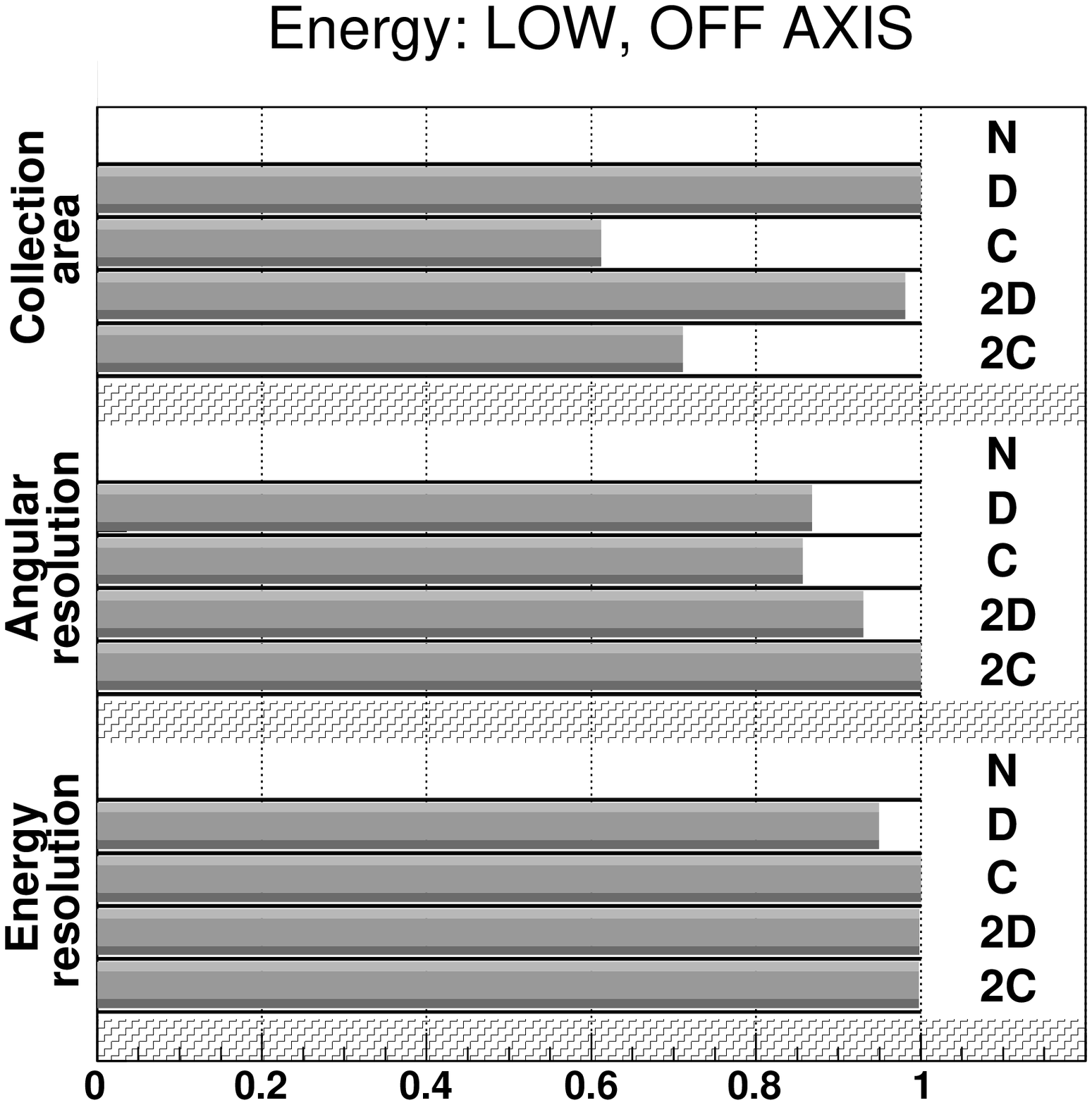}}%
  \scalebox{0.95}{%
    \includegraphics[trim = 3mm 2mm 2mm 0mm, clip, width=4.9cm,height=4.9cm]{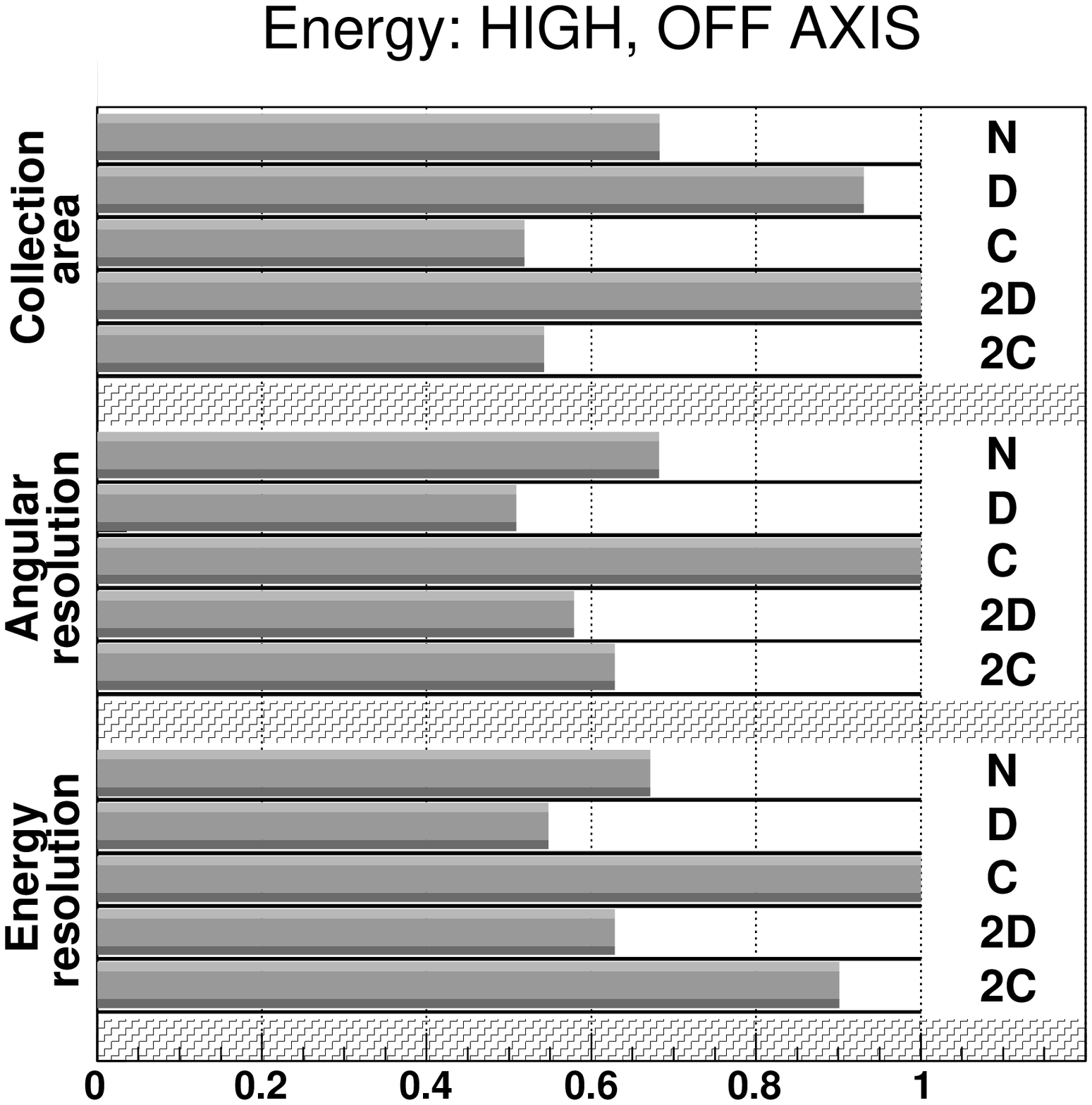}}
  \scalebox{0.95}{%
    \includegraphics[trim = 3mm 2mm 2mm 10mm, clip, width=4.9cm,height=2.0cm]{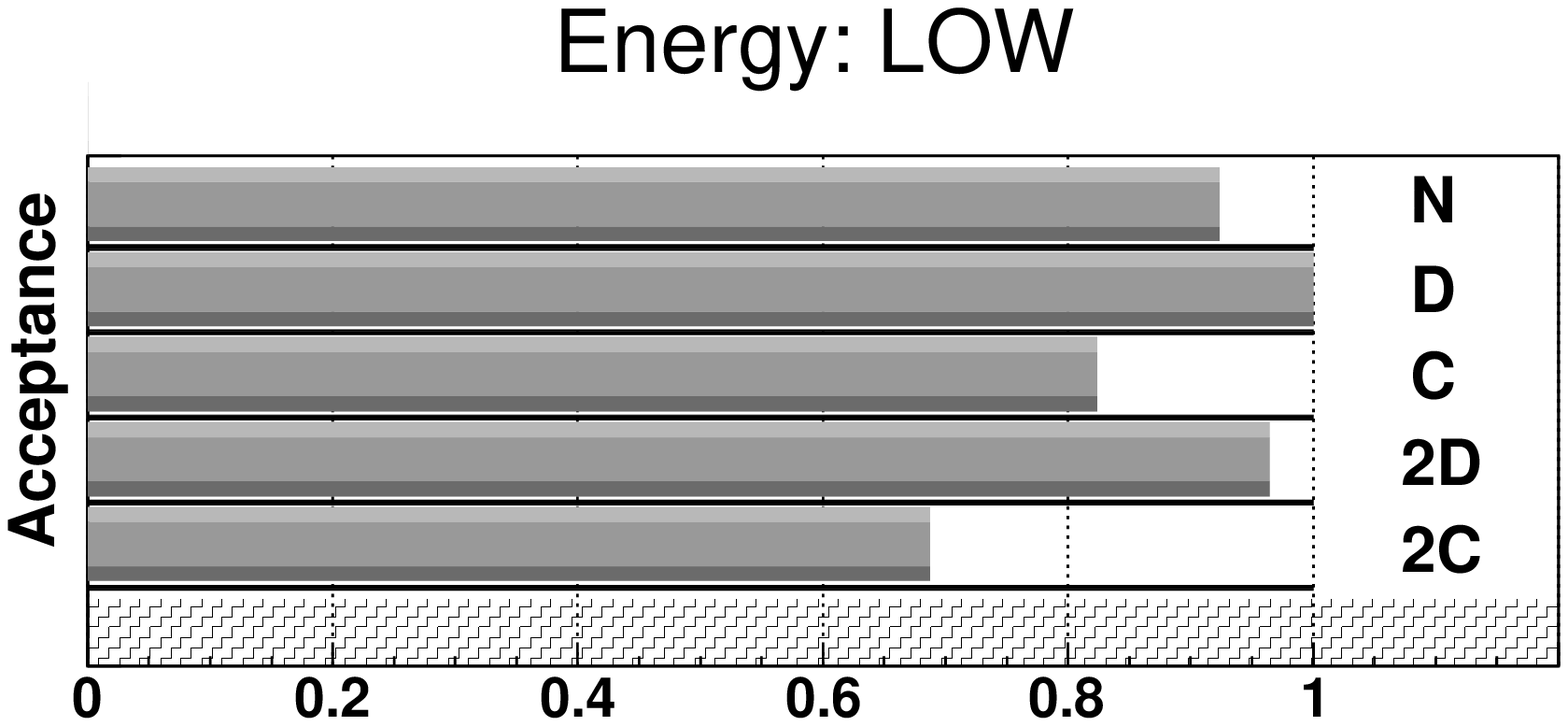}}%
  \scalebox{0.95}{%
    \includegraphics[trim = 3mm 2mm 2mm 10mm, clip, width=4.9cm,height=2.0cm]{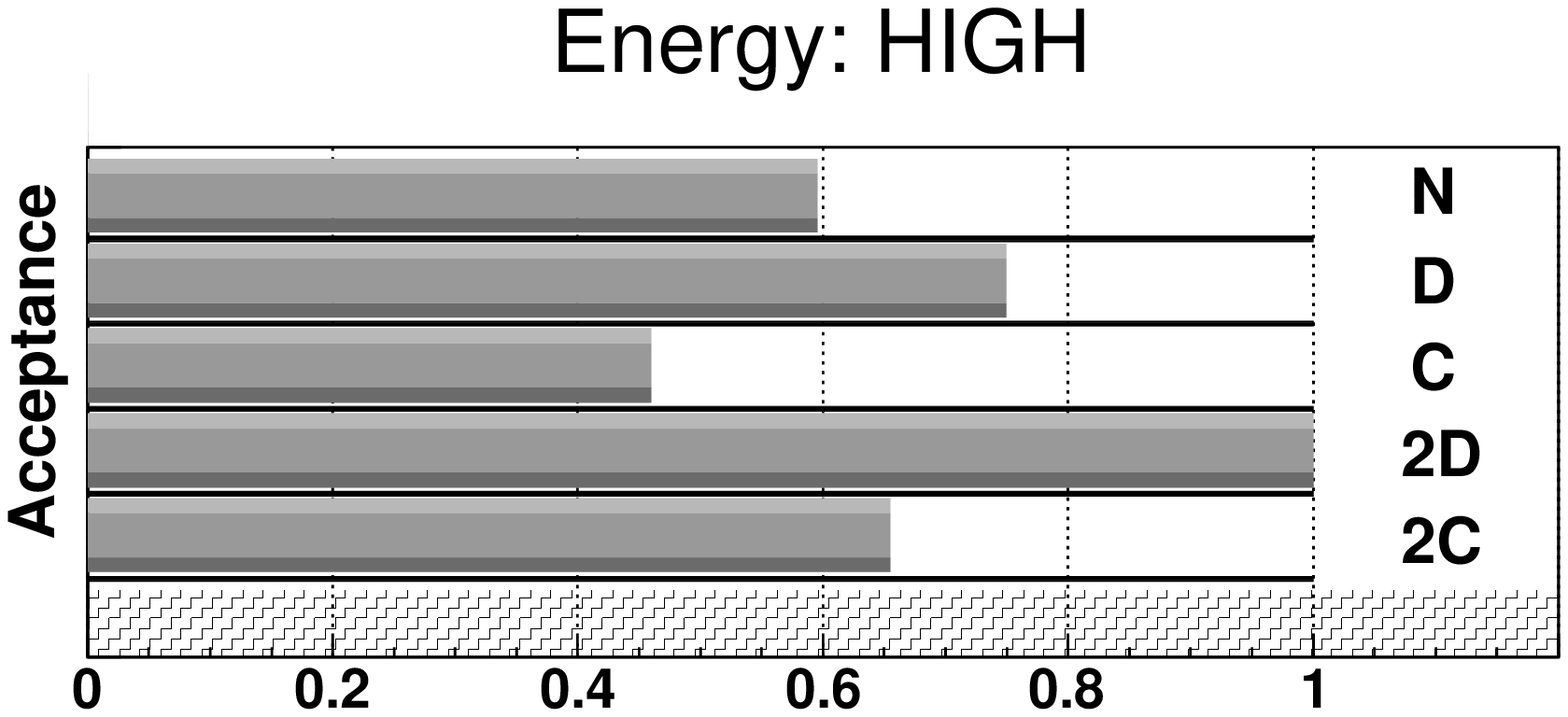}}

\caption{Summary of performance parameters for the five modes at the trigger level for $E=70$ GeV (left column) and 3 TeV (right column). The top (on-axis) panels are for $\phi_{\rm off}=0^{\circ}$ and the middle (off-axis) for $\phi_{\rm off} = 4.9^{\circ}$. For each parameter, $x$, its best value in the five modes, $x_{\rm b}$, is used to scale the parameter values so the charts show $x/x_{\rm b}$ for the collection area and acceptance and $x_{\rm b}/x$ for the angular and energy resolution.}
\label{fig:summit}
\end{center}
\end{figure}

\begin{figure}[t!]
\begin{center}
  \scalebox{0.95}{%
    \includegraphics[trim = 3mm 2mm 2mm 0mm, clip, width=4.9cm,height=4.9cm]{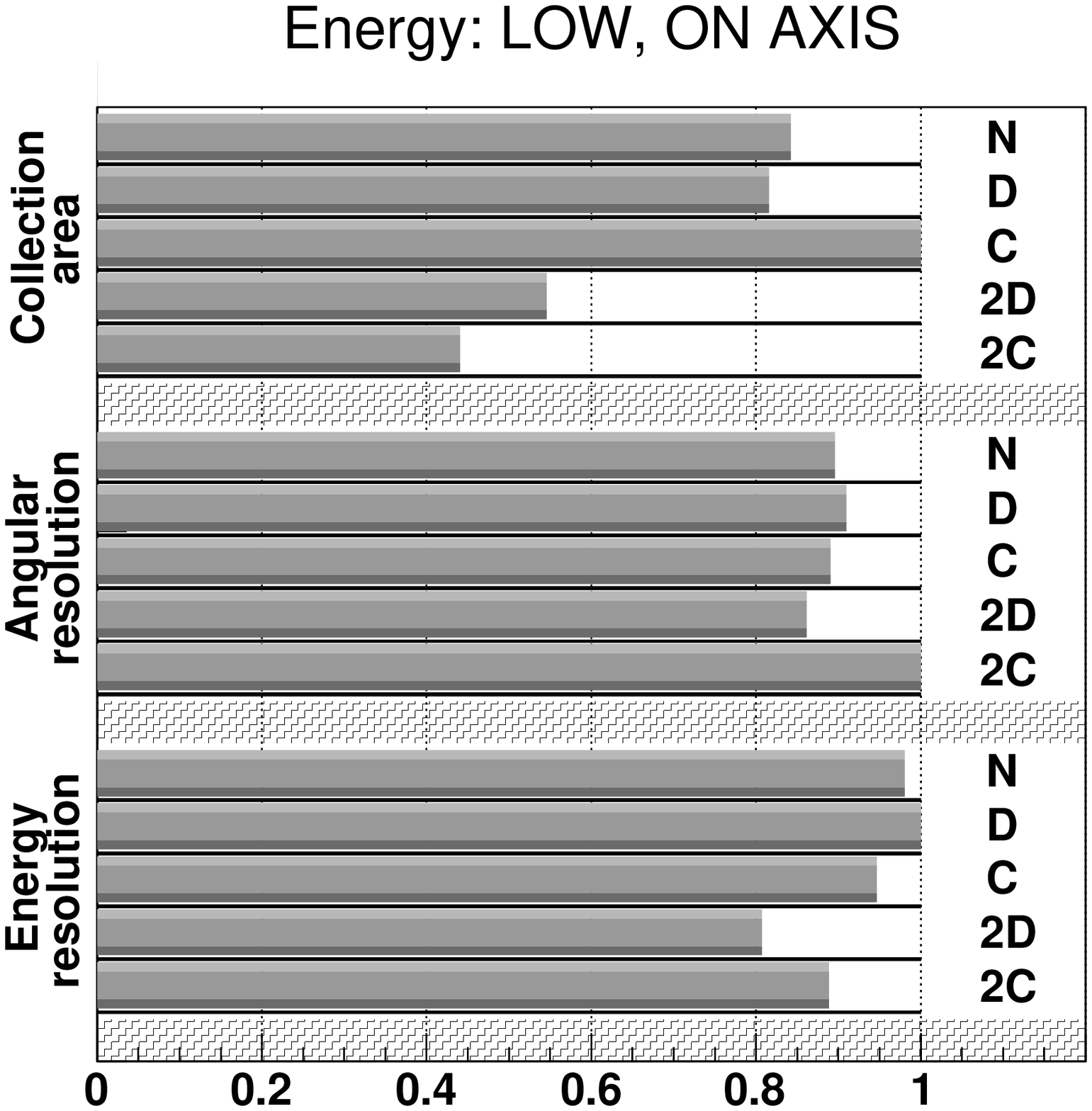}}%
  \scalebox{0.95}{%
    \includegraphics[trim = 3mm 2mm 2mm 0mm, clip, width=4.9cm,height=4.9cm]{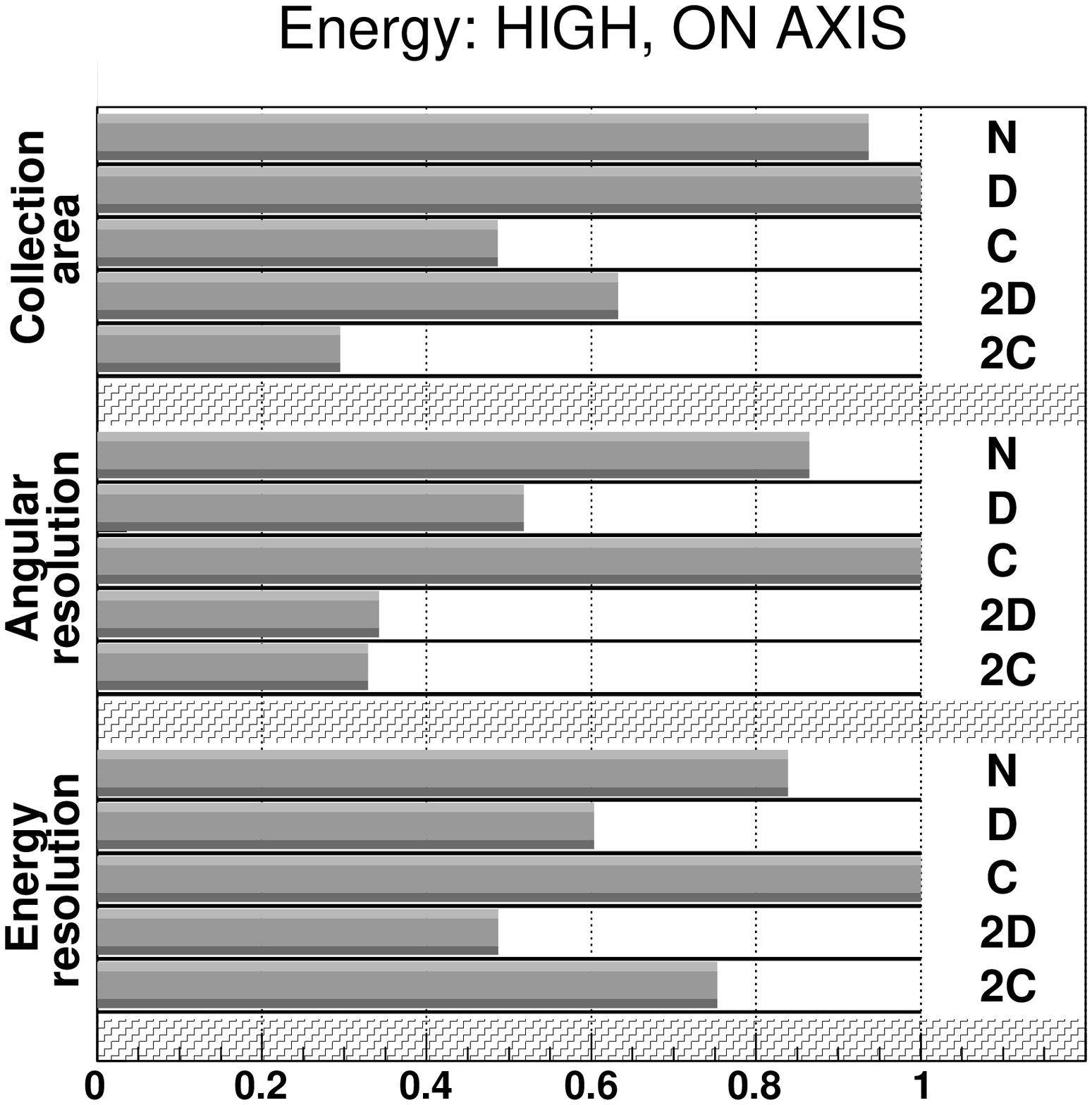}}
  \scalebox{0.95}{%
    \includegraphics[trim = 3mm 2mm 2mm 0mm, clip, width=4.9cm,height=4.9cm]{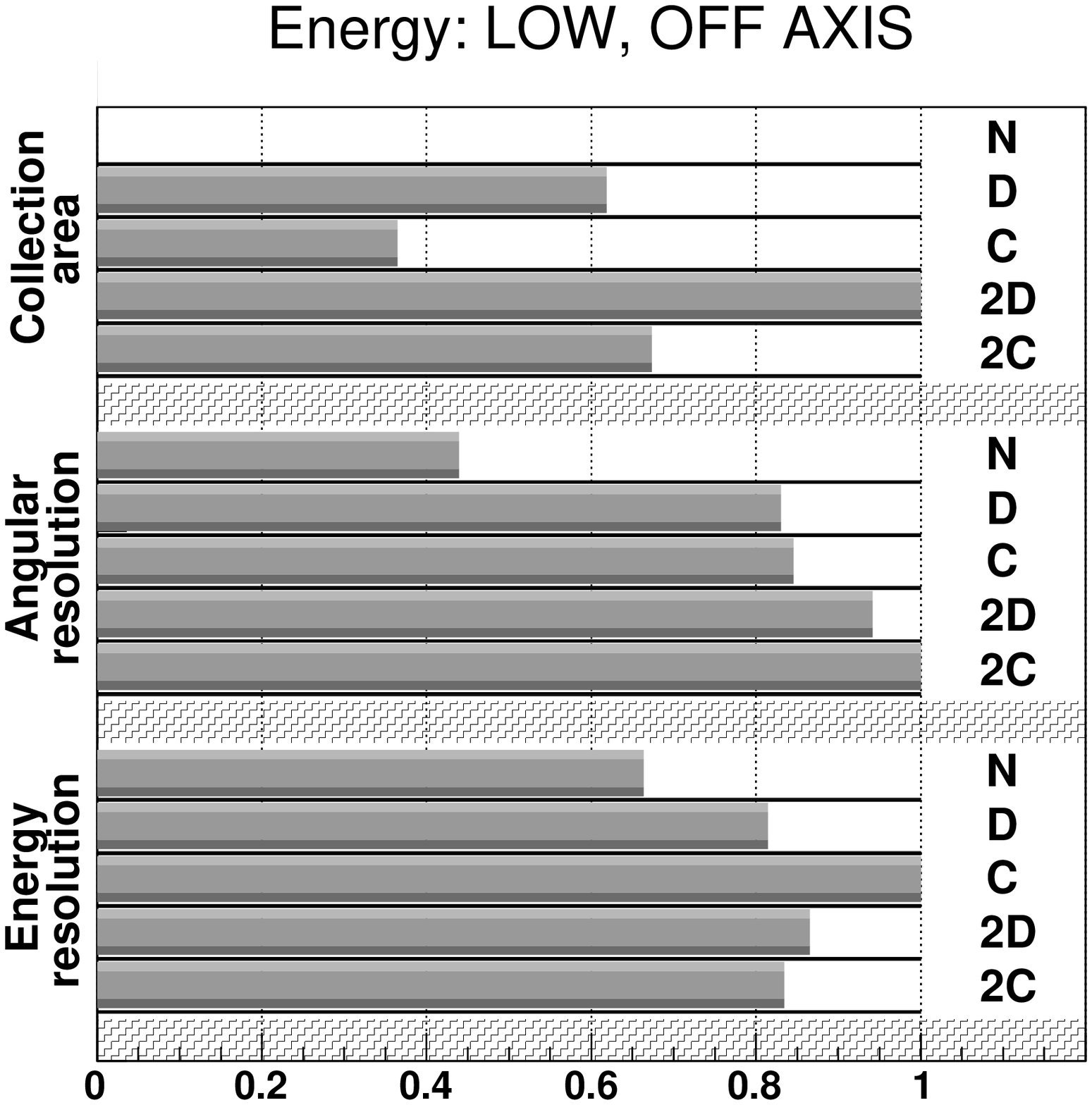}}%
  \scalebox{0.95}{%
    \includegraphics[trim = 3mm 2mm 2mm 0mm, clip, width=4.9cm,height=4.9cm]{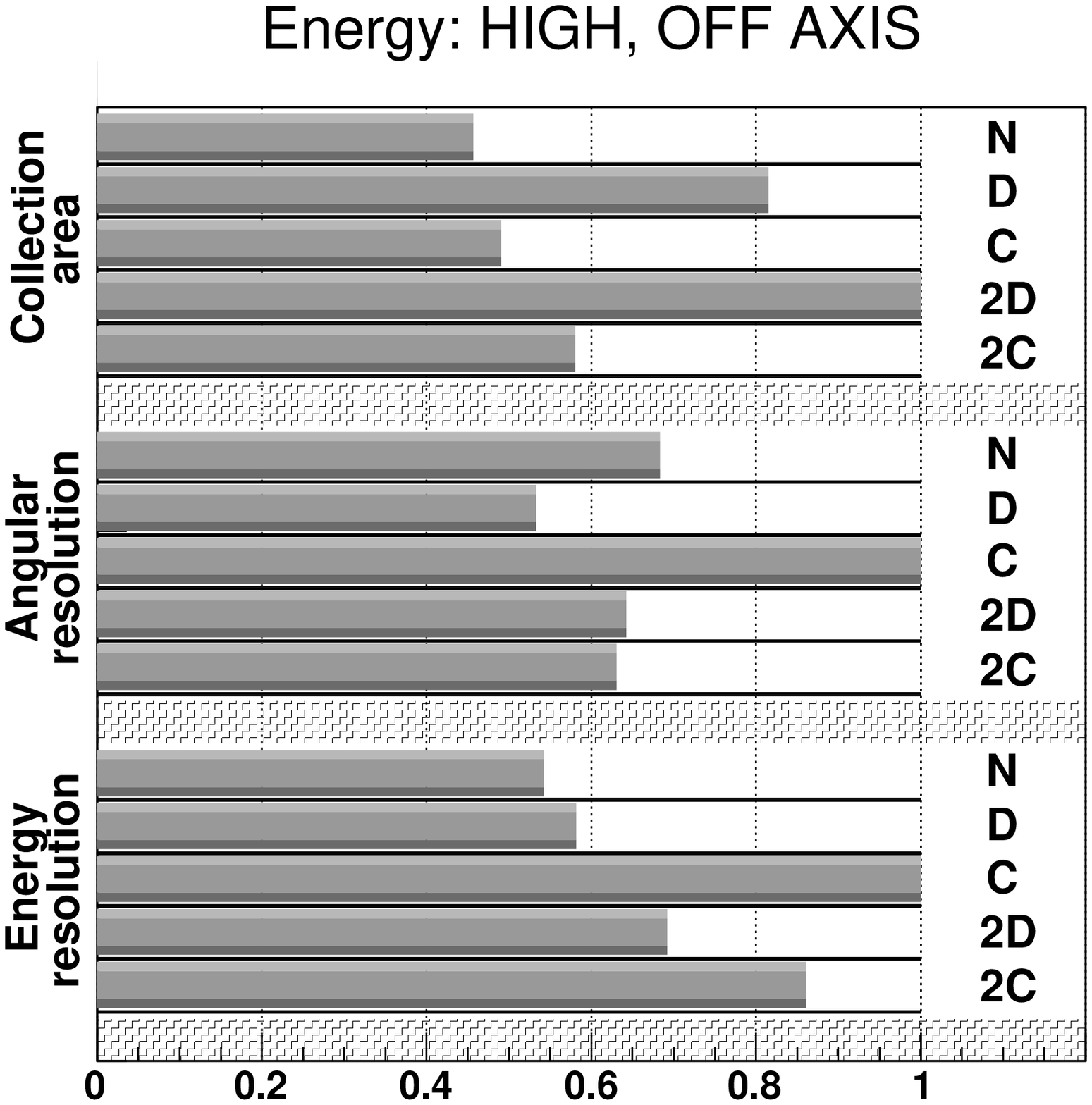}}
  \scalebox{0.95}{%
    \includegraphics[trim = 3mm 2mm 2mm 10mm, clip, width=4.9cm,height=2.0cm]{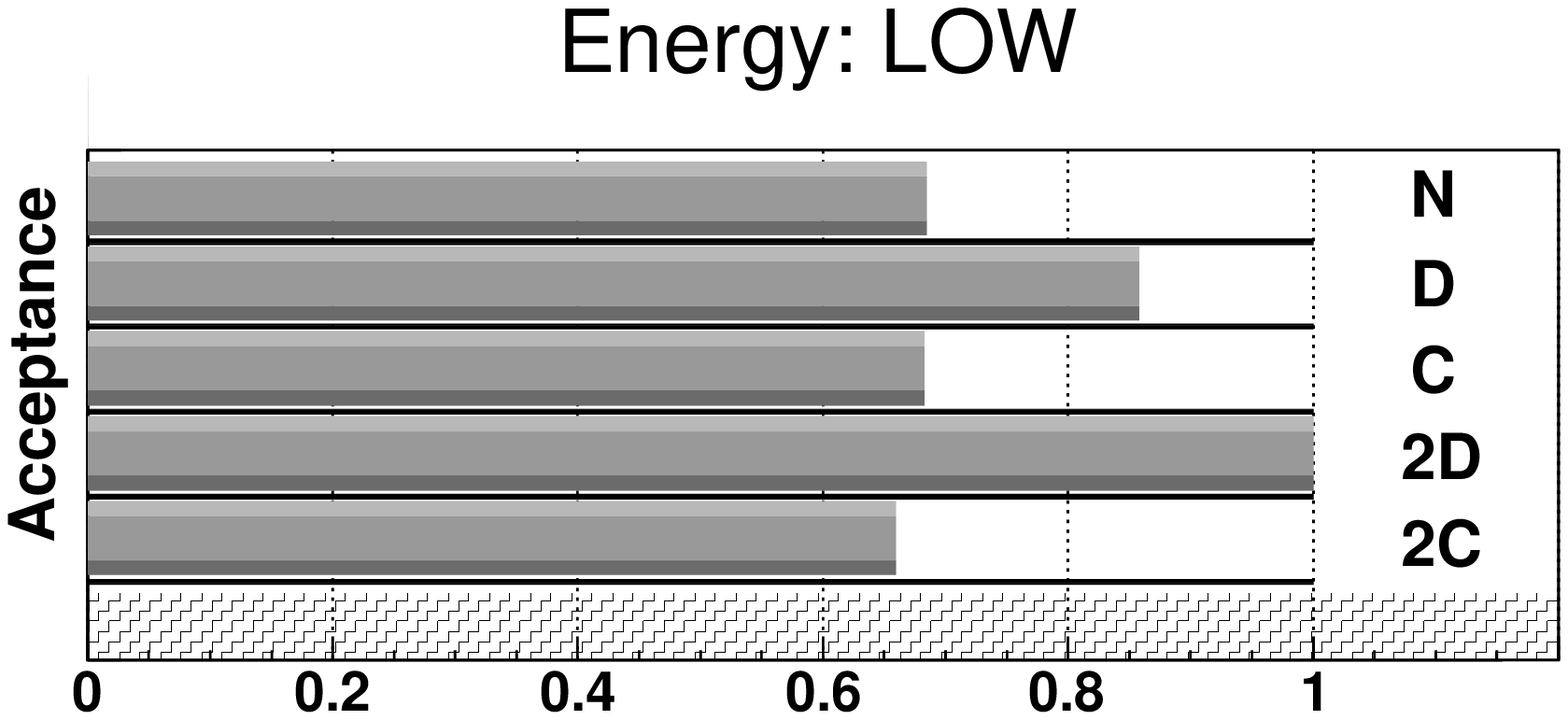}}%
  \scalebox{0.95}{%
    \includegraphics[trim = 3mm 2mm 2mm 10mm, clip, width=4.9cm,height=2.0cm]{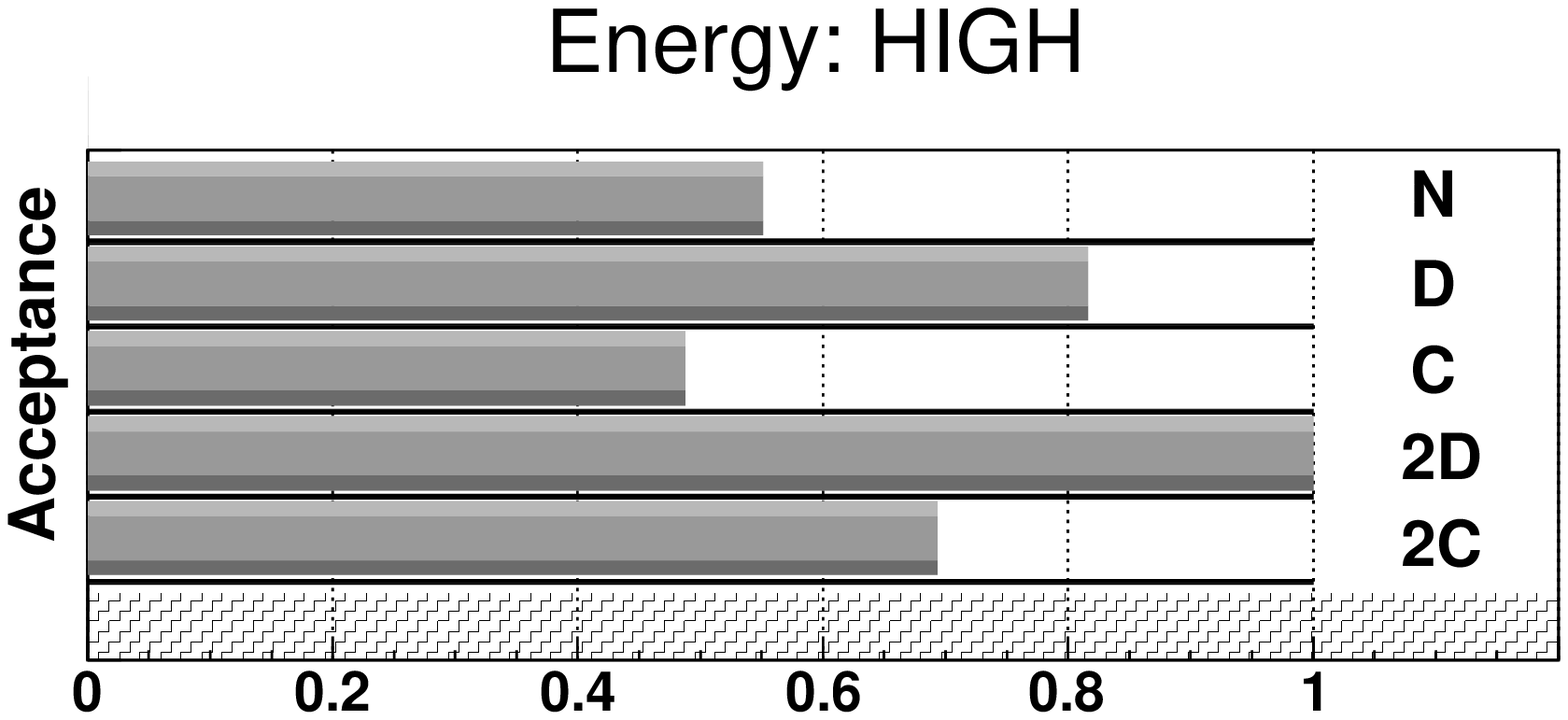}}
\caption{Summary of performance parameters for the five modes at the analysis level for $E=150$ GeV (left column) and 3 TeV (right column). The top (on-axis) panels are for $\phi_{\rm off} = 0^{\circ}$ and the middle (off-axis) for $\phi_{\rm off} = 4.9^{\circ}$. Values of parameters are scaled by the best value, as in Fig.\ \ref{fig:summit}.}
\label{fig:summit_cuts}
\end{center}
\end{figure}

As a summary of our detailed results, 
charts in Figs.\ \ref{fig:summit} and \ref{fig:summit_cuts}
compare the performance parameters for the five modes at the trigger and analysis levels.

The parameters are given for low and high energies; for the latter $E = 3\;{\rm TeV}$ is used at both levels. For low energies, we choose $E$ around the energy threshold, which is different at both levels and then we use $E = 70\;{\rm GeV}$ for the trigger level (Fig.\ \ref{fig:summit}) and $E = 150\;{\rm GeV}$ for the analysis level (Fig.\ \ref{fig:summit_cuts}).

 Dependence on the source position with respect to the axis of the array is illustrated by using two values of $\phi_{\rm off}=0^{\circ}$ and
  $4.9^{\circ}$. The former value is less important for the sky scans, because small  $\phi_{\rm off}$ contribute negligibly to ${\rm Acc}$, however, it allows to illustrate the performance of different modes in deep observations on-source. The latter value is crucial for sky scans, as the largest contribution to  ${\rm Acc}$ at high energies comes from $\phi_{\rm off} \simeq (4-6)^{\circ}$ (except for mode N) and at low energies in modes 2D and 2C from $\phi_{\rm off} \simeq (3-5)^{\circ}$.
For an easier comparison, 
for each value of $E$ and $\phi_{\rm off}$ the best value of a given parameter among the five modes is used to scale the parameters of all modes, so that the best value of each parameter corresponds to 1.

As we can see, in both ranges of $E$ the largest collection areas at large  $\phi_{\rm off}$ correspond to mode 2D, which property is reflected also in the largest values of ${\rm Acc}$ for this mode.  Mode D has a similarly large $A(\phi_{\rm off}=4.9^{\circ})$ at the trigger level, however, it has a less efficient gamma/hadron separation at low energies, leading to a larger loss of gamma-ray events, and then at the analysis level its parameters are worse than in 2D. Also at all energies 2D has a larger collection area at large $\phi_{\rm off}$, $\sim 6^{\circ}$, which results in larger Acc in this mode. For modes N, C and 2C, collection areas at large  $\phi_{\rm off}$, as well as acceptance values,  are much lower which disfavors these modes for sky scans. This conclusion is further confirmed by the mean sensitivities of scans (Section 4.2.4), which indicate the best scan performance in modes D and 2D and, at intermediate energies, the overall best performance of mode 2D.

In contrast, the parameters for $\phi_{\rm off}=0^{\circ}$ favor mode C for on-source observation. This mode has the best angular and energy resolution at high energies, and at low energies its resolutions are similar to other modes. In addition, for the energies around the analysis energy threshold ($\approx 150\; {\rm GeV}$) mode C offers the less biased energy reconstruction (see Fig.~\ref{fig:energy_res}c).  It also has the largest collection area for low $E$ and, although at high $E$ it has a small $A$, the rejection of background events is very efficient in this mode for $\phi_{\rm off}=0^{\circ}$ and then for on-axis observations  mode C typically gives  the best sensitivities.

The standard mode N appears disfavored for the types of observation studied in this work. Nevertheless we do not regard the N mode as an entirely inefficient. In the wobble mode of observation, in which the telescopes are slightly offset from the source position, the N mode might appears as an optimal configuration for a particular offsets.  Somewhat speculatively we note also that for the sky surveys using the full CTA, i.e.\ including 
also the  LST subarray (4 telescopes planned in CTA array E) and the small sized telescopes subarray (SST; 32 telescopes in array E), a configuration of each subarray in a different mode can be considered. Namely, the 2D mode, offering the most efficient detection of sources, has a rather poor reconstruction performance, especially at high energies. Then, to allow for more accurate spectral and morphological studies of the discovered high-energy sources, the SST subarray could be set in the C mode, which at high $E$ has the best reconstruction at any $\phi_{\rm off}$. An assessment of the feasibility of such a strategy obviously requires explicit computations.

\section*{Acknowledgements}
\label{sec:acknowledgments}
\noindent We thank anonymous Referee of the Astroparticle Physics Journal for useful suggestions concerning this work.\\
We thank Konrad Bernl\"ohr for the review of this work and Jean-Pierre Ernenwein and Abelardo Moralejo (internal referees of the CTA Collaboration) for many useful comments.\\

\noindent This work was supported by the NCBiR grant ERA-NET-ASPERA/01/10 and NCN grant UMO-2011/01/M/ST9/01891.

\bibliographystyle{elsarticle-num}
\bibliography{sky_cta}

\end{document}